\documentclass[a4paper,12pt]{article}
\pdfoutput=1
\usepackage{feynmp-auto,expdlist}
\usepackage{amsmath, amsfonts, amssymb}
\usepackage{graphicx}
\usepackage{enumerate}
\usepackage{caption}
\usepackage{latexsym}
\usepackage{hepnicenames}
\usepackage{enumerate}
\usepackage[utf8]{inputenc}
\usepackage{soul}
\usepackage[normalem]{ulem}
\usepackage{bbold}
\usepackage{wasysym} 
\usepackage{makecell}
\RequirePackage[ colorlinks=true,urlcolor=blue,anchorcolor=blue,citecolor=blue,filecolor=blue,
               linkcolor=blue,menucolor=blue,linktocpage=true,pdfproducer=medialab,pagebackref=true]{hyperref}
 \hypersetup{
 colorlinks   = true, 
 urlcolor     = PineGreen, 
 linkcolor    = blue, 
  citecolor   = blue 
}

\font\tenrsfs=rsfs10 at 12pt
\font\sevenrsfs=rsfs7
\font\fiversfs=rsfs5
\newfam\rsfsfam
\textfont\rsfsfam=\tenrsfs
\scriptfont\rsfsfam=\sevenrsfs
\scriptscriptfont\rsfsfam=\fiversfs

\numberwithin{equation}{section}

\usepackage{mathrsfs}
\usepackage{braket}
\usepackage{titling}
\usepackage{amsmath}
\usepackage{caption}
\usepackage{subcaption}
\usepackage{slashed}
\usepackage{amssymb}
\usepackage{epsfig}
\usepackage{graphicx}
\usepackage{verbatim}
\usepackage{color}
\usepackage{rotating}
\usepackage[margin=1.in]{geometry}
\usepackage[table,xcdraw,dvipsnames]{xcolor}
\usepackage[compress,numbers,sort]{natbib}
\usepackage{colortbl}
\usepackage{pdflscape}
\usepackage{color}
\usepackage{mathtools}
\usepackage{colortbl}
\definecolor{Gray}{gray}{0.95}
\definecolor{RGray}{gray}{0.85}
\definecolor{CGray}{gray}{0.92}

\newcommand{\M}{{\cal M}}

\definecolor{nicered}{rgb}{0.7,0.1,0.1}
\definecolor{nicegreen}{rgb}{0.1,0.5,0.1}
\definecolor{red}{rgb}{1.0, 0, 0}
\definecolor{niceblue}{rgb}{0,0,0.8}
\definecolor{red}{rgb}{1.0, 0, 0}
\allowdisplaybreaks

\definecolor{rosso}{cmyk}{0,1,1,0.4}
\definecolor{rossos}{cmyk}{0,1,1,0.55}
\definecolor{rossoc}{cmyk}{0,1,1,0.2}
\definecolor{blu}{cmyk}{1,1,0,0.3}
\definecolor{blus}{cmyk}{1,1,0,0.6}
\definecolor{bluc}{cmyk}{1,1,0,0.1}
\definecolor{verde}{cmyk}{0.92,0,0.59,0.25}
\definecolor{verdec}{cmyk}{0.92,0,0.59,0.15}
\definecolor{verdes}{cmyk}{0.92,0,0.59,0.4}

\definecolor{orangeQCD}{rgb}{0.9725490196078431,0.6274509803921569,0.49411764705882355}
\definecolor{colZN}{rgb}{0.968735,0.465621,0.212103}
\definecolor{greenAstro}{rgb}{0.000000,0.427451,0.172549}
\definecolor{greenAstroL}{rgb}{0.0, 0.66, 0.42}
\definecolor{blueEDM}{rgb}{0.031373,0.317647,0.611765}
\definecolor{BlueC0}{rgb}{0.000000,0.447059,0.698039}
\definecolor{GreenC1}{rgb}{0.000000,0.619608,0.450980}
\definecolor{OrangeC2}{rgb}{0.835294,0.368627,0.000000}

\def\det{\mbox{det}\,}

\renewcommand{\bar}{\overline}

\newcommand{\beq}{\begin{equation}}
\newcommand{\eeq}{\end{equation}}
\newcommand{\bea}{\begin{eqnarray}}
\newcommand{\eea}{\end{eqnarray}}

\renewcommand{\[}{\left[}

\usepackage[capitalise]{cleveref}
\usepackage{nicefrac}
\usepackage{subcaption}\small 
\newcommand{\email}[1]{\href{mailto:#1}{\tt #1}}

\begin{document}
\vspace{1.5cm}
 
{\flushright
{DESY-23-140} \\
\hfill 
}

\begin{center}
{\Large\LARGE\Huge\bf\color{blus} 
Inflation and Higgs Phenomenology in a Model Unifying the DFSZ Axion with the Majoron}\\[1cm]
{\bf Michael Matlis$^a$, Juhi Dutta$^b$, Gudrid Moortgat-Pick$^{a,c}$, Andreas Ringwald$^{a}$}\\[7mm]

{\it $^a$Deutsches Elektronen-Synchrotron DESY, \\ 
Notkestr.\ 85, 22607 Hamburg, Germany}\\[1mm]
{\it $^b$ Homer L. Dodge Department of Physics and Astronomy,
University of Oklahoma, \\ Norman, OK 73019, USA}\\[1mm]
{\it $^c$II. Institut für Theoretische Physik,
Universit\"at Hamburg, \\ 
Luruper Chaussee 149, 22761 Hamburg, Germany}\\[1mm]

\vspace{0.5cm}
\begin{abstract}
The Two-Higgs-Doublet-Standard Model-Axion-Seesaw-Higgs-Portal inflation (2hdSMASH) model consisting of two Higgs doublets, a Standard Model (SM) singlet complex scalar and three SM singlet right-handed neutrinos can embed axion dark matter, neutrino masses and address inflation. We report on an investigation of the inflationary aspects of 2hdSMASH and its subsequent impact on low energy phenomenology. In particular, we identify inflationary directions for which the parameter values required for successful inflation do not violate perturbative unitarity and boundedness-from-below conditions. 
By analyzing the renormalization-group flow of the parameters we identify the necessary and sufficient constraints for running all parameters perturbatively and maintaining stability from the electroweak to the PLANCK scale. We observe that stringent constraints arise on the singlet scalar self coupling from inflationary constraints, i.e, $\lambda_S\sim 10^{-10}$. Further, we find that all theoretical and experimental constraints are satisfied if the portal couplings are typically in the range $(\frac{v}{v_S})$ and $(\frac{v}{v_S})^2$ (where $v, v_S$ refer to the electroweak and singlet scalar vacuum expectation value respectively). As a consequence, inflation is realized in a variety of field space directions in the effective single field regime.  Finally we  provide testable benchmark scenarios at colliders.  

\end{abstract}

\begin{minipage}[l]{.9\textwidth}
{\footnotesize \vspace{1cm}
\begin{center}
\textit{E-mail:} 
\email{michael.maxim.matlis@gmx.de}\,,
\email{juhi.dutta@ou.edu}\,, 
\email{gudrid.moortgat-pick@desy.de}\,, 
\email{andreas.ringwald@desy.de}
\end{center}}
\end{minipage} 
\thispagestyle{empty}
\bigskip

\end{center}

\setcounter{footnote}{0}

\newpage
\tableofcontents


\section{Introduction}
\label{sec:introduction}
The discovery 
by the ATLAS and CMS collaborations in the year 2012
of a new particle 
that is, within the current theoretical and experimental uncertainties,  consistent with predictions of the Higgs boson in the Standard Model (SM) of particle physics 
is by no means a completion of the SM. In fact, 
the discovered Higgs particle is still compatible with a wide range of Beyond Standard Model (BSM) physics models that are addressing in addition 
a number of facts for which the SM is lacking an explanation:  {\em i)} the non-observation of strong CP-violation, {\em ii)} the evidence for non-baryonic dark matter (DM), {\em iii)} the evidence for neutrino masses and mixing, {\em iv)} the evidence for a baryon asymmetry of the universe (BAU), and {\em v)} the circumstantial evidence for an era of inflationary expansion of the universe before the hot thermal radiation dominated era.

Remarkably, all five shortcomings can be solved in one smash~\cite{Ballesteros:2016euj,Ballesteros:2016xej} in a minimal extension of the SM by three SM singlet right-handed neutrinos and a Kim-Shifman-Vainshtein-Zakharov (KSZV) type axion model~\cite{Kim:1979if,Shifman:1979if} involving an exotic quark and a complex SM singlet scalar. The model, dubbed SM-Axion-Seesaw-Higgs Portal inflation (SMASH), 
solves {\em i)} by the Peccei-Quinn (PQ) mechanism~\cite{Peccei:1977hh}, {\em ii)} by axion dark matter~\cite{Preskill:1982cy,Abbott:1982af,Dine:1982ah}, {\em iii)} by the seesaw mechanism~\cite{Minkowski:1977sc,Gell-Mann:1979vob,Yanagida:1979as,Mohapatra:1979ia},  {\em iv)} by thermal leptogenesis~\cite{Fukugita:1986hr}, and {\em v)} by Higgs portal inflation~\cite{Bezrukov:2007ep,Lebedev:2011aq}. 
SMASH is very predictive and constitutes a kind of minimal, but complete model of particle physics, up to the PLANCK scale, and of cosmology, back to inflation.

Possible variants of SMASH have been discussed superficially in Ref.~\cite{Ballesteros:2019tvf}. In this paper, we concentrate on one of them, namely,  the Two-Higgs-Doublet-SM-Axion-Seesaw-Higgs-Portal inflation (2hdSMASH)~\cite{Matlis:2022iwr} model\footnote{For some supersymmetric models addressing inflation see~\cite{Choudhury:2011jt, Hollik:2020plc}.}.
The 2hdSMASH model consists of two Higgs doublets, a complex SM singlet PQ scalar and three SM singlet right-handed Majorana neutrinos. 
It differs from SMASH by replacing in the latter the KSVZ axion model by a Dine-Fischler-Srednicki-Zhitnitsky (DFSZ) axion model~\cite{Zhitnitsky:1980tq,Dine:1981rt} and does not feature an exotic quark. 
The field content of the 2hdSMASH model was originally proposed in refs.~\cite{Volkas:1988cm,Clarke:2015bea} in order to explain neutrino masses, baryogenesis, the strong CP problem, and dark matter. It has been dubbed `Neutrino-DFSZ' ($\nu$DFSZ) model in Ref.~\cite{Clarke:2015bea} where the authors discuss the high scale validity and the technical naturalness of the model, while  Ref.~\cite{Espriu:2015mfa} focuses on the low energy phenomenology of the Higgs sector but does not account for the impact of inflationary constraints on the model nor does it involve high scale validity of the model. In this paper,  we combine both directions to study the impact of inflationary  constraints on low-energy Higgs phenomenology.

Recent work, Ref.~\cite{Sopov:2022bog}, also extends the $\nu$DFSZ model to address inflationary directions. Our work has several differences from theirs. Ref.~\cite{Sopov:2022bog} addresses inflation in a similar model (VISH$\nu$) but with a trilinear U(1) breaking term for non-minimal couplings in the large regime ($\xi_1, \xi_2 \gg 1$) and hierarchical regime ($\xi_S \gg \xi_1, \xi_2$) while neglecting the U(1) breaking coupling.  It also addresses the domain wall problem by having only the top quark carrying PQ charge, resulting in a domain wall number 
$N_{DW}$ = 1, rendering domain walls cosmologically unstable.  
In our work  on 2hdSMASH, we discuss the inflationary directions in the presence of a quartic U(1) breaking term ($\nu$DFSZ) consistent with all theoretical constraints (including perturbative unitarity, boundedness-from-below, high scale  validity). 
Our main focus is on the hierarchical regime including the effect of small portal couplings ($\lambda_{1S}, \lambda_{2S}$ and $\lambda_{12S}$) thereby allowing for inflation in the  effective singlet field directions including singlet and singlet-doublet directions as discussed in sec.\ref{sec:PQ-and mixed inflation} (see Ref.~\cite{Matlis:2022iwr} for more details). 
For completeness,  the 2HDM directions are discussed in the appendix~\ref{app:2hdm inflation}. We address the domain wall issue by assuming that the Peccei-Quinn symmetry arises as an accidental symmetry from an exact discrete $Z_N$ symmetry for large $N\geq 9$. This in turn also addresses the axion quality problem (see appendix~\ref{app:axion_bounds}).

Further, we also investigate the connection between inflation its impact on BAU, thermal leptogenesis and its influence on the low-energy Higgs phenomenology and provide suitable benchmark points for future collider studies.
While Ref.~\cite{Sopov:2022bog} also deals with a top-specific model where the second Higgs doublet couples only to the  top quarks,  we make no such specifications but focus on the phenomenology of the full flavour sector and no FCNC decays are allowed by construction at tree-level from the $Z_2$ conserving Type II 2HDM-like Higgs potential structure. Therefore, 2hdSMASH \cite{Matlis:2022iwr} yields another viable complete model for particle physics and cosmology, complementary to SMASH and VISH$\nu$. 

The paper is structured as follows: in section \ref{sec:model} and \ref{sec:mass_spec} we introduce the field content of 2hdSMASH, its theoretical constraints and its scalar mass spectrum. In section \ref{sec:inflation} we describe inflation in 2hdSMASH and its theoretical predictions for inflationary observables. In section \ref{sec:infl_implications_portal_couplings} we discuss the connection of the inflationary epoch with particle phenomenology by investigating whether all relevant constraints are fulfilled to accommodate successful inflation while remaining perturbative and stable. In particular, we discuss in this context 
the stability of the renormalization-group-flow of the PQ-scalar self-coupling $\lambda_{S}$ 
and its implications 
for successful inflation, generation of the BAU via thermal leptogenesis, the interconnection of portal couplings, 
stability analysis on the effective low energy theory and benchmark points acquired from the accumulated analysis. In section \ref{sec:allowedhiggs} we present the parameter space regions allowed by both theoretical and experimental constraints and summarise our main conclusions in section \ref{sec:conclusions}.

\section{\boldmath The 2hdSMASH Model}
\label{sec:model}
 
In this section, we introduce the 2hdSMASH model by specifying its field content, its theoretical constraints and its scalar mass spectrum. 

\subsection{\boldmath Field content and PQ symmetry}
\label{subsec:fieldcontent}
In the 2hdSMASH model, one adds to the SM three right-handed neutrinos $N_{R,i}$ 
and 
a complex SM-singlet scalar field $S$. Furthermore, one extends the Higgs sector to a Type-II 2HDM, involving two $SU(2)_L$ Higgs doublets $\Phi_i$, $i=1,2$,  and imposes a global $U(1)_{\text{PQ}}$-symmetry, 
whose charges are distributed among the fields as shown in table~\ref{TabPQ}.
\begin{table}[t]
\begin{center}
 \begin{tabular}{|c||c|c|c|c|c|c|c|c|c|}
   \hline\hline
  Field & $S$ & $\Phi_1$ & $\Phi_2$ & $q_L$ & $u_R$ & $d_R$ & $l_L$ & $N_R$ & $e_R$ \\
   \hline
  Charge & $X_S$ & $X_1$ & $X_2$ & $X_q$ & $X_u$ & $X_d$ & $X_l$ & $X_N$ & $X_e$ \\
   \hline
Value & $1$ & $-\frac{2x}{x-1}$ & $\frac{2}{1-x}$ & $0$ & $\frac{2}{1-x}$ & $\frac{2x}{x-1}$ & $\frac{3}{2}-\frac{2}{1-x}$ & -$\frac{1}{2}$ & $\frac{7}{2} -\frac{4}{1-x}$ \\
   \hline\hline
 \end{tabular}
\end{center}
 \caption{Charges of fields under the PQ symmetry~\cite{Clarke:2015bea}. Here $x\neq 1$.
}
 \label{TabPQ}
\end{table}
The most general renormalizable scalar potential invariant under this symmetry is then given by~\cite{Clarke:2015bea}:
\begin{align}
 V(\Phi_1,\Phi_2,S) =&\,
  M^2_{11}\, \Phi_1^\dagger \Phi_1
  + M^2_{22}\, \Phi_2^\dagger \Phi_2 
  +M_{SS}^2\, S^*S
  \nonumber\\ 
&
  + \frac{\lambda_1}{2} \left( \Phi_1^\dagger \Phi_1 \right)^2
  + \frac{\lambda_2}{2} \left( \Phi_2^\dagger \Phi_2 \right)^2
  + \frac{\lambda_S}{2}\left(S^*S\right)^2
  \nonumber\\ 
&
  + \lambda_3\, \left(\Phi_1^\dagger \Phi_1\right) \left(\Phi_2^\dagger \Phi_2\right)
  + \lambda_4\, \left(\Phi_1^\dagger \Phi_2\right) \left(\Phi_2^\dagger \Phi_1\right)
  \nonumber\\   
&   +\lambda_{1S}\left(\Phi_1^\dagger\Phi_1\right)\left(S^*S\right)
        +\lambda_{2S}\left(\Phi_2^\dagger\Phi_2\right)\left(S^*S\right)-\lambda_{12S}\left(\Phi_2^\dagger\Phi_1 S^2+h.c.\right)\,,
\label{EqnuDFSZ}
\end{align}
while 
the most general Yukawa interactions of the fermions in the model read~\cite{Clarke:2015bea}: 
\begin{equation}
  -\mathcal{L}_Y =  Y_u\overline{q_L}\tilde\Phi_2 u_R + Y_d\overline{q_L}\Phi_1 d_R 
    + Y_e \overline{l_L}\Phi_1 e_R + Y_\nu \overline{l_L}\tilde\Phi_1 N_R 
    + \frac12 y_N \overline{(N_R)^c}S N_R + {\rm h.c.}\,,
\label{EqYuk}
\end{equation}
where family indices are implied and $\tilde{\Phi}_i\equiv i\tau_2\Phi_i^*$.
These Yukawa interactions 
correspond to the ones of a Type-II 2HDM extended by Majorana and Dirac neutrino interactions\footnote{Choosing another charge assignment to the right-handed charged leptons, $e_R$, one may also realise a Flipped 2HDM in 2hdSMASH~\cite{Clarke:2015bea}.}.

\subsection{\boldmath Vacuum structure, particle content and masses}
\label{subsec:vacuumstructureandmasses}

It is assumed that the parameters in the scalar potential are such that its minimum is attained at the vacuum expectation values (VEVs) $\langle \Phi_i \rangle \equiv \left(0,v_i/\sqrt{2}\right)^T$ and $\langle S \rangle \equiv v_S/\sqrt{2}$, 
where $\sqrt{v_1^{2}+v_2^{2}} \equiv v\simeq 246$ GeV and 
$v_S\gg v$.

After symmetry breaking, the model features two $SU(2)_L$ doublets and a complex scalar singlet, so a total of $4+4+2=10$ spin-zero particle excitations around the vacuum. Three of these excitations are eaten by the $W^\pm$ and $Z$ bosons 
and $7$ of them are left in the spectrum: two charged Higgses, $H^\pm$, two 
CP-odd states, $a$ and $A$, and three neutral CP-even states, $h$, $H$, and $s$. 
We defer the calculation of their  masses at tree-level  to appendix~\ref{app:deriv_scalar_masses} (see also refs.~\cite{Volkas:1988cm,Espriu:2015mfa}). Here we just quote the results. 

\subsubsection{Mass of charged Higgs bosons}

The tree-level masses of the charged Higgs bosons are obtained as 
\begin{align}
m_{H^{\pm}}^{2}=\frac{1}{2}  \left(\frac{\left(t_{\beta}^2+1\right) \lambda_{12S} }{t_{\beta}}-\frac{\lambda_{4} v^2}{v_{S}^2}\right)\,v_{S}^2\,,\label{mHpm}
\end{align}
where $t_\beta \equiv \tan\beta =v_2/v_1$. 

\subsubsection{Masses of CP-odd scalars}

One of the two CP-odd scalars is massless at tree level. It is the axion, $a$, the Nambu-Goldstone boson from the breaking of the $U(1)_{\rm PQ}$ 
symmetry\footnote{
In 2hdSMASH, the PQ symmetry is at the same time a lepton symmetry (cf. table~\ref{TabPQ}). Correspondingly, the axion is at the same time also the majoron, see also~\cite{Berezhiani:1989fp}.}~\cite{Weinberg:1977ma,Wilczek:1977pj}.  
Non-perturbatively, it  
acquires a small mass through mixing with the neutral pion,    
\begin{equation}
m_a \simeq  \frac{\sqrt{z}}{1+z}\frac{m_\pi f_\pi}{f_a}\simeq 0.57\ {\rm meV} \left( \frac{10^{10}\ {\rm GeV}}{f_a} \right) \,,
\end{equation}
where $z=m_u/m_d$ is the ratio of the masses of the up and the down quark, and $m_\pi$ and $f_\pi$ are the mass and the decay constant of the neutral pion, respectively, and 
$f_a$ is the axion decay constant: 
\begin{align}
f_a
=\dfrac{\sqrt{v_{S}^{2}+4\frac{v_{1}^{2}v_{2}^{2}}{v^{2}}}}{6}\,.
\end{align}
Important constraints on the axion decay constant, $f_a$, and therefore on the PQ scale, 
$v_S\simeq 6 f_a$, come from astrophysics and cosmology. 
A lower bound arises from the measured duration of the neutrino signal of the supernova 1987A and an upper bound from the requirement that there is not too much dark matter. In appendix~\ref{app:axion_bounds} we argue that the PQ scale in 2hdSMASH lies preferably in the range 
\begin{eqnarray}
\nonumber
3.1\times 10^9\,{\rm GeV}\lesssim v_S\lesssim 5.9\times 10^{10}\,{\rm GeV}\,,\      {\rm for}\ \tan\beta \lesssim 0.5\,,\\
5.9\times 10^9\,{\rm GeV}\lesssim v_S\lesssim 5.9\times 10^{10}\,{\rm GeV}\,,\      {\rm for}\ \tan\beta \gtrsim 5\,.
\label{eq:preferred_v_S range}
\end{eqnarray}

The second CP-odd scalar boson, $A$, has the tree-level mass

\begin{align}
m_{A}^{2}=\frac{2\lambda_{12S} }{1+t_{\beta}^{2}}  \left(\frac{\left(1+t_{\beta}^2\right)^{2}}{4 t_{\beta}} +\frac{v^2}{v_{S}^{2}} t_{\beta}\right)\,v_{S}^{2}\,.
\label{mA}
\end{align}
To avoid a tachyonic mass for $m_A$, we require
\begin{align}
\lambda_{12S}\geq0\text{ .}
\end{align}

\subsubsection{Masses of CP-even scalars}

The analytic form of the expression for the masses of the three CP-even scalars is not very enlightening. Here, we give the mass squares, in units of $v_S^2$, as power series in $v/v_{S}\ll 1$: 
\begin{align}
\frac{m_{h}^{2}}{v_S^2}&=  \frac{1}{\left(1+t_{\beta}^2\right)^2} \left[ {\lambda_{1}+ t_{\beta}^4\, \lambda_{2}+2\, t_{\beta}^2\, \lambda_{34}}{}
-\dfrac{\left(\lambda_{1S}+t_{\beta}^2\,\lambda_{2S} -2 t_{\beta} \lambda_{12S} \right)^2}{\lambda_{S}}\right]\left(\frac{v}{v_S}\right)^2     
 + {\mathcal O}\left( \left( \frac{v}{v_S} \right)^4\right)
\,,
\label{mh1 full}\\
\frac{m_{H}^{2}}{v_S^2}&= \dfrac{\left(1+t_{\beta}^2\right)\lambda_{12S}}{2t_{\beta}}
+ \frac{t_{\beta}}{\left(1+t_{\beta}^{2}\right)^{2}}\,\Biggl[ 
 \dfrac{2\left(\left(\lambda_{1S}-\lambda_{2S}\right)t_{\beta}+\lambda_{12S}\left(1-t_{\beta}^{2}\right)\right)^{2}}{\lambda_{12S}\left(1+t_{\beta}^2\right)-2 t_{\beta}\lambda_{S} }
\label{mh2 full}\\
&+\left(\lambda_{1}+\lambda_{2}-2\lambda_{34}\right)t_{\beta}\Biggr]
\left(\frac{v}{v_S}\right)^2
+ {\mathcal O}\left( \left( \frac{v}{v_S} \right)^4\right) \,,\notag\\
\frac{m_s^{2}}{v_S^2}&= \lambda_{S}  + 
\frac{t_{\beta}}{\left(1+t_{\beta}^{2}\right)^{2}}\,\Biggl[\dfrac{\left(\lambda_{1S}+\lambda_{2S}t_{\beta}^{2}-2t_{\beta}\lambda_{12S}\right)^{2}}{\lambda_{S}}
\label{mh3 full}\\
&- \dfrac{2 t_{\beta}^{2}\left(\left(\lambda_{1S}-\lambda_{2S}\right)t_{\beta}+\lambda_{12S}\left(1-t_{\beta}^{2}\right)\right)^{2}}{\lambda_{12S}\left(1+t_{\beta}^2\right)-2 t_{\beta}\lambda_{S} }
\Biggr]\left(\frac{v}{v_S}\right)^2 + {\mathcal O}\left( \left( \frac{v}{v_S} \right)^4\right)\,.
\notag
\end{align}
In the series for $m_{h}^{2}/v_S^2$, the leading contribution, $\mathcal O((v/v_{S})^{0})$, 
vanishes, while the latter is present for $m_{H}^{2}/v_S^2$ and $m_s^{2}/v_S^2$. 
Correspondingly, $m_{h}$ is  $\mathcal O(v)$, and we can associate $h$ with the SM-like Higgs Boson which is constrained by collider searches to $125.25\pm 0.17$~GeV~\cite{Workman:2022ynf}, while the masses of $H$ and $s$ are of order 
$v_S$. We defer the discussion of the characteristics of the scalar mass spectrum in 
2hdSMASH to sec.~\ref{sec:mass_spec}. 

\subsubsection{Masses of neutrinos}

After symmetry breaking, the last two terms in eq. \eqref{EqYuk} give rise to the aforementioned Majorana- and Dirac neutrino mass terms, realizing the Type I 
seesaw mechanism~\cite{Minkowski:1977sc,Gell-Mann:1979vob,Yanagida:1979as,Mohapatra:1979ia}. The neutrino mass matrix reads: 
\begin{align}
M_{\nu}=\begin{pmatrix}
\mathbf{0_{3\times 3}}&M_{D}\\M_{D}^{T}&M_{M}
\end{pmatrix}
=\dfrac{1}{\sqrt{2}}\begin{pmatrix}
\mathbf{0_{3\times 3}}&Y_{\nu}v_{1}\\Y_{\nu}^{T}v_{1}&\frac{1}{2}Y_{N}v_{S}
\end{pmatrix} \,,
\end{align}
where $M_{M}$ represents the Majorana neutrino mass matrix and $M_{D}$ represents the Dirac neutrino mass matrix. The smallness of the masses of the SM active neutrinos is thus 
explained by the hierarchy $v_S\gg v_1$: 
\begin{align}
m_{\nu}=-M_{D}M_{M}^{-1}M_{D}^{T}=-\dfrac{Y_{\nu}Y_{N}^{-1}Y_{\nu}^{T}}{\sqrt{2}}\dfrac{v_{1}^{2}}{v_{S}}\,.
\label{eq:neutrino_mass_matrix}
\end{align}

\subsection{Theoretical constraints}
\label{subsec:theoretical_const}

In this subsection, we give theoretical constraints regarding vacuum stability by considering the Boundedness-from-Below (BfB) conditions and perturbative unitarity conditions.

The BfB conditions guarantee that the scalar potential remains positive in all field directions for large field values. This prevents unphysical lower minima to develop and keeps the vacuum stable. The necessary and sufficient Boundedness-from-Below conditions (BfB) are based on copositivity criteria\footnote{The derivation of the necessary \& sufficient BfB conditions is given in Appendix \ref{app:bfb}.}. We start with the necessary BfB conditions:
\begin{align}
&\lambda_{1}>0~~,~~\lambda_{2}>0~~,~~\lambda_{3}>-\sqrt{\lambda_{1}\lambda_{2}}~~,~~\lambda_{34}>-\sqrt{\lambda_{1}\lambda_{2}}\,,\nonumber\\
&\lambda_{S}>0~~,~~\sqrt{\lambda_{1}\lambda_{S}}>\lambda_{1S}>-
\sqrt{\lambda_{1}\lambda_{S}}~~,~~\sqrt{\lambda_{2}\lambda_{S}}>\lambda_{2S}>-\sqrt{\lambda_{2}\lambda_{S}}\,,\nonumber\\
&\lambda_{S}\lambda_{34}-\lambda_{1S}\lambda_{2S}
+\sqrt{\left(\lambda_{1}\lambda_{S}-\lambda_{1S}^{2}\right)\left(\lambda_{2}\lambda_{S}-\lambda_{2S}^{2}\right)}>0\,,
\label{eq:nec_bfb_conditions} \\
&\lambda_{1S}+\lambda_{2S}>0~~,~~\lambda_{1S}\lambda_{2S}-\lambda_{12S}^{2} >0\,,\nonumber
\end{align}
where $\lambda_{34}\equiv \lambda_3 + \lambda_4$. And the sufficient BfB conditions are given by:
\begin{align}
&\lambda_{1S}>0, ~~\lambda_{2S}>0, ~~\lambda_{1S}\lambda_{2S}-\lambda_{12S}^{2}>0\,.
\end{align}
The sufficient BfB conditions are very restrictive and can be softened. This is done by using the \texttt{Mathematica} package \texttt{BFB} \cite{Ivanov:2018jmz} where the BfB conditions are solved numerically by using the Resultants method. If a parameter point is positive (semi-)definite then it is allowed, otherwise, it is dismissed. This will be useful when considering inflation in section \ref{sec:inflation}.

The perturbativity conditions on the quartic couplings for 2hdSMASH ensure that our theory remains finite and perturbative at low- and high energies. This avoids divergencies that make the theory unpredictable as for any finite quantum field theory. The perturbative unitarity conditions for 2hdSMASH are given by\footnote{The conditions on perturbative unitarity are derived in Appendix \ref{app:perturbative_unitarity}.}:
\begin{align}
&\left|\lambda_{1,2,1s,2s}\right|<8\pi\label{pert1},\\
&\left|\lambda_{3}\pm\lambda_{4}\right|<8\pi\label{pert2},\\
&\left|\frac{1}{2}\left(\lambda_{1}+\lambda_{2}+\sqrt{\left(\lambda_{1}-\lambda_{2}\right)^{2}+4\lambda_{4}^{2}}\right)\right|<8\pi\label{pert3},\\
&\left|\frac{1}{2}\left(\lambda_{1s}+\lambda_{2s}+ \sqrt{16\lambda_{12S}^{2}+\left(\lambda_{1s}-\lambda_{2s}\right)^{2}}\right)\right|<8\pi\label{pert4},\\
&\left|\frac{1}{2}\left(\lambda_{3}+2\lambda_{4}+\lambda_{s}\pm \sqrt{16\lambda_{12S}^{2}+\left(\lambda_{3}+2\lambda_{4}-\lambda_{s}\right)^{2}}\right)\right|<8\pi\label{pert5},\\
&\frac{1}{2}\left|a_{1,2,3}\right|<8\pi\label{pert6}.
\end{align}

 The ratio of VEVs $\tan\beta\equiv v_2/v_1$ is bounded by perturbative unitarity of the Yukawa couplings $y_{u,d,e}$ given for any Type-II 2HDM by~\cite{Chen:2013kt}:
\begin{align}
0.28\lesssim\tan\beta\lesssim 140.
\end{align}

\section{Characteristics of the Scalar Mass Spectrum}
\label{sec:mass_spec}

The hierarchy of scales, $v/v_{S}\ll 1$, required to satisfy astrophysical and cosmological constraints, cf. eq.~\eqref{eq:preferred_v_S range}, implies a 
hierarchical scalar spectrum: {\em i)} the very light axion, $a$, with $m_a\propto 1/v_S$, {\em ii)}
the CP-even Higgs $h$, to be identified with the Higgs boson detected at the LHC, with $m_h \propto v$, and {\em iii)} the CP-odd Higgs boson $A$, the  
CP-even Higgs $H$, the charged Higgses $H^{\pm}$, and the PQ scalar boson 
$s$, with masses proportional to the PQ scale $v_S$. 

The 2HDM quartic couplings 
$\lambda_i$, $i=1,2,34$, the Higgs portal couplings $\lambda_{iS}$, $i=1,2,12$, the PQ scalar quartic coupling $\lambda_S$, and $\tan\beta$, all appearing in \eqref{mh1 full}, have to be chosen such that they result in an $m_h$ which is within 2$\sigma$ of its measured value. 
The spectrum of the remaining scalars, on the other hand, grossly depends only on the value of $\tan\beta$, $\lambda_{12S}$, $v_S$, and 
$\lambda_S$ (see also ref.~\cite{Espriu:2015mfa}). 
This is seen by observing that in a huge range of parameter space, namely for 
\begin{equation}
\lambda_{12S}, \lambda_S \gg (v/v_S)^2\simeq 6\times 10^{-16} \left( \frac{10^{10}\,{\rm GeV}}{v_S}\right)^2 ,  
\end{equation} 
the leading terms on the right-hand sides of Eqs.~\eqref{mHpm}, \eqref{mA}, \eqref{mh2 full}, \eqref{mh3 full} proportional to  $v_S^2$  dominate over the next-to-leading terms.
This results in an almost degenerate mass spectrum for the scalars $A$, $H^\pm$ and 
$H$, whose masses are determined mainly by $\tan\beta$, $\lambda_{12S}$ and 
$v_S$, while the mass of the $s$ is determined by $\lambda_S$ and $v_S$:
\begin{equation}
m_A^2\approx m_{H^\pm}^2 \approx m_H^2 \approx  \frac{1}{2} \dfrac{t_{\beta}^2+1}{t_{\beta}}\,  \lambda_{12S}\,v_{S}^2 ; 
\hspace{3ex} m_s^2 \approx \lambda_S v_S^2 .
\label{eq:scalar_mass_spectrum}
\end{equation}

For illustration, we present in table~\ref{tab:bp for illustration}  four benchmark examples illustrating the 
cross features of the scalar mass spectrum for different values of $\lambda_{12S}$ and 
$\lambda_S$. For these examples, the masses have been calculated using the full
tree level expressions, going beyond the expansion in $v/v_S$. Moreover, the parameters have been 
chosen to obey the BfB and perturbative unitarity constraints. 
The mass spectrum~\eqref{eq:scalar_mass_spectrum} seems to 
reproduce nicely the numerical results, even in the case when $\lambda_{12S}$ is approaching the boundary value $\lambda_{12S}\sim (v/v_S)^2$ behind the 
approximate relation~\eqref{eq:scalar_mass_spectrum}.

\begin{table}[ht]
\begin{center}
 \begin{tabular}{|c|c|c|c|c|}
  \hline
  Parameters &\textbf{BP1}&\textbf{BP2}&\textbf{BP3} &\textbf{BP4} \\
  \hline
  $\lambda_1$ & 0.1 & 0.1& 0.1 & 0.1  \\
  $\lambda_2$ & 0.258 & 0.258& 0.258 & 0.258 \\
  $\lambda_3$& 0.54 & 0.54& 0.54 & 0.54  \\
  $\lambda_4$&-0.14 &-0.14&-0.14 &-0.14 \\
  $\lambda_S$&1.0&1.0&1.0 &$10^{-10}$ \\
  $\lambda_{1S}$&$10^{-15}$&$10^{-15}$&$10^{-15}$ &$10^{-15}$  \\
  $\lambda_{2S}$&$10^{-15}$&$10^{-15}$&$10^{-15}$&$10^{-15}$ \\
  $\lambda_{12S}$&0.1&$2.5\times 10^{-8}$&$2.5\times 10^{-16}$& $2.5\times 10^{-16}$  \\
  $\tan \beta$ &26&26&26&26 \\
  $v_S$ &$3.0\times 10^{10}$&$3.0\times 10^{10}$&$3.0\times 10^{10}$&$3.0\times 10^{10}$   \\
  \hline
  $m_h\left(\text{GeV}\right)$ &125.1&125.1&125.1& 125.1  \\
  $m_H\left(\text{GeV}\right)$ &$3.4\times 10^{10}$&$1.7\times 10^{7}$&1711.5&1711.5\\
  $m_{s}\left(\text{GeV}\right)$&$3.0\times 10^{10}$&$3.0\times 10^{10}$&$3.0\times 10^{10}$&$3.0\times 10^{5}$ \\
  $m_A\left(\text{GeV}\right)$ &$3.4\times 10^{10}$&$1.7\times 10^{7}$&1711.5&1711.5 \\
  $m_{H^{\pm}}\left(\text{GeV}\right)$ &$3.4\times 10^{10}$&$1.7\times 10^{7}$&1712.8 &1712.8  \\  
  \hline
 \end{tabular}
 \caption{List of benchmarks considering cases a)-d).}
 \label{tab:bp for illustration}
\end{center}
\end{table}

Benchmark point 1 ({\bf BP1}) illustrates that for $\lambda_{12S},\lambda_S={\mathcal O}(1)$ one obtains the extreme decoupling limit where all extra Higgses are at the PQ-symmetry breaking scale and decoupled from the SM. In this region of parameter space, 2hdSMASH  cannot be tested at the LHC or future colliders.

{\bf BP2} shows that for $\lambda_{12S}\sim (v/v_S)\sim 10^{-8}$ and 
$\lambda_S\sim 1$, the extra Higgses have a mass in the range 
$\sim 10^{3}- 10^{4}$ TeV. Only the PQ-scalar $s$ is still decoupled and has a mass around the PQ-symmetry breaking scale. 

{\bf BP3} is phenomenologically more interesting. It illustrates that a portal coupling as low as $\lambda_{12S}\sim (v/v_S)^2\sim 10^{-16}$ pushes the masses of the extra Higgses all the way down to the electroweak scale~\cite{Volkas:1988cm,Espriu:2015mfa}. 
Observable effects at the HL-LHC and future colliders may be a result of this case~\cite{Espriu:2015mfa}. 

In {\bf BP4} we present a case where $\lambda_S\sim 10^{-10}$, in addition to 
$\lambda_{12S}\sim (v/v_S)^2\sim 10^{-16}$. In this case, the mass of the 
$s$ is pushed to intermediate scales between the electroweak and the PQ scale. Such a tiny PQ self-coupling is required for inflation, as we will see in the next section.

Intriguingly, portal couplings in the range exploited in {\bf BP3} and {\bf BP4}, 
leading to possible signatures at colliders, are also favored by theoretical considerations.  
In fact, tiny values of the portal couplings can be motivated by the fact that 
they provide a technically natural stabilisation of the required hierachy between the electroweak and the PQ scale, $v/v_S\gg 1$~\cite{Volkas:1988cm,Foot:2013hna,Clarke:2015bea}. 

The argument goes as follows: 
The portal couplings are responsible for the communication between the electroweak sector and the PQ sector, 
\begin{equation}
-\mathcal L_{\rm portal} = 
\lambda_{1S}\left(\Phi_1^\dagger\Phi_1\right)\left(S^*S\right)
        +\lambda_{2S}\left(\Phi_2^\dagger\Phi_2\right)\left(S^*S\right)-\lambda_{12S}\left(\Phi_2^\dagger\Phi_1 S^2+h.c.\right) \,.
\end{equation}
Clearly, the hierarchy $v/v_S\gg 1$ is preserved at tree level, if the portal couplings are parametrically suppressed, 
 \begin{equation}
|\lambda_{1S}|, |\lambda_{2S}|, |\lambda_{12S}| \lesssim 
\left(\frac{v}{v_S}\right)^2\simeq 6\times 10^{-16} \left( \frac{10^{10}\,{\rm GeV}}{v_S}\right)^2 \ll 1. 
\label{eq:portal_couplings_techn_nat} 
\end{equation}
Tiny values for the portal couplings are technically natural because 
the 2HDM sector and the PQ sector decouple from each other in this limit, giving an enhanced Poincare symmetry. The portal couplings are also controlling 
the radiative corrections to the masses of the electroweak mass scalars from
the PQ sector. Correspondingly, their masses are also not destabilized by radiative corrections, if the portal couplings are in the range~\eqref{eq:portal_couplings_techn_nat}.  In addition, this ensures that tiny values of $\lambda_S$, as required from inflation, are also technically natural as will be  shown in sec~\ref{subsec:stability_analysis_of_ls}. 

\section{Inflation in 2hdSMASH}
\label{sec:inflation}   
Inflation in 2hdSMASH is described by chaotic inflation. Any theory with a plateau-like scalar potential at sufficiently high field values which hosts a slow-roll regime for the fields involved can give rise to chaotic inflation. This idea was first introduced by Andrei Linde in Ref.\cite{LINDE1986395}. In 2hdSMASH chaotic inflation is an automatic feature  where the Higgs doublets $\Phi_{1}$, $\Phi_{2}$ and the PQ-singlet $S$ are non-minimally coupled to the Ricci scalar $R$~\cite{Spokoiny:1984bd,Futamase:1987ua,Salopek:1988qh,Fakir:1990eg,Bezrukov:2007ep}. At operator mass dimension four we show the action of 2hdSMASH in Jordan frame:
\begin{align}
S_{\text{2hdSMASH}}\supset 
- \int d^4x \sqrt{-g}
\left(\dfrac{M^{2}}{2}+ \xi_{1}\left|\Phi_{1}\right|^{2}+\xi_{2}\left|\Phi_{2}\right|^{2}+\xi_{S}\left|S\right|^{2}\right)R ,
\label{eq:2hdSMASH_Jordan_action}
\end{align}
where $\xi_i$, $i=1,2,S$, are the dimensionless non-minimal couplings and $M$ is the mass which is related to the reduced Planck mass, $M_{p}\equiv 1/\sqrt{8\pi G}$, given by 
\begin{align}
M^{2}=M_{p}^{2}+\xi_{1} v_{1}^{2}+\xi_{2} v_{2}^{2}+\xi_{S} v_{S}^{2} .
\end{align}
By means of metric- and scalar field transformation via the so-called Weyl transformation, we transform the action of eq. \eqref{eq:2hdSMASH_Jordan_action} from Jordan to Einstein frame. In Einstein frame the non-minimal couplings cause the quartic potential to be asymptotically flat and convex such that a plateau-like region is created suitable for inflation. This is true for any quartic potential and is preferred by current cosmic microwave background (CMB) measurements ~\cite{Planck:2018jri}. We restrict ourselves to the neutral part of the two $SU(2)_{L}$ doublets $\Phi_{1,2}$ and define the scalar fields involved in inflation as:
\begin{align}
\Phi_{1}^{0}=\dfrac{h_{1}}{\sqrt{2}}~e^{i\cdot \theta_{1}}~~,~~\Phi_{2}^{0}=\dfrac{h_{2}}{\sqrt{2}}~e^{i\cdot\theta_{2}}~~,~~S=\dfrac{s}{\sqrt{2}}~e^{i\cdot \theta_{S}}\,,
\end{align}
where the angular fields are expressed by $\theta_{i}\equiv a_{i}/v_{i}$.
The Weyl transformation is given by the frame function
\begin{equation}
\Omega^2(x)= 1+\frac{\xi_1 h_{1}^2(x) +\xi_2 h_{2}^2(x) +\xi_{S}s^2(x)}{M_{\text{ P}}^2} \,.
\label{conf_fac}
\end{equation}
Thus, we transform the metric into Einstein frame via
\begin{equation} \label{weyl}
\tilde g_{\mu \nu}(x) = \Omega^{2}(h_1(x),h_2(x),s(x))\, g_{\mu \nu} (x) ,
\end{equation}
for which we obtain the action in Einstein frame relevant for inflation 
\begin{align} 
\label{Eact}
S_{\text{2hdSMASH}}^{\text{(E)}}\supset {\int} d^4x\sqrt{-\tilde{g}}\left[-\frac{M_P^2}{2}\tilde{R}+\frac{1}{2}\sum_{i,j}\mathcal{G}_{ij}\tilde g^{\mu\nu}\partial_\mu\phi_i\partial_\nu\phi_j-\tilde{V}(\phi_i)\right]\,,
\end{align}
with fields $\phi = (\phi_1,\phi_2,\phi_3,\phi_4,\phi_5,\phi_6)=(h_1,h_2,s,\theta_1, \theta_2, \theta_S)$, Weyl-transformed metric $\tilde g^{\mu\nu}(x)=\Omega^{2}(x) g^{\mu\nu}(x)$, canonical Einstein-Hilbert action given by the gravitational term $\frac{M_P^2}{2}\tilde{R}$, induced field space metric $\mathcal{G}_{ij}$ and scalar potential $\tilde{V}(\phi_i)$. The scalar potential is transformed under Weyl transformation as follows
\begin{align}
\label{VEpot}
 \tilde{V} (\phi_i) 
= \frac{1}{\Omega^4(\phi_i)} V(\phi_i) \,,
\end{align}
where a field-dependent factor of $\Omega^{-4}$ rescales the Jordan frame scalar potential into the Einstein frame and makes it flat for large field values. Therefore, we neglect the quadratic part and only consider the quartic part of the scalar potential at large field values
\begin{align}
\label{VEpot_quartic}
&\tilde{V}_{\text{quartic}}(h_{1},h_{2},s,\tilde{\theta}_1) \\
&=\dfrac{\lambda_{1}h_{1}^{4}+\lambda_{2}h_{2}^{4}+\lambda_{S}s^{4}+2\left(\lambda_{34}h_{1}^{2}h_{2}^{2}+\lambda_{1S}h_{1}^{2}s^{2}+\lambda_{2S}h_{2}^{2}s^{2}-2\lambda_{12S}h_{1}h_{2}s^{2}\cos(\tilde{\theta})\right)}{8\left( 1+\frac{\xi_1 h_{1}^2 +\xi_2 h_{2}^2 +\xi_{S}s^2}{M_{p}^2}\right)^2}\,,
\notag
\end{align}
where $\lambda_{34}\equiv \lambda_3+\lambda_4$. We note that we have four dynamical fields, namely $h_{1}$, $h_{2}$, $s$ and an effective angle $\tilde{\theta}$ which is defined as follows
\begin{align}
\tilde{\theta}\equiv 2\theta_S +\theta_1 - \theta_2\,,
\end{align}
where orthogonal directions to $\tilde{\theta}$ are omitted since they correspond to flat directions. The cosine of the $\lambda_{12S}$-term can take extrema in the interval $\left[-1,1\right]$. We make the restriction $\tilde{\theta}\in\left[0,\pi\right]$ and determine the extrema by taking the partial derivative of $\tilde{V}_{\text{quartic}}$ w.r.t. $\tilde{\theta}$:
\begin{align}
\dfrac{\partial \tilde{V}_{\text{quartic}}}{\partial \tilde{\theta}}\overset{!}{=}0~~\Rightarrow~~ \tilde{\theta}_{0}=\left\lbrace 0,\pi \right\rbrace\,.
\label{eq:theta_extrema}
\end{align}
The sufficient conditions for $\tilde{\theta}$ at its extrema are calculated by taking the second partial derivative of $\tilde{V}_{\text{quartic}}$ w.r.t. $\tilde{\theta}$
\begin{align}
\dfrac{\partial^{2} \tilde{V}_{\text{quartic}}}{\partial \tilde{\theta}^{2}}=
\begin{cases}
\dfrac{\lambda_{12S}h_{1}h_{2}s^{2}M_{p}^4}{8\left( M_{p}^2+\xi_1 h_{1}^2 +\xi_2 h_{2}^2 +\xi_{S}s^2\right)^2}\geq 0 & ,~\tilde{\theta}_{0}=0\\
~\\
\dfrac{-\lambda_{12S}h_{1}h_{2}s^{2}M_{p}^4}{8\left( M_{p}^2+\xi_1 h_{1}^2 +\xi_2 h_{2}^2 +\xi_{S}s^2\right)^2}\leq 0 & ,~\tilde{\theta}_{0}=\pi
\end{cases}\,.
\end{align}
Since $\lambda_{12S}\geq 0$ and the product $h_{1}h_{2}s^{2}$ is rotation invariant, $\tilde{\theta}$ is stabilized in its minimum at $\tilde{\theta}_{0}=0$. Hence, the potential reads:  
\begin{align}
\label{VEpot_quartic2}
\tilde{V}_{\text{quartic}}(h_{1},h_{2},s)= \dfrac{\lambda_{1}h_{1}^{4}+\lambda_{2}h_{2}^{4}+\lambda_{S}s^{4}+2\left(\lambda_{34}h_{1}^{2}h_{2}^{2}+\lambda_{1S}h_{1}^{2}s^{2}+\lambda_{2S}h_{2}^{2}s^{2}-2\lambda_{12S}h_{1}h_{2}s^{2}\right)}{8\left( 1+\frac{\xi_1 h_{1}^2(x) +\xi_2 h_{2}^2(x) +\xi_{S}s^2(x)}{M_{p}^2}\right)^2}\,.
\end{align}
There is now a three-dimensional induced field space metric in Einstein frame  spanned by $\phi=\left(h_{1},h_{2},s\right)$ which is calculated via: 
\begin{align} \nonumber
\mathcal{G}_{ij}&= \frac{\delta_{ij}}{\Omega^{2}}+\frac{3}{2}M_{p}^2\frac{\partial\log\Omega^2}{\partial {\phi_i}}\frac{\partial\log\Omega^2}{\partial {\phi_j}},\label{eq:field_space}\\
&=\dfrac{1}{\Omega^{2}}
\begin{pmatrix}
 1  + 6 \xi_{1}^2 \frac{h_{1}^2}{\Omega^2 M_{p}^2}  & 6 \xi_{1} \xi_{2} \frac{h_{1} h_{2}}{\Omega^2 M_{p}^2}  & 6 \xi_{1} \xi_{S} \frac{h_{1}s}{\Omega^2 M_{p}^2} \\
 6\xi_{1} \xi_{2} \frac{h_{1} h_{2}}{\Omega^2 M_{p}^2}  & 1 + 6 \xi_{2}^2\frac{h_{2}^2}{\Omega^2 M_{p}^2}  & 6 \xi_{2} \xi_{S} \frac{h_{2} s}{\Omega^2 M_{p}^2} \\
 6 \xi_{1} \xi_{S}\frac{h_{1}  s}{\Omega^2 M_{p}^2}  & 6 \xi_{2} \xi_{S} \frac{h_{2} s}{\Omega^2 M_{p}^2}    & 1 + 6 \xi_{S}^2 \frac{s^2}{\Omega^2 M_{p}^2}
\end{pmatrix}\,.\notag
\end{align}
The scalar potential is thus symmetric under $h_1 \to - h_1$, $h_2 \to - h_2$, and $s \to -s$. Therefore, we use spherical field space coordinates as parametrization for $h_{1}$, $h_{2}$ and $s$:
\begin{align}
h_1(x)={\phi(x)\cos\vartheta\sin\gamma}, \hspace{3ex}
h_2(x)={\phi(x)\sin\vartheta\sin\gamma}, \hspace{3ex}
s(x)={\phi(x)\cos\gamma} \,.
\end{align}
During inflation the scalar potential given in eq. \eqref{VEpot_quartic2} becomes a constant by cancellation since the numerator and denominator scale as $\phi^{4}$. Hence, we can make the following approximation
\begin{align}
    {\phi^2(x)} \gg   \frac{M_{p}^2}{\left(\xi_{1}\cos^{2}\vartheta\sin^{2}\gamma+\xi_{2}\sin^{2}\vartheta\sin^{2}\gamma+\xi_{S}\cos^{2}\gamma\right)}  \,,           
\end{align}
which allows us to express the scalar potential solely by angles $\vartheta$ and $\gamma$:
\begin{align}
\label{VEpot_quartic_angles}
&\tilde{V}_{\text{quartic}}(\vartheta,\gamma)\simeq \\ &M_{p}^{4}~\dfrac{t_{\gamma}^4 \left(\lambda_{1}+\lambda_{2} t_{\vartheta}^4+2 \lambda_{34} t_{\vartheta}^2\right)
+2 t_{\gamma}^2 \left(t_{\vartheta}^2+1\right) \left(\lambda_{1S}+\lambda_{2S} t_{\vartheta}^2-2 \lambda_{12S} t_{\vartheta}\right)
+\lambda_{S} \left(t_{\vartheta}^2+1\right)^2}{8 \left(t_{\gamma}^{2}(\xi_{1}+\xi_{2}t_{\vartheta}^{2})+\xi_{S}(1+t_{\vartheta}^{2})\right)^2} \,,
\nonumber
\end{align}
where $t_{x}\equiv \tan x$. Furthermore, we require the portal couplings to be tiny in order to avoid large radiative corrections which is technically natural and associated with an enhanced Poincaré symmetry \footnote{This was introduced by Ref. \cite{Clarke:2015bea} in the context of the $\nu$DFSZ model which we adopted for 2hdSMASH and discussed in section \ref{sec:mass_spec}.} \cite{Clarke:2015bea}. Therefore we consider two portal coupling regimes
\begin{align}
& \lambda_{12S}\ll \lambda_{1,2,34,1S,2S,S}\,, &&\text{(I)}\notag\\
& \lambda_{12S}\lesssim \lambda_{1S}\sim \lambda_{2S}\ll \lambda_{1,2,34,S}\,, &&\text{(II)}\notag
\end{align}
where cases with $\lambda_{12S}\gg \lambda_{iS}$ are neglected due to scale-invariance, and cases with $\lambda_{1S,2S}\ll\lambda_{2S,1S}$ are neglected due to the intricacies of RG running\footnote{We comment on the features of RG running with portal couplings in section \ref{sec:infl_implications_portal_couplings}.}. Case (I) is the most general case  to consider, whereas case (II) can be understood as the limit $\lambda_{1S,2S}\to 0$ by going from (I) to (II). In (II), all of the portal couplings are tiny, so that the PQ-sector decouples completely from the 2HDM sector. In order to find the most general description, we will focus on (I) and note when (II) can be applied. The scalar potential for case (I) is therefore given by
\begin{align}
\label{VEpot_quartic_angles_c1}
&\tilde{V}_{\text{quartic}}(\vartheta,\gamma)\simeq \\ &M_{p}^{4}~\dfrac{t_{\gamma}^4 \left(\lambda_{1}+\lambda_{2} t_{\vartheta}^4+2 \lambda_{34} t_{\vartheta}^2\right)
+2 t_{\gamma}^2 \left(t_{\vartheta}^2+1\right) \left(\lambda_{1S}+\lambda_{2S} t_{\vartheta}^2\right)
+\lambda_{S} \left(t_{\vartheta}^2+1\right)^2}{8 \left(t_{\gamma}^{2}(\xi_{1}+\xi_{2}t_{\vartheta}^{2})+\xi_{S}(1+t_{\vartheta}^{2})\right)^2} \,.
\nonumber
\end{align}
From this expression, we are able to determine the minima and thus the effective single-field trajectories. This is done by considering the Jacobian of eq. \eqref{VEpot_quartic_angles_c1} in two-dimensional field space :
\begin{align}
J\left(\vartheta,\gamma\right)=\begin{pmatrix}
\dfrac{\partial \tilde{V}_{\text{quartic}}(\vartheta,\gamma)}{\partial \vartheta} & \dfrac{\partial \tilde{V}_{\text{quartic}}(\vartheta,\gamma)}{\partial \gamma}
\end{pmatrix}^{2}\,.
\label{eq:jacobian_Vpot_angles}
\end{align}
With $J_{2}\equiv\frac{\partial \tilde{V}_{\text{quartic}}}{\partial \gamma}=0$ we determine the extrema of $\gamma$:
\begin{align}
\gamma_{0,i}=
\begin{cases}
\gamma_{\text{THI}}=\dfrac{\pi}{2}\\
\gamma_{\text{PQI}}=0\\
\gamma_{\text{PQTHI}}=\gamma_{\text{PQTHI}}(\vartheta)
\end{cases}\,,
\label{eq:gamma_extrema}
\end{align}
which correspond to the three coarse field space directions for inflation, i.e. 2HDM-inflation denoted by THI, PQ-inflation denoted by PQI and mixed PQ-2HDM-inflation denoted by PQTHI. We have omitted for now the detailed expression of $\gamma_{\text{PQTHI}}$ which will be mentioned in section \ref{sec:PQ-and mixed inflation}. There are in principle seven inflationary directions, namely three 2HDM-field directions ($h_1$, $h_2$, $h_{12}$), one PQ-field direction $s$ and three mixed PQ-2HDM directions ($sh_1$, $sh_2$, $sh_{12}$). All of these field directions are effective single-field trajectories that omit multi-field effects.  
As mentioned above, we will use case (I) as the most general description and take the limit to case (II) for THI and PQI since both sectors decouple from each other and are thus technically natural \cite{Clarke:2015bea}. In the following, we will start 
the discussion on PQI and PQTHI (see appendix~\ref{app:2hdm inflation} for the discussion on THI). This is done by considering the $\gamma$-directions given by $\gamma_{\text{THI}}$, $\gamma_{\text{PQ}}$ and $\gamma_{\text{PQTHI}}(\vartheta)$, respectively.

\subsection{PQ- and PQ-2HDM-inflation in 2hdSMASH}
\label{sec:PQ-and mixed inflation}
\begin{figure}[htb!]
\centering
\begin{subfigure}{.5\textwidth}
  \centering
  \includegraphics[width=.9\linewidth]{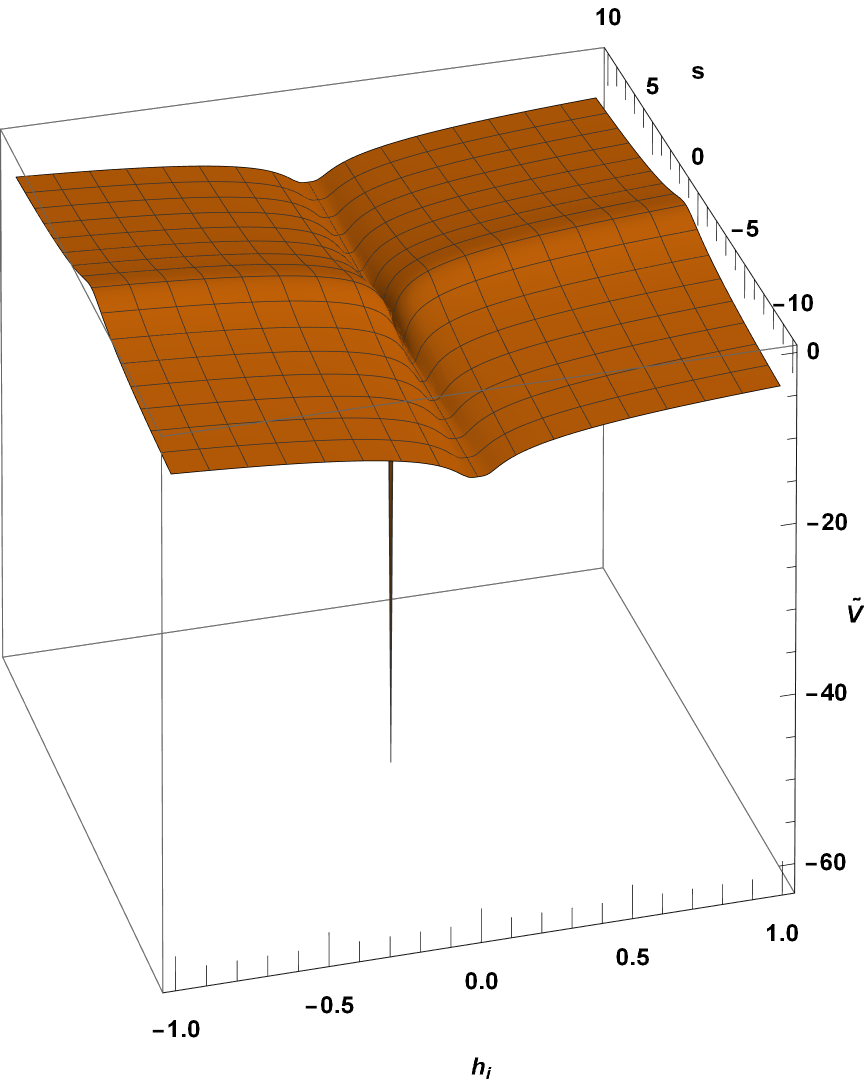}
\end{subfigure}%
\begin{subfigure}{.5\textwidth}
  \centering
  \includegraphics[width=.9\linewidth]{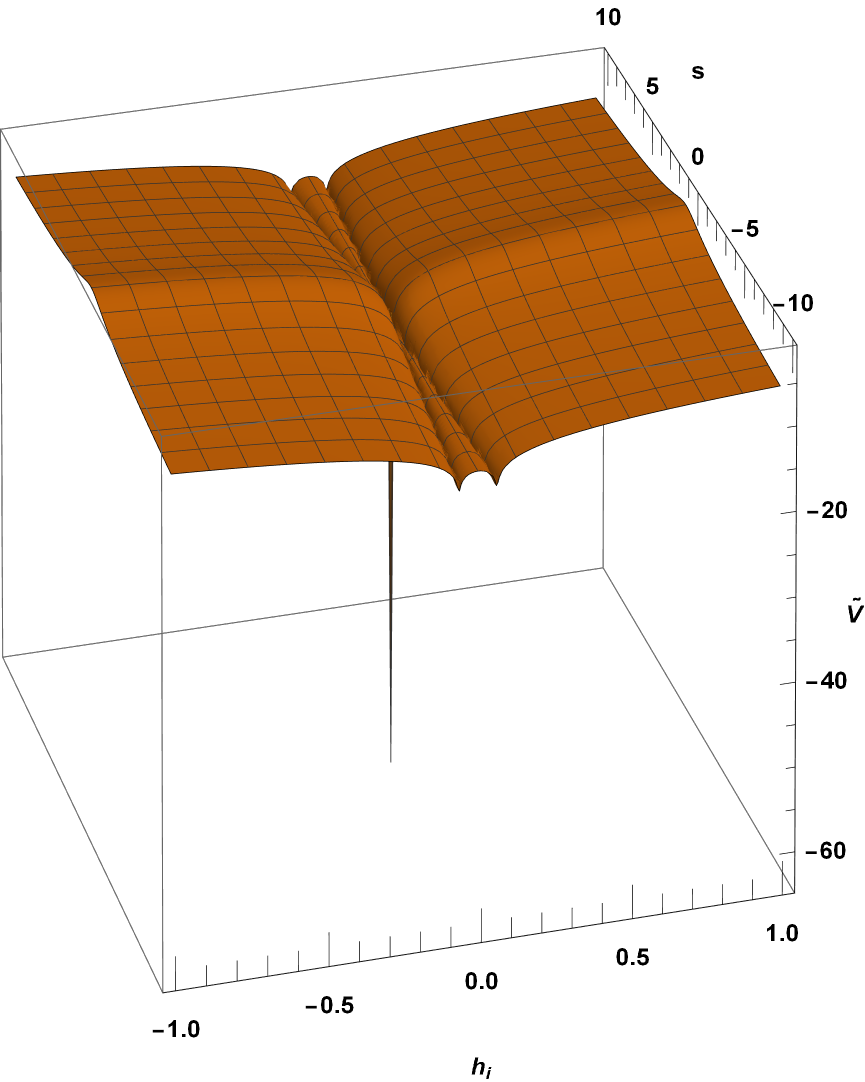}
\end{subfigure}
\caption{Decadic log Einstein frame scalar potential depicting PQI (left) and PQTHI (right) in units of $M_p$ as a function of the PQ-scalar $s$ and the two Higgs fields $h_1$ and $h_2$.}
\label{fig:2hdsmash_pots}
\end{figure}

\begin{figure}
\centering
\begin{subfigure}{.5\textwidth}
  \centering
  \includegraphics[width=0.93\linewidth]{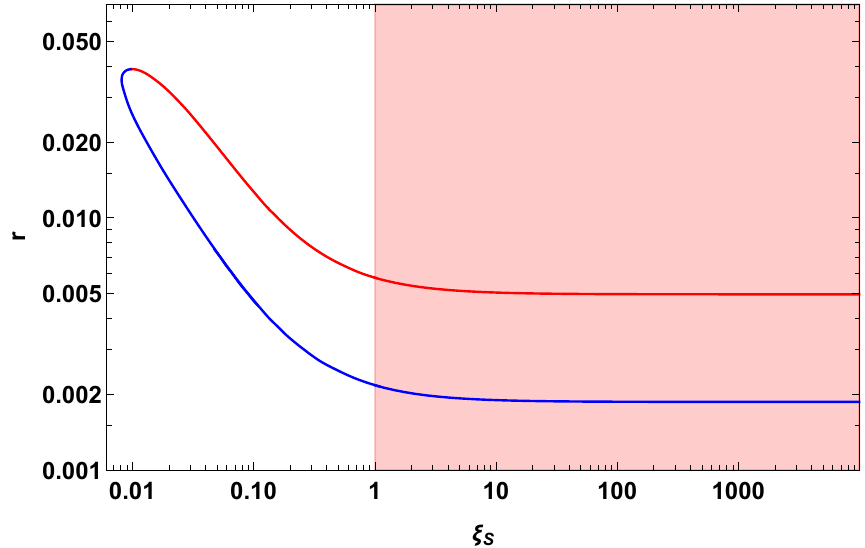} 
\end{subfigure}%
 \begin{subfigure}{.5\textwidth}
  \centering
  \includegraphics[width=0.93\linewidth]{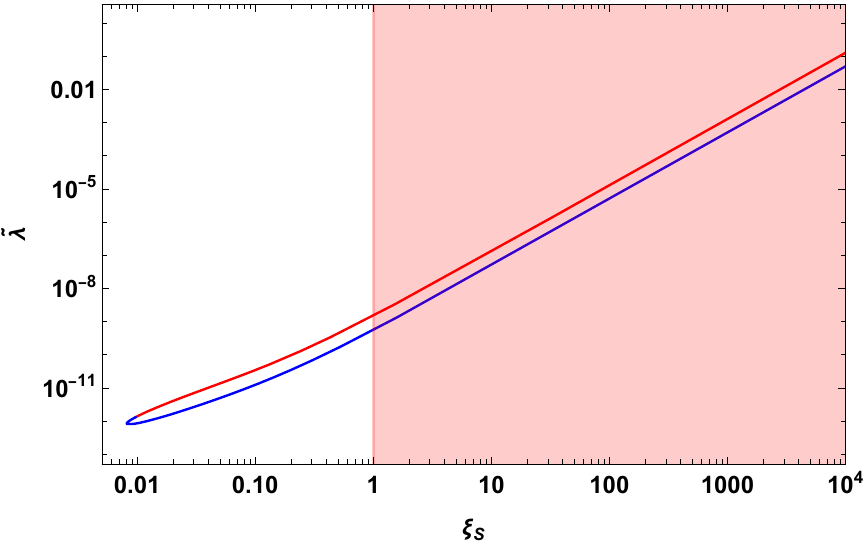}
\end{subfigure}
\caption{Shown are the 95\% C.L. contours of $r$ vs. $\xi_{S}$ (left) and the effective quartic coupling for inflation $\tilde{\lambda}$ vs. $\xi_{S}$ at the pivot scale $k_{\ast}=0.002$ Mpc$^{-1}$. The red shaded region given by $\xi_S >1$ indicates the unnatural regime according to our naturalness philosophy. The red and blue curves indicate constraints given by the redder spectrum and the bluer spectrum of $n_{s}$ \cite{Planck:2018vyg,BICEP:2021xfz}.}
\label{fig:r_vs_xi_and_lambda_vs_xi}
\end{figure}

In this section, we discuss inflation in the PQ- and PQ-2HDM directions. This corresponds to the field directions determined by $\gamma_{\text{PQI}}$ and $\gamma_{\text{PQTHI}}$ respectively. These types of inflationary field directions have been discussed in the context of a KSVZ-type model, dubbed SMASH \cite{Ballesteros:2016xej}, where the number of field directions is drastically reduced compared to 2hdSMASH.                   
 In the context of the DFSZ-type model, inflation has been previously  discussed in the 2HDM direction where quartic self-couplings were taken to be of order $\mathcal{O}(1)$ (see 
 Ref.~\cite{Nakayama:2015pba}) and  recently in an extension of the $\nu$DFSZ model with a trilinear $U(1)$ breaking term,  VISH$\nu$ \cite{Sopov:2022bog}. Ref.~\cite{Sopov:2022bog} has considered inflation along all possible directions including the 2HDM, singlet and mixed directions for non-minimal couplings in the large regime ($\xi_1, \xi_2 \gg 1$) and hierarchical regime ($\xi_S \gg \xi_1, \xi_2$) while neglecting the U(1) breaking coupling. 
 Since both studies differ in the structure of the scalar potential, it is instructive to study the inflationary conditions for the quartic U(1) breaking case. In this work, we focus  on the hierarchical regime including the effect of small portal couplings ($\lambda_{1S},\lambda_{2S}$ and $\lambda_{12S}$) that contribute to the singlet and singlet-doublet mixed inflation directions. For completeness,   the 2HDM directions are discussed in the appendix~\ref{app:2hdm inflation}  for which THI requires $\lambda_i^{\text{THI}}\lesssim 10^{-10}$ for $\xi_{1,2}\lesssim 1$ \cite{Giudice:2010ka,Barbon:2015fla,Ballesteros:2016xej}. According to our naturalness philosophy non-minimal couplings should be radiatively generated and are thus an automatic feature in the very early universe. Accounting for RG running of the 2HDM self-couplings, i.e. $\beta_{\lambda_1}$ and $\beta_{\lambda_2}$, this would spoil the picture since either coupling would make the other one large in order to satisfy Higgs phenomenology. Hence, we consider PQI and PQTHI for which we implement a hierarchy of non-minimal couplings, i.e. $\xi_{S}\gg\xi_{1,2}$, in order to effectively decouple the PQ- from the 2HDM-sector. This makes PQTHI parametrically close to the PQI which results in the suppression of large self-coupling contributions from the 2HDM-sector at the Planck scale. Therefore, we require at the Planck scale $\xi_{S}\lesssim 1$ and thus $\lambda_{\text{PQI, PQTHI}}\lesssim 10^{-10}$. These small self-couplings can be generated quite naturally for PQI and PQTHI since their RGE's are effectively decoupled from the 2HDM-sector and thus hidden from low energy phenomenology. We discuss these intricacies in section \ref{sec:infl_implications_portal_couplings} and explain in the following our naturalness philosophy regarding the portal couplings and which role they play for decoupling.\\
\\
We consider for PQI and PQTHI case (I) but note for PQI the limit to case (II). This consideration can be understood by the decoupling of the 2HDM- from the PQ- sector by going from case (I) to (II). For PQTHI, however, the mixing between these two sectors is allowed to a certain extend. In fact, the mixing is sufficiently suppressed which refines case (I) by
\begin{align}
\lambda_{12S}\ll\lambda_{S}\lesssim \lambda_{1S,2S}\ll \lambda_{1,2,34}~~~\text{with}~~~\lambda_{iS}^{2}/\lambda_S\ll \lambda_{1,2,34} &&\text{(I')}\notag
\end{align}
and denotes a mild decoupling compared to case (II). Therefore, case (I') decouples the two sectors as well, which corresponds to an enhanced Poincaré symmetry, i.e. $\mathcal{G}_{P}^{\text{2HDM}}\times \mathcal{G}_{P}^{\text{PQ}}$. This is due to the fact, that all 2HDM couplings are much greater than the portal couplings which accounts to a technically natural limit where radiative corrections are negligible to low energy physics.\\
\\
Hence, we will start with the mixed PQ-2HDM directions before we proceed with the PQ-direction. As mentioned in section \ref{sec:inflation}, we will keep the discussion on PQI as general as possible, i.e. use case (I), but mention the limiting case by going from (I) to (II).\\
\\
Thus, we obtain the following scalar potential
\begin{align}
\label{VEpot_quartic_angles3}
\tilde{V}(\vartheta,\gamma)&\simeq \\ &M_{p}^{4}~\dfrac{t_{\gamma}^4 \left(\lambda_{1}+\lambda_{2} t_{\vartheta}^4+2 \lambda_{34} t_{\vartheta}^2\right)
+2 t_{\gamma}^2 \left(t_{\vartheta}^2+1\right) \left(\lambda_{1S}+\lambda_{2S} t_{\vartheta}^2\right)
+\lambda_{S} \left(t_{\vartheta}^2+1\right)^2}{8 \xi_{S}^{2}(1+t_{\vartheta}^{2})^2}\,.
\nonumber
\end{align}
The Jacobian for the PQTHI is given by
\begin{align}
    J(\vartheta,\gamma_{\text{PQTHI}})&=\begin{pmatrix}
    \dfrac{\partial \tilde{V}_{\text{quartic}}}{\partial \vartheta} &\dfrac{\partial \tilde{V}_{\text{quartic}}}{\partial \gamma}\end{pmatrix}^{T},\\
    &=\begin{pmatrix}
    \frac{\left(t_{\vartheta}^3+t_{\vartheta}\right) \left(\lambda_{1S}+\lambda_{2S} t_{\vartheta}^2\right) \left(-\lambda_{1} \lambda_{2S}+\lambda_{1S} \lambda_{34}+t_{\vartheta}^2 (\lambda_{1S}
   \lambda_{2}-\lambda_{2S} \lambda_{34})\right)}{2 \xi_{S}^2 \left(\lambda_{1}+\lambda_{2} t_{\vartheta}^4+2 \lambda_{34} t_{\vartheta}^2\right)^2}\\ 0
   \end{pmatrix}\,,\notag
\end{align}
where we used $\gamma_{\text{PQTHI}}(\vartheta)$ for the PQ-2HDM direction which is given by
\begin{align}
\gamma_{\text{PQTHI}}(\vartheta)=
\text{arctan}\left(\sqrt{\dfrac{\left(t_{\vartheta}^2+1\right) \left(-\lambda_{1S}-\lambda_{2S} t_{\vartheta}^2\right)}{\lambda_{1}+\lambda_{2} t_{\vartheta}^4+2 \lambda_{34} t_{\vartheta}^2}}\right)\,.
\end{align}
The extrema for $\vartheta$ are thus obtained via $J_{1}(\vartheta,\gamma_{\text{PQTHI}})\overset{!}{=}0$ 
\begin{align}
\vartheta_{\text{PQTHI}}=
\begin{cases}
\vartheta_{s h_{1}}=0 \\
\vartheta_{s h_{2}}=\dfrac{\pi}{2}\\
\vartheta_{s h_{12}}= \arctan\left(\sqrt{\dfrac{\lambda_{1} \lambda_{2S}-\lambda_{1S} \lambda_{34}}{\lambda_{1S} \lambda_{2}-\lambda_{2S} \lambda_{34}}}\right)
\end{cases}\,,
\end{align}
which corresponds to the following $\gamma_{\text{PQTHI}}$-values
\begin{align}
\gamma_{\text{PQTHI}}=
\begin{cases}
\gamma_{sh_{1}}=\arctan\left(\sqrt{-\frac{\lambda_{1S}}{\lambda_{1}}}\right)\\
\gamma_{sh_{2}}=\arctan\left(\sqrt{-\frac{\lambda_{2S}}{\lambda_{2}}}\right)\\
\gamma_{sh_{12}}=\arctan\left(\sqrt{-\frac{\lambda_{1S}\left(\lambda_{2}-\lambda_{34}\right)+\lambda_{2s}\left(\lambda_{1}-\lambda_{34}\right)}{\lambda_{1}\lambda_{2}-\lambda_{34}^{2}}}\right)
\end{cases}
\end{align}
for the three possible field directions, i.e. $s h_{1}$, $s h_{2}$ and $s h_{12}$. We can now determine the inflationary vacuum energies in the PQ-2HDM direction  which need to be positive in order to avoid tachyonic vacuum states
\begin{align}
V_{0}^{\text{PQTHI}}\geq 0 ~~~\Leftrightarrow ~~~\dfrac{1}{8\xi_{S}^{2}}
\begin{cases}
\lambda_{S}-\frac{\lambda_{1S}^2}{\lambda_{1}}\geq 0\\
~\\
\lambda_{S}-\frac{\lambda_{2S}^2}{\lambda_{2}}\geq 0\\
~\\
\lambda_{S}-\frac{\lambda_{1} \lambda_{2S}^2+\lambda_{1S}^2 \lambda_{2}-2 \lambda_{1S} \lambda_{2S} \lambda_{34}}{\lambda_{1} \lambda_{2}-\lambda_{34}^2}\geq 0
\end{cases}\,.
\end{align}
In the following, we give the minimum conditions for the PQ-2HDM direction, i.e. for $s h_{1,2,12}$, in order to determine whether the extrema correspond to inflationary valleys while other directions correspond to inflationary ridges. As in section \ref{app:2hdm inflation}, the minima conditions are determined via the hessian and obey the conditions given by eq. \eqref{eq:minima_conditions} for the PQ-2HDM direction, i.e. $s h_{1,2,12}$.
\begin{align*}
&\textbf{PQTHI-$\left(s h_{1}\right)$:}\\
&\kappa_{1s}\equiv\lambda_{1S}\leq 0\,,\\
&\kappa_{s1}\equiv\lambda_{2S}\lambda_{1}-\lambda_{1S}\lambda_{34}\geq 0\,.\\
&\\
&\textbf{PQTHI-$\left(s h_{2}\right)$:}\\
&\kappa_{2s}\equiv\lambda_{2S}\leq 0\,,\\
&\kappa_{s2}\equiv\lambda_{1S}\lambda_{2}-\lambda_{2S}\lambda_{34}\geq 0\,.\\
&\\
&\textbf{PQTHI-$\left(s h_{12}\right)$:}\\
&\kappa _{s 1} \kappa _{s 2} \left(\kappa _{s 1}+\kappa _{s 2}\right)\leq 0\,,\\
&\left(\kappa _{s 1}+\kappa _{s 2}\right) \left(\lambda_{1S} \kappa _{s 2}+\lambda_{2S} \kappa _{s 1}\right)\leq 0\,.
\end{align*}
Most importantly, the inflationary conditions are given by the minimum conditions which are supplemented by the maximum conditions of other field directions in order to classify inflationary valleys and ridges accordingly. Similar to our discussion for 2HDM field space directions of appendix \ref{app:2hdm inflation}, the PQTHI-$sh_{12}$ direction already contains this feature automatically since PQTHI-$sh_{12}$ is a mixture of all three fields. However, the conditions of the PQTHI-$sh_{12}$ direction need some refining. The first of the two conditions state
\begin{align}
\kappa_{s1}\leq 0~~~,~~~\kappa_{s2}\leq 0\,.
\end{align}
This leads to the second condition to become 
\begin{align}
\kappa_{1s}\equiv\lambda_{1S}>0~~~,~~~\kappa_{2s}\equiv\lambda_{2S}>0\,.
\end{align}
We list the complete set of inflationary conditions for the 2HDM-$h_{12}$ direction in Table \ref{tab:inflation_cond}.\\
\\ 
Considering that PQTHI is composed of two or three field directions, i.e. $s h_{1,2,12}$, we need to examine whether orthogonal field directions contribute to inflation. Therefore, we provide the field space metric in these three inflationary directions

\begin{align}
\mathcal{G}_{ij}^{sh_{i}}\simeq \dfrac{b_i}{\Omega_{sh_i}^{2}}
\begin{pmatrix}
1 & 0 & 0 \\
 0 & 1 & 0 \\
 0 & 0 & \frac{\Omega_{sh_i}^{2}+6 \xi_{S}^{2}\frac{\phi ^2}{M_{p}^2}}{\Omega_{sh_i}^{2}} 
\end{pmatrix}\,,
\end{align}
where $b_i$ are the mixing parameters determined via $\sin^{2}\gamma_{sh_i}=1-b_{i}^{-1}$,
\begin{align}
&b_{1}\equiv 1+\left|\dfrac{\lambda_{1S}}{\lambda_{1}}\right|\,,\\
&b_{2}\equiv 1+\left|\dfrac{\lambda_{2S}}{\lambda_{2}}\right|\,,\\
&b_{12}\equiv 1+\left|\dfrac{\lambda_{2S} (\lambda_{1}-\lambda_{34})+\lambda_{1S} (\lambda_{2}-\lambda_{34})}{\lambda_{1} \lambda_{2}-\lambda_{34}^2}\right|
\end{align}
and $\Omega_{sh_i}^{2}$ are the frame functions given by
\begin{align}
\Omega_{sh_i}^{2}=b_i+6\xi_S\frac{s^{2}}{M_{p}^{2}}\,.
\end{align}
We can see from the field space metric and its corresponding frame functions that 
the mixing parameter determine whether inflation occurs in the PQ-2HDM ($b_i\neq 1$) or in the PQ direction ($b_i=1$) resembling case (II). In order to examine whether the orthogonal field space directions contribute to PQTHI, we compute the instantaneous masses
\begin{align}
\left. m_{\phi_{i}}^{2}\right|_{\substack{ \vartheta=\vartheta_{sh_i}\\ \gamma=\gamma_{sh_i}}}\simeq 
\begin{cases}
-\frac{\lambda_{1S}}{\xi_{S}}~~,~~\frac{\kappa_{s2}}{2\lambda_{2}\xi_{S}}~~,~~-\frac{\kappa_{s2} \lambda_{1}}{\xi_{S} \left(\lambda_{1} \lambda_{2}-\lambda_{34}^2\right)} & \left(h_1\right)\\
~\\
-\frac{\lambda_{2S}}{\xi_{S}}~~,~~\frac{\kappa_{s1}}{2\lambda_{1}\xi_{S}} ~~,~~ -\frac{\kappa_{s1} \lambda_{2}}{\xi_{S} \left(\lambda_{1} \lambda_{2}-\lambda_{34}^2\right)} & \left(h_2\right)\\
~\\
\frac{\lambda_{1S}^{2}}{\lambda_{1}\xi_{S}\left(1+6\xi_{S}\right)}~~,~~\frac{\lambda_{2S}^{2}}{\lambda_{2}\xi_{S}\left(1+6\xi_{S}\right)}~~,~~\frac{\kappa_{s1} \lambda_{2S}+\kappa_{s2} \lambda_{1S}}{\xi_{S} (6 \xi_{S}+1) \left(\lambda_{1} \lambda_{2}-\lambda_{34}^2\right)} & \left(s\right)
\end{cases}\,,
\end{align}
where the masses are given for $s h_{1}$, $s h_{2}$ and $s h_{12}$ directions from left to right. These results are now related to the Hubble rate $\mathcal{H}^{2}\approx \tilde{V}/3M_{p}^{2}$ 

\begin{align}
&\left.\frac{m_{s}^{2}}{\mathcal{H}^{2}}\right|_{\substack{ \vartheta=\vartheta_{sh_i}\\ \gamma=\gamma_{sh_i}}}\simeq \frac{24\xi_{S}}{1+6\xi_S}\,\frac{\delta \lambda_S}{\tilde{\lambda}}\lesssim 1\,,\\
&\left.\frac{m_{h_i}^{2}}{\mathcal{H}^{2}}\right|_{\substack{ \vartheta=\vartheta_{sh_i}\\ \gamma=\gamma_{sh_i}}}\simeq \frac{12\xi_{S}}{\tilde{\lambda}}
\begin{cases}
2\left|\lambda_{iS}\right|\gtrsim 1 & \text{($sh_i$)}\\
~\\
\left(\lambda_{iS}+(b_{j}-1)\lambda_{34}\right)\gtrsim 1 & \text{($sh_j$)}\\
~\\
\frac{2\lambda_1\lambda_2(\lambda_{iS}+(b_{j}-1)\lambda_{34})}{\lambda_{1}\lambda_{2}-\lambda_{34}^{2}}\gtrsim 1 & \text{($s h_{12}$)}
\end{cases}\,,
\end{align}
with $\xi_S \sim 2\times 10^{4}\sqrt{\tilde{\lambda}}$ and $\tilde{\lambda}=\lambda_S-\delta\lambda_S\gtrsim 0$. This shows that all masses of orthogonal directions are stabilized while the inflaton remains dynamical at the inflationary valley.\\
\\
For the PQ-direction, i.e. $\gamma_{\text{PQ}}=0$, we acquire the following minimum condition given by the single component hessian
\begin{align}
\dfrac{\xi_{S} \left(\lambda_{1S}+\lambda_{2S} t_{\vartheta}^{2}-2 \lambda_{12S} t_{\vartheta} \right)-\lambda_{S} \left(\xi_{1}+\xi_{2} t_{\vartheta}^{2}\right)}{2 \xi_{S}^3 \left(1+t_{\vartheta}^{2}\right)}\geq 0 \,.
\label{eq:HSI_infl_cond_x11}
\end{align}
In order to obtain minimum conditions w.r.t. all field directions, we simply apply the $\vartheta$-extrema of the 2HDM- and PQ-2HDM directions, i.e. $\vartheta_{\text{THI}}$ and $\vartheta_{\text{PQTHI}}$. This amounts to a total of four minimum condition for the PQ-direction:
\begin{align*}
&\textbf{PQI:}\\
&\kappa_{1s}\equiv \lambda_{1S}\geq 0\,,\\
&\kappa_{2s}\equiv \lambda_{2S}\geq 0\,,\\
&\left(\kappa _{s 1} +\kappa _{s 2}\right)\left(\lambda_{1S} \kappa _{s 2}+\lambda_{2S} \kappa _{s 1}\right)\geq 0\,,\\
&\left(\kappa _{s 1} + \kappa _{s 2}\right)\left(\lambda_{1}+\lambda_{2}-2\lambda_{34}\right)\geq 0\,,
\end{align*}
where the first two correspond to the 2HDM-$h_{1,2}$ directions and the last two conditions correspond to the PQTHI-$sh_{12}$ direction and to the 2HDM-$h_{12}$ direction, respectively. By considering the last two conditions, we cannot make a conclusive statement whether $\kappa_{s1}$ and $\kappa_{s2}$ are positive or negative. This can only be determined if either $\lambda_{1,2}\geq \lambda_{34}$ or $\lambda_{1,2}\leq \lambda_{34}$ is given. Thus, the last two conditions can be neglected. Moreover, we note that these conditions vanish by taking the limit to case (II). The absence of the portal couplings and the non-minimal coupling hierarchy, i.e. $\xi_{S}\gg \xi_{1,2}$, are mainly responsible for an inflationary valley to exist in the PQ-direction. For completeness we refer to the above minimum conditions as the inflationary conditions and add case (II) as a another condition in our model.\\
\\
Since we are considering four inflationary trajectories, we need an adequate description of the inflaton field $\phi$ which we identify as the PQ-scalar $s$. Therefore, we canonically normalize the inflaton field $s$ with the following field redefinitions for inflation in the $s$-, $sh_{1}$-, $sh_{2}$- and $sh_{12}$ direction, respectively
\begin{align}
&\Omega^{2}\dfrac{d\chi_{s}}{d s}=\sqrt{\Omega^{2}+6\xi_{S}^{2}\dfrac{s^{2}}{M_{p}^{2}}}\,,
\label{chi_rho}\\
&\Omega^{2}\dfrac{\chi_{sh_i}}{d s}=\sqrt{b_{i}\left(\Omega_{i}^{2}+6\xi_{S}^{2}\dfrac{s^{2}}{M_{p}^{2}}\right)}\,.
\label{chi_rho_h_i}
\end{align}
\begin{table}[htb!]
\begin{center}
\scalebox{0.70}{%
\begin{tabular}{ |c||c|c|c| }
 \hline
 inflation along & Potential \eqref{VEpot_quartic_angles3} minimized at & Inflationary conditions & Einstein frame slow roll potential  \\
 \hline \hline
$s\,h_{1}$& \begin{tabular}{@{}c@{}}$\gamma_{0} =\arctan\left(\sqrt{-\frac{\lambda_{1S}}{\lambda_{1}}}\right)$\\~\\ $\vartheta_{0}=0$ \end{tabular}&  \begin{tabular}{@{}c@{}}\\ $\kappa_{s1}\geq 0$~,~ $\kappa_{s2}\leq 0$\\~\\
$\kappa_{1s}\leq 0$~,~ $\kappa_{2s}\geq 0$\\~\end{tabular}
& $\dfrac{\lambda_{sh_1}}{8}s^4\left( 1 + \xi_S\frac{s^2}{M_P^2}\right)^{-2}$ \\ 
\hline 
$s\, h_{2}$& \begin{tabular}{@{}c@{}}$\gamma_{0} =\arctan\left(\sqrt{-\frac{\lambda_{2S}}{\lambda_{2}}}\right)$\\~\\ $\vartheta_{0}=\dfrac{\pi}{2}$ \end{tabular}
&  \begin{tabular}{@{}c@{}}\\
$\kappa_{s1}\leq 0$~,~ $\kappa_{s2}\geq 0$\\~\\
$\kappa_{1s}\geq 0$~,~ $\kappa_{2s}\leq 0$\\~\end{tabular}& $\dfrac{\lambda_{sh_2}}{8}s^4\left( 1 + \xi_S\frac{s^2}{M_P^2}\right)^{-2}$ \\ 
\hline 
$s\,h_{12}$&\begin{tabular}{@{}c@{}}$\gamma_{0} =\arctan\left(\sqrt{-\frac{\kappa_{s 2}+\kappa_{s 2}}{\lambda_{1}\lambda_{2}-\lambda_{34}^2}}\right)$\\~\\ $\vartheta_{0}=\arctan\left(\sqrt{\frac{\kappa_{s1}}{\kappa_{s2}}}\right)$ \end{tabular}
&  \begin{tabular}{@{}c@{}}\\ $\kappa_{s1}\leq 0$~,~ $\kappa_{s2}\leq 0$\\~\\
$\kappa_{1s}\leq 0$~,~ $\kappa_{2s}\leq 0$\\~
   \end{tabular}
& \begin{tabular}{@{}c@{}}\\
 $\dfrac{\lambda_{sh_{12}}}{8}s^4\left( 1 +\xi_{2}\frac{s^2}{M_P^2}\right)^{-2}$ \\~
\end{tabular}\\
\hline
$s$&\begin{tabular}{@{}c@{}}$\gamma_{0} =0$\\~\\$\vartheta_{0}=\left\lbrace 0,\frac{\pi}{2}\right\rbrace$\end{tabular}
&  \begin{tabular}{@{}c@{}}\\ $\kappa_{1s}\geq 0$~,~ $\kappa_{2s}\geq 0$\\~\\
$\vee ~\lambda_{1S,2S}\ll \lambda_{S}$\\~
   \end{tabular}
& \begin{tabular}{@{}c@{}}\\
 $\dfrac{\lambda_{S}}{8}s^4\left( 1 +\xi_{S}\frac{s^2}{M_{p}^2}\right)^{-2}$ \\~
\end{tabular}\\
\hline  
\end{tabular}}
\end{center}
\captionsetup{justification=centering}
\caption{Conditions and characteristics for PQI and PQTHI, i.e. $s$- and $s h_{1,2,12}$-inflation, with $\xi_{S}\gg\xi_{1,2}$.}
\label{tab:inflation_cond}
\end{table}
By integration we obtain the canonically normalized fields $\chi_{s, sh_1, sh_2,sh_{12}}$
\small{
\begin{align}
&\frac{\sqrt{\xi_{S}}}{M_{p}}\chi_{s}=\sqrt{1+6 \xi_{S}}\, \text{arcsinh}\left(\sqrt{1+6 \xi_{S}}\, u(s)\right)-\sqrt{6\xi_{S}}\, \text{arctanh}\left(\frac{\sqrt{6\xi_{S}}\, u(s)}{\sqrt{1+(1+6 \xi_{S})\, u^2(s)}}\right)\,,\\
&\frac{1}{M_{p}}\sqrt{\frac{\xi_{S}}{b_i}}\chi_{sh_i}= \sqrt{1+6 \xi_{S}}\, \text{arcsinh}\left(\sqrt{\frac{1+6 \xi_{S}}{b_i}}u(s)\right)-\sqrt{6 \xi_{S}}\,  \text{arctanh}\left(\frac{\sqrt{6\xi_{S}}\, u(s)}{\sqrt{b_i+(1+6 \xi_{S})\, u^2(s)}}\right)
\end{align}}
with $u(s)\equiv \sqrt{\xi_{S}}s/M_{p}$. By taking the inverse of the canonically normalized fields, we can write the inflationary scalar potential in Einstein frame for PQI and PQTHI in the usual form:
\begin{align}
V_{s,sh_{i}}(\chi_i)=\dfrac{\tilde{\lambda}}{8} \dfrac{s^{4}\left(\chi_i\right)}{\left(1+\xi_{S}\frac{s^{2}\left(\chi_i\right)}{M_{p}^{2}}\right)^{2}}\,,
\label{scal.pot. phi chi_rho}
\end{align}
where $\tilde{\lambda}$ is given by the following inflationary directions:
\begin{align}
\tilde{\lambda}=
\begin{cases}
\lambda_{S} & \text{(PQI)}\\
\lambda_{S}-\dfrac{\lambda_{iS}^{2}}{\lambda_{i}} & \text{(PQTHI-$s h_{i}$)}\\
\lambda_{S}-\dfrac{\lambda_{1} \lambda_{2S}^2+\lambda_{1S}^2 \lambda_{2}-2 \lambda_{1S} \lambda_{2S} \lambda_{34}}{\lambda_{1} \lambda_{2}-\lambda_{34}^2} & \text{(PQTHI-$s h_{12}$)}
\end{cases}\,.
\label{eq:effective_lambda_s}
\end{align}
In Table \ref{tab:inflation_cond} we summarize the extrema, inflationary conditions and Einstein frame slow-roll potential for PQI and PQTHI. We show the two inflationary scenarios in figure \ref{fig:2hdsmash_pots} which represents PQI (left) and PQTHI (right).
As discussed in the appendix ~\ref{app:2hdm inflation}, the inflationary predictions $A_s$, $n_s$ and $r$ are constrained by PLANCK/BICEP data \cite{Planck:2018vyg,Planck:2018jri,BICEP:2021xfz}. According to our naturalness philosophy, we demand the non-minimal coupling to be constrained by $\xi_S \lesssim 1$ which is shown in figure \ref{fig:r_vs_xi_and_lambda_vs_xi}. Thus, $A_s$ sets the following limits to $\xi_S$ and $\tilde{\lambda}$
\begin{align}
&8.5\times 10^{-3}\lesssim\xi\lesssim 1~~\text{implying}~~9\times 10^{-10}\gtrsim \tilde{\lambda}\gtrsim 4.5\times 10^{-13}\,.
\end{align}
In order to determine $n_s$ and $r$ we require the number of e-folds from some time $t_\ast$ where the scale $k_{\ast}=a_{\ast}\mathcal{H}_{\ast}$ exited the horizon to the time where inflation ended denoted by $k_{\text{end}}=a_{\text{end}}\mathcal{H}_{\text{end}}$. This is defined by
\begin{align}
N_{\ast}\equiv \log\left(\dfrac{a_{\text{end}}}{a_{k_{\ast}}}\right)=\int_{\chi_{I}}^{\chi_{\text{end}}}\dfrac{d\chi}{\sqrt{2\epsilon}}\,,
\end{align}
which can be solved exactly by the Klein-Gordon eq.
\begin{align}
\dfrac{d^{2}\chi}{dN^{2}}+3\dfrac{d\chi}{dN}-\dfrac{1}{2 M_{p}^{2}}\left(\dfrac{d\chi}{dN}\right)^{3}+\sqrt{2\epsilon}\left(3M_{p}-\dfrac{1}{2M_{p}}\left(\dfrac{d\chi}{dN}\right)^{2}\right)\,.
\label{eq:kein_gordon_PQ_mixed}
\end{align}
During inflation, the largest scales exit the horizon at $k_{\ast}$ which re-enter the horizon at a later time, i.e. at matter or radiation domination, cf. Ref. \cite{Kamionkowski:2015yta}. 
In fact, the largest scales are the last to re-enter the horizon which corresponds to scales of our current horizon, i.e. $k_{0}=a_{0}\mathcal{H}_{0}$. The required amount of e-folds for solving the horizon problem is therefore related to the complete expansion history of the universe and is given by \cite{Liddle:2003as} 
\begin{align}
\frac{k_{\ast}}{k_{0}}\equiv\frac{k_{\ast}}{a_0 \mathcal{H}_0} = \frac{a_{k_{\ast}} \mathcal{H}_{k_{\ast}}}{a_0 \mathcal{H}_0} = \frac{a_{k_{\ast}}}{a_{\text{end}}} \, \frac{a_{\text{end}}}{a_{\text{eq}}}  \, 
\frac{\mathcal{H}_{k_{\ast}}}{\mathcal{H}_{\text{eq}}} \,\frac{a_{\text{eq}}\mathcal{H}_{\text{eq}}}{a_0 \mathcal{H}_0} \,,
\label{eq:N_scale_factor_relation}
\end{align}
where the subscripts "eq" and "0" denote matter-radiation equality and the present time, respectively. With the ratio $a_{k_{\ast}}/a_{\text{end}}=\exp(-N_{\ast})$, we identify the number of e-folds during inflation $N_{\ast}$ by reformulating eq. \eqref{eq:N_scale_factor_relation} to find \cite{Liddle:2003as}
\begin{align}
N_{\ast}=-\log\frac{k_{\ast}}{a_0 \mathcal{H}_{0}}+\log\frac{a_{\text{end}}}{a_{\text{eq}}}+\log\frac{\mathcal{H}_{k_{\ast}}}{\mathcal{H}_{\text{eq}}}+\log\frac{a_{\text{eq}}\mathcal{H}_{\text{eq}}}{a_{0}\mathcal{H}_{0}}\,.
\end{align}
By using the slow-roll approximation for the Hubble rate during inflation, i.e.\\ $\mathcal{H}_{k_{\ast}}\simeq \sqrt{V_{k_{\ast}}/3\,M_{p}^{2}}$, we obtain \cite{Liddle:2003as}
\begin{align}
N_{\ast}=-\log\frac{k_{\ast}}{a_0 \mathcal{H}_{0}}+\log\frac{a_{\text{end}}}{a_{\text{eq}}}+\log\sqrt{\frac{V_{k_{\ast}}}{3~M_{p}^{2}}}\frac{1}{\mathcal{H}_{\text{eq}}}+\log\frac{a_{\text{eq}}\mathcal{H}_{\text{eq}}}{a_{0}\mathcal{H}_{0}}\,.
\end{align}
In 2hdSMASH we adopt the simplicity of the expansion history of the universe from SMASH \cite{Ballesteros:2016xej}. Based on the smallness of the non-minimal couplings, they can be neglected by the end of inflation. Correspondingly, preheating and reheating occur in an approximate quartic potential, where the universe is radiation-dominated. This has been worked out for SMASH in detail by elaborate lattice simulations where it was shown that after a few oscillations the scalar potential is approximately quartic \cite{Ballesteros:2016xej,2022arXiv220300621R}. 

\begin{figure}[htb!]
\centering
\includegraphics[width=1\linewidth]{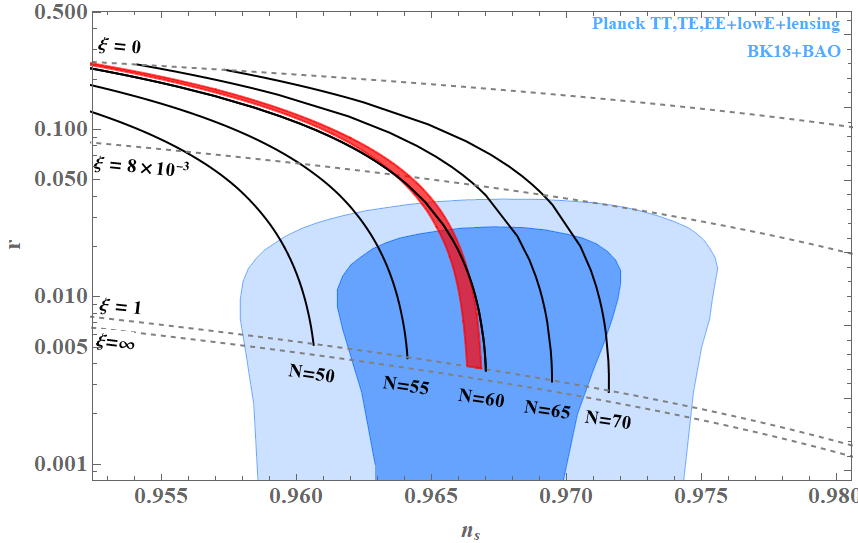}
\caption{Shown are the 95\% and 68\% C.L. contours of $r$ vs. $n_{s}$ constrained by PLANCK/BICEP \cite{Planck:2018vyg,Planck:2018jri,BICEP:2021xfz} (shaded blue),  the isocontours of constant $\xi$ (gray dashed), the isocontours of constant $N$ (solid black) and the inflationary predictions by 2hdSMASH for PQI/PQTHI taking into account the numerically determined expansion history of the universe (thick red).}
\label{fig:ns_r}
\end{figure}

By using the relation
\begin{align}
\log\frac{a_{\text{end}}}{a_{\text{eq}}}=\log\frac{a_{\text{end}}}{a_{0}}+\log\frac{a_{0}}{a_{\text{eq}}}\,,
\end{align}
we can simplify the number of e-folds as follows
\begin{align}
N_{\ast}=-\log\frac{k_{\ast}}{a_0 \mathcal{H}_{0}}+\log\frac{a_{\text{end}}}{a_{0}}+\log\sqrt{\frac{V_{k_{\ast}}}{3~M_{p}^{2}}}+\log\frac{1}{\mathcal{H}_{0}}\,.
\end{align}
The ratio $a_{\text{end}}/a_{0}$ relates the scale of the end of inflation with the scale of the present time which is given by \cite{Figueroa:2017vfa}
\begin{align}
\frac{a_{\text{end}}}{a_0} = \frac{a_{\text{end}}}{a_{\text{RD}}}\frac{a_{\text{RD}}}{a_{\text{end}}} =  \left(\frac{g_{\ast s,\text{RD}}}{g_{\ast s,0}}\right)^{-1/3}\left(\frac{g_{\ast \rho,\text{RD}}}{g_{\ast \rho,0}}\right)^{1/4}\left(\frac{\rho_{0,\text{RD}}}{\rho_{\text{end}}}\right)^{1/4}\left(\frac{a_{\text{end}}}{a_{\text{RD}}}\right)^{(1-3w)}\,,
\label{eq:aend_a0}
\end{align}
where the subscript "RD" denotes radiation domination, $g_{\ast,\rho}$ and $g_{\ast,s}$ denote the number of relativistic degrees of freedom for the energy density and entropy, respectively. Since the universe is radiation-dominated by the end of inflation, the energy of state parameter $w=p/\rho$ approaches $1/3$ instantaneously in the epoch of preheating for which we can approximate eq. \eqref{eq:aend_a0}
\begin{align}
\frac{a_{\text{end}}}{a_0} \simeq  \left(\frac{g_{\ast s}(T_R)}{g_{\ast s}(T_0)}\right)^{-1/3}\left(\frac{g_{\ast \rho}(T_R)}{g_{\ast \rho}(T_0)}\right)^{1/4}\left(\frac{\rho_{0,\text{RD}}}{V_{\text{end}}}\right)^{1/4}\,,
\end{align}
where the subscript "$R$" denotes reheating. The corresponding number of relativistic degrees of freedom are given by \cite{Ringwald:2020vei}
\begin{align}
g_{\ast s}(T_0)\simeq 3.91~,~g_{\ast \rho}(T_0)\simeq 2~,~g_{\ast s}(T_R)\simeq g_{\ast \rho}(T_R)\simeq 124.5 
\end{align}
and the present time energy density of radiation $\rho_{0,\text{RD}}$ is given by \cite{Ringwald:2020vei}
\begin{align}
\rho_{0,\text{RD}}=\frac{\pi^{2}T_{0}^{4}}{15}\simeq 2.02\times 10^{-15}~\text{eV}^{4}\,. 
\end{align}
Furthermore, we give the updated values of the present time Hubble parameter $\mathcal{H}_{0}$ and Hubble scale $k_{0}=a_{0}\mathcal{H}_{0}$ by Ref. \cite{Zyla:2020zbs}
\begin{align}
&\mathcal{H}_{0}\simeq 5.9\times 10^{-61}\, h\, M_{p}~~~~~~~~\text{with}~~~ h\simeq 0.674\,,\label{eq:infl_constants}\\
&a_{0}\mathcal{H}_{0}\simeq 22.47\times 10^{-5}\, \text{Mpc}^{-1}\,,\notag  
\end{align}
where $h$ is the Hubble constant. The number of e-folds can now be approximated and reads\\ \cite{Liddle:2003as,Liddle:1993fq}
\begin{align}
N_{\ast}&\simeq 61.25 -\log\frac{k_{\ast}}{a_{0}\mathcal{H}_{0}}-\log\frac{10^{16}\,\text{GeV}}{V_{k_{\ast}}^{1/4}}+\log\frac{V_{k_{\ast}}^{1/4}}{V_{\text{end}}^{1/4}}\label{eq:N_thermal}\\
&\simeq 59 -\log\left(\frac{k_{\ast}}{10^{-3}~\text{Mpc}^{-1}}\right)-\log\left(\frac{10^{-4}~\text{Mpc}^{-1}}{a_{0}\mathcal{H}_{0}}\right)-\log\frac{10^{16}\,\text{GeV}}{V_{k_{\ast}}^{1/4}}+\log\frac{V_{k_{\ast}}^{1/4}}{V_{\text{end}}^{1/4}}\,,\notag
\end{align}
where we included the maximum horizon of the observable universe given by the upper bound $V_{k_{\ast}}^{1/4}\lesssim 10^{16}$ GeV, cf. \cite{Ballesteros:2016euj,Ballesteros:2016xej,Ringwald:2020vei}. The scale of horizon exit is $k_{\ast}=0.002$ Mpc$^{-1}$ which corresponds to the largest observable scales in the CMB measured by Planck/BICEP \cite{Planck:2018vyg,BICEP:2021xfz}. With the constants of eq. \eqref{eq:infl_constants} we obtain
\begin{align}
N_{\ast}\simeq 59.06 -\log\frac{10^{16}\,\text{GeV}}{V_{k_{\ast}}^{1/4}}+\log\frac{V_{k_{\ast}}^{1/4}}{V_{\text{end}}^{1/4}}\,,
\label{eq:N_thermal2}
\end{align} 
which illustrates that only the energy scales $V_{k_{\ast}}^{1/4}$ and $V_{\text{end}}^{1/4}$ are relevant. By utilizing the the Planck constraints to fit the quartic coupling $\tilde{\lambda}$ to $A_s$ we can determine $N_{\ast}$ of eq. \eqref{eq:N_thermal2} which must match the number of e-folds obtained from solving the inflaton's eom from eq. \eqref{eq:kein_gordon_PQ_mixed}. Correspondingly, the energy scales $V_{k_{\ast}}^{1/4}$ and $V_{\text{end}}^{1/4}$ are determined. 
\begin{figure}
\centering
  \includegraphics[width=.8\linewidth]{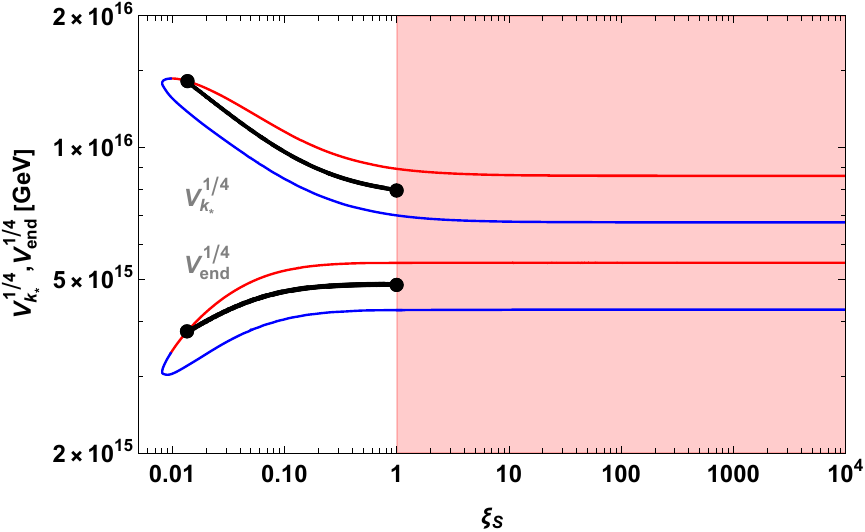}
\caption{Shown are the 95\% C.L. contours of $V_{k_{\ast}}^{1/4}$ and $V_{\text{end}}^{1/4}$ vs. $\xi_{S}$ at the pivot scale $k_{\ast}=0.002$ Mpc$^{-1}$. The red shaded region given by $\xi_S >1$ indicates the unnatural regime according to our naturalness philosophy. The red and blue curves indicate constraints given by the redder spectrum and the bluer spectrum of $n_{s}$ \cite{Planck:2018vyg,BICEP:2021xfz}. The black solid curves take the 2hdSMASH expansion history of our universe into account where inflation is followed by immediate radiation-domination. Correspondingly, the black dots specify the range of validity for a consistent expansion history.}
\label{fig:vk_vend}
\end{figure}
In figure \ref{fig:vk_vend} we show this correspondence of $V_{k_{\ast}}^{1/4}$ and $V_{\text{end}}^{1/4}$ as a function of the non-minimal coupling $\xi_{S}$ where the solid black curve manifests this result. Typical values of $V_{k_{\ast}}^{1/4}$ and $V_{\text{end}}^{1/4}$ are in the range of
\begin{align}
&8\times 10^{15}~\text{GeV}\lesssim V_{k_{\ast}}^{1/4} \lesssim 1.4 \times 10^{16}~\text{GeV}\,,\\
&3.8\times 10^{15}~\text{GeV}\lesssim V_{\text{end}}^{1/4} \lesssim 4.8 \times 10^{15}~\text{GeV} 
\end{align} 
for
\begin{align}
1.35\times 10^{-2}\lesssim \xi_{S} \lesssim 1~~~\text{implying}~~~ 9\times 10^{-10}\gtrsim \tilde{\lambda}\gtrsim 2.2\times 10^{-12}\,.
\end{align}
These ranges are indicated by the black dots in figure \ref{fig:vk_vend} for which we can compute the corresponding range of the number of e-folds $N_{\ast}$ from eq. \eqref{eq:N_thermal2}
\begin{align}
59.3\lesssim N_{\ast}\lesssim 60.7\,.
\label{eq:N_range}
\end{align}
We show our result in figure \ref{fig:ns_r} where the thick red curve is obtained by eq. \eqref{eq:N_thermal2} which accounts for the numerically determined expansion history\footnote{We utilized a numerical code from Ref. \cite{Ballesteros:2016xej} to obtain the red curve shown in figure \ref{fig:ns_r}.}. The black solid lines have been acquired by solving the Klein-Gordon equation \eqref{eq:kein_gordon_PQ_mixed} without taking the expansion history of our universe into account. The red curve is close to $N=60$ as indicated in eq. \eqref{eq:N_range}.
\\
Correspondingly, we provide the range of $n_s$ and $r$ for PQI and PQTHI in 2hdSMASH
\begin{align}
&\xi_S\simeq 1 \Rightarrow 
\begin{cases}
0.9664\lesssim n_s \lesssim 0.9668\\
0.0037\gtrsim r\gtrsim 0.0036
\end{cases}~~~,~~~
\xi_S\simeq 1.35\times 10^{-2} \Rightarrow 
\begin{cases}
0.9646\lesssim n_s \lesssim 0.9651\\
0.037\gtrsim r\gtrsim 0.036
\end{cases}\,.
\end{align} 
\section{Connecting Inflation with TeV Scale Particle Physics}
\label{sec:infl_implications_portal_couplings}
This section is dedicated to make the connection between inflationary constraints and low energy constraints. 
For this reason, we consider the one-loop RGEs (see appendix \ref{app:rges}) which run the 2hdSMASH parameters from low- to high-scale and meet the constraints at their respective scales. In an effective field theory approach, the PQ-scalar $s$ can be integrated out below the matching scale $m_{s}\sim \sqrt{\lambda_{S}}v_{S}$ where 2hdSMASH is matched to its low-energy theory, i.e. the $\nu$2HDM. Hence, the PQ self-coupling $\lambda_{S}$ will determine the matching scale for fixed PQ-scale $v_{S}$.\\  
\\
In order to ensure successful inflation at the Planck scale, i.e. $\lambda_{S}(M_{P})\lesssim 10^{-10}$, the parameters need to evolve stable from the matching scale to $M_{P}$. To this effect, constraints on the  portal couplings and the neutrino Yukawa couplings ensure the required stability as we will see in sec. \ref{subsec:stability_analysis_of_ls}. Moreover, there are two inflationary scenarios realized in 2hdSMASH, namely PQI and PQTHI, which differ by sign and size of the portal couplings. These constraints must be met at the Planck scale. In order to account for thermal leptogenesis and BAU, we obtain further constraints on the neutrino Yukawa couplings $Y_\nu$ and $Y_N$ affecting the RG evolution of the portal couplings. This effect will only consider PQI while PQTHI scenarios will face the challenge to run the portal couplings stable all the way up to the Planck scale. From a naturalness point of view, we further demand that the portal couplings do not impact Higgs phenomenology through large radiative corrections giving rise to an enhanced Poincaré symmetry $\mathcal{G}_{P}^{2\textit{HDM}}\times \mathcal{G}_{P}^{S}$ 
(cf. sec.~\ref{sec:mass_spec}) 
where $\lambda_{1S,2S,12S},Y_{N}\to 0$ \cite{Volkas:1988cm,Clarke:2015bea}. Hence, the portal couplings underlie tight constraints for connecting inflation with particle phenomenology. Along these lines, all parameters need to run stable and perturbative accounted by the vacuum stability and perturbative unitarity conditions (see sec. \ref{subsec:theoretical_const}). High-scale validity is ensured by obeying the BfB and perturbative unitarity conditions for all energy scales until the Planck scale. By breaking perturbative unitarity, Landau poles emerge which lead to a breakdown of predictivity. If vacuum stability is not guaranteed, our universe would not correspond to the universe we know today. We will discuss these intricacies in the following subsections.\\
\\
We structure the sections as follows: In section \ref{subsec:stability_analysis_of_ls} we will discuss the stability requirements for $\lambda_{S}$ by considering the RG running effects of the right handed neutrino Yukawa and portal couplings. In section \ref{subsec:bau_thermal_lepto} we discuss the constraints from thermal leptogenesis and BAU and their impact on RG running. In section \ref{subsec:Infl_particle_pheno} we will analyze the running of the portal couplings and give conditions which motivate PQI and PQTHI. Therefore, we give three benchmark points which underline our considerations and illustrate the interconnection of inflation and Higgs phenomenology. In section \ref{subsec:stability_analysis_of_2hdm} we discuss vacuum-stability, perturbativity and high-scale validity by identifying the effective 2HDM model in our RG-analysis. Since all other couplings are fixed, i.e. portal couplings $\lambda_{1S,2S,12S}$, neutrino Yukawa couplings $Y_{\nu,N}$ and PQ self-coupling $\lambda_S$, we consider the remaining 2HDM couplings $\lambda_{1,2}$ and $\lambda_{34}$. In particular we pay attention to the stability of $\lambda_{2}$ which tends to run negative. We will argue that $\lambda_3$ and $\lambda_{34}$ can cure this instability while keeping $\lambda_{1}$ fixed. This is illustrated by representative benchmarks. In section \ref{subsec:benchmark_points} we accumulate all of the analysis and provide benchmark points which accommodate inflation, theoretical and experimental constraints while presenting interesting Higgs phenomenology relevant for HL-LHC and future colliders.
\subsection{Stability analysis of $\lambda_{S}$}
\label{subsec:stability_analysis_of_ls}
In this section we want to outline the stability condition on the running of $\lambda_{S}$ for successful inflation. This in turn will lead to conditions on the Higgs portal couplings $\lambda_{1S}$, $\lambda_{2S}$, $\lambda_{12S}$ and the right handed Yukawa couplings $Y_N$. We give to one-loop accuracy, the renormalization group evolution of $\lambda_S(\mu )$ (see Appendix~\ref{app:rges})
\begin{align}
&(4\pi)^2\frac{ d}{{ d}\ln\mu} \lambda_S = \label{eq:rge_lambdaS}\\
& 10\lambda_S^2 + 4\lambda_{1S}^2 + 4\lambda_{2S}^2 + 8\lambda_{12S}^2 -2\text{Tr}\left( Y_N^\dagger Y_N Y_N^\dagger Y_N \right) +2\lambda_S \text{Tr}\left( Y_N^\dagger Y_N\right).\notag
\end{align}
As discussed in section \ref{sec:PQ-and mixed inflation}, we require for successful inflation $\lambda_S(M_p)\lesssim 10^{-10}$~ \cite{Ballesteros:2016xej}. Such a small quartic coupling can in fact suffer from RG-running instability, i.e. large enhancements of $\lambda_S$ at high energies, thus spoiling inflation. Therefore, we require $\lambda_S(m_s)\sim \lambda_S(M_{p})$ for stability reasons. Imposing this requirement on eq. \eqref{eq:rge_lambdaS} leads to constraints on the squared  Higgs portal couplings $\lambda_{1S,2S,12S}^{2}$ and on the trace of right handed neutrino Yukawas $\text{Tr} \left(Y_{N}^{\dagger}Y_{N}Y_{N}^{\dagger}Y_{N}\right)$, as we will argue now.\\ 
\\
As a first observation, we can neglect the terms proportional to $\lambda_S$ on the right hand side of eq.~\eqref{eq:rge_lambdaS}. This is per our request for negligible running of $\lambda_S$. Furthermore, the RG running effects of the portal couplings and the right handed neutrino Yukawas are taken to be negligible due to RG stability. With these reasonable assumptions, we can integrate eq.~\eqref{eq:rge_lambdaS} and find
\begin{align}\label{eq:integ_ls}
&\frac{\lambda_{S}(\mu)}{ \lambda_{S}(m_s)}\approx \\&1+\frac{1}{4\pi^2  \lambda_{S}(m_s)}
\left[\lambda_{1S}^2(m_s) + \lambda_{2S}^2(m_s) + 2\lambda_{12S}^2(m_s) -\frac{1}{2}\text{Tr}\left( Y_N^\dagger Y_N Y_N^\dagger Y_N \right)_{m_s} \right]
\log \frac{\mu}{m_s }\label{eq:integ_ls}\,. \notag
\end{align}
From eq. \eqref{eq:integ_ls} we can see that the portal couplings enhance, while the right handed neutrino Yukawas diminish $\lambda_S$. This endangers vacuum stability (cf. Ref.~\cite{Ballesteros:2016xej}) and can be avoided in one of two ways. Either the portal coupling contribution is greater than the Yukawa contribution or both contributions are negligible compared to $\lambda_S$. By choosing the latter, we avoid tremendous fine tuning at the matching scale $m_s$ which leads to the following constraints 
\begin{equation}\label{eq:ls_upper_bound}
\left.
\begin{array}{l}
|\lambda_{1S}(m_s )| \\
|\lambda_{2S}(m_s )| \\
|\lambda_{12S}(m_s )| \\
\sqrt{\text{Tr}\left( Y_N^\dagger Y_N Y_N^\dagger Y_N \right)_{|m_s}}
\end{array}
\right\}
\ll \sqrt{\lambda_S(m_s )}\approx 10^{-5}
\,,
\end{equation}
implying $\lambda_S(M_P)\approx \lambda_S(m_s)\approx 10^{-10}$ as expected. Note that the small value of $\lambda_S$ is technically natural when the portal couplings are also small, as shown in eq.(\ref{eq:ls_upper_bound}). The portal couplings are technically natural due to the enhanced Poincare symmetry as shown in sec.~\ref{sec:mass_spec} and also pointed out in Ref.~\cite{Clarke:2015bea, Espriu:2015mfa}.

\subsection{RG-Analysis with BAU and thermal leptogenesis}
\label{subsec:bau_thermal_lepto}
\begin{figure}[htb!]
\centering
  \includegraphics[width=0.7\linewidth]{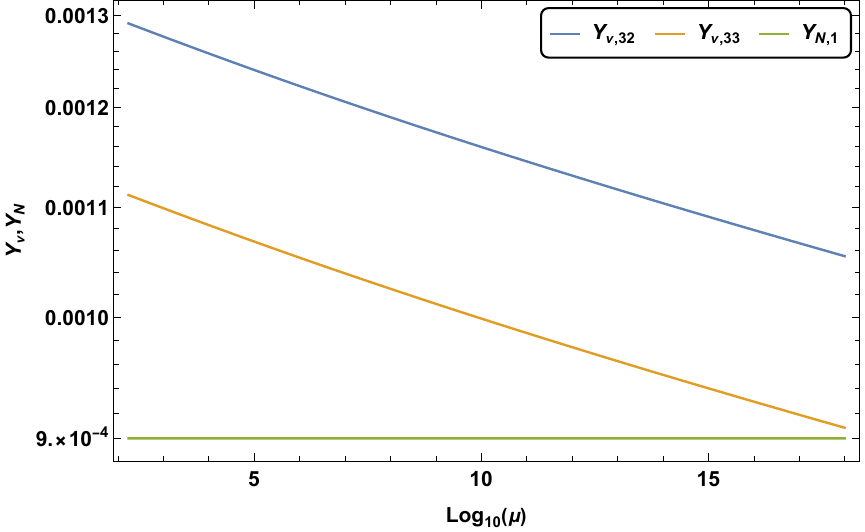}
\caption{Shown are the the RG-running of the largest components of the Dirac neutrino couplings $Y_{\nu,32,33}$ and the main Majorana neutrino coupling $Y_{N,1}$ with $t_{\beta}=5.5$, $\lambda_S=10^{-10}$ and $v_S=10^{10}$ GeV.  The renormalization scale is set by the top pole mass $m_{t}=\left(172.5\pm 0.7\right)$ GeV given by Ref. \cite{Zyla:2020zbs} to run from low- to high scale.}
\label{fig:yn_ynu_running}
\end{figure}

In the previous section we derived an upper bound on $\text{Tr}\left(Y_{N}^{\dagger}Y_{N}Y_{N}^{\dagger}Y_{N}\right)$ at the scale $m_{s}$ (see eq. \eqref{eq:ls_upper_bound}). This upper bound can be further specified for higher energy domains, e.g. $\mu \simeq 30 M_{p}$, in order to investigate the minimum condition of $\lambda_{S}$~\cite{Ballesteros:2016xej}. Therefore, by considering $\mu \simeq 30 M_{p}$ we obtain
\begin{align}
Y_{N}\lesssim \left(\dfrac{8\pi^{2}\lambda_{S}}{163\log\left(\frac{30 M_{\text{P}}}{m_{s}}\right)}\right)^{1/4}\,,
\label{eq:yn_upper_bound}
\end{align}
where $Y_{N,33}=Y_{N,22}=3Y_{N,11}$ accommodates vanilla leptogenesis with hierarchical right-handed neutrinos. By considering the right-handed neutrino masses given by
\begin{align}
M_{N,i}=\dfrac{Y_{N,i}v_{S}}{\sqrt{2}}
\end{align} 
with $M_{N,3}=M_{N,2}=3M_{N,1}$, we can convert the upper bound of eq. \eqref{eq:yn_upper_bound} to an upper bound on $M_{N,1}$:
\begin{align}
M_{N,1}\lesssim \dfrac{\sqrt{2}}{v_{S}}\left(\dfrac{8\pi^{2}\lambda_{S}}{163\log\left(\frac{30 M_{\text{P}}}{m_{s}}\right)}\right)^{1/4}\,.
\end{align}
The CP-violating and out-of-equilibrium decay of the lightest right-handed neutrino $N_{1}\to l\Phi_{1}$ produces BAU via thermal leptogenesis \cite{Clarke:2015bea,Clarke:2015hta,Ballesteros:2016xej}. This is quantified by the CP-asymmetry $\epsilon_{1}$ which is given by \cite{Buchmuller:1999cu,Buchmuller:2004nz,Covi:1996wh,Davidson:2002qv}:
\begin{align}
\epsilon_{1}&=\dfrac{\Gamma_{D}\left(N_{1}\to l\Phi_{1}\right)-\Gamma_{D}\left(N_{1}\to l^{\ast}\Phi_{1}^{\ast}\right)}{\Gamma_{D}\left(N_{1}\to l\Phi_{1}\right)+\Gamma_{D}\left(N_{1}\to l^{\ast}\Phi_{1}^{\ast}\right)}\simeq \dfrac{1}{8\pi}\dfrac{\sum_{j}\text{Im}\left[\left(Y_{\nu}Y_{\nu}^{\dagger}\right)_{1j}^{2}\right]}{\left(Y_{\nu}Y_{\nu}^{\dagger}\right)_{11}}~g\left(x_{1j}\right),
\end{align}
where $g_{1j}\left(x_{1j}\right)$ is given as
\begin{align}
g_{1j}\left(x_{1j}\right)=\sqrt{x_{1j}}\left(\dfrac{2-x_{1j}}{x_{1j}-1}+(1+x_{1j})\log\left(\dfrac{1+x_{1j}}{x_{1j}}\right)\right)~~~\text{with}~~~x_{1j}=\dfrac{M_{j}^{2}}{M_{1}^{2}}\,.
\end{align}
For $x_{1j}>1$ we can approximate the CP-asymmetry as follows \cite{Davidson:2002qv}:
\begin{align}
\epsilon_{1}&\simeq -\dfrac{3}{16\pi}\dfrac{\sum_{j}\text{Im}\left[\left(Y_{\nu}Y_{\nu}^{\dagger}\right)_{1j}^{2}\right]}{\left(Y_{\nu}Y_{\nu}^{\dagger}\right)_{11}}\left(\dfrac{M_{1}}{M_{j}}\right)=-\dfrac{3 M_{1}}{16\pi}\dfrac{\sum_{j}\text{Im}\left[\left(Y_{\nu}m_{\nu}^{\dagger}Y_{\nu}^{\dagger}\right)_{1j}^{2}\right]}{\left(Y_{\nu}Y_{\nu}^{\dagger}\right)_{11}}\,,
\end{align}
where we used eq. \eqref{eq:neutrino_mass_matrix} to substitute the light neutrino mass matrix $m_{\nu}$. Based on the observed BAU we need to place an upper bound on $\epsilon_{1}$. Therefore, we further simplify $\epsilon_{1}$ by using the Casas-Ibarra parametrization for the light neutrino Yukawa coupling $Y_{\nu}$:
\begin{align}
Y_{\nu}=\dfrac{\sqrt{2}}{v_{1}}D_{\sqrt{M}}~R~D_{\sqrt{m}}~U_{\nu}^{\dagger}\,,
\end{align}
where $D_{\sqrt{A}}\equiv \text{diag}\left(\sqrt{a_{1}},\sqrt{a_{2}},\sqrt{a_{3}}\right)$ and $R$ is an orthogonal (complex) matrix with $R^{\dagger}R=\mathbb{1}$. In particular, we use the fact that we sum over two right-handed neutrinos and use the orthogonality condition $\sum_j R_{1j}^{2}=1$. We can therefore approximate the upper bound as:
\begin{align}
\left|\epsilon_{1}\right|\lesssim \dfrac{3}{16\pi}\dfrac{M_{1}}{v_{2}^{2}}\left(m_{3}-m_{1}\right)\,.
\end{align}
For simplification we set $m_{1}\simeq 0$ and therefore get:
\begin{align}
\left|\epsilon_{1}\right|\lesssim \dfrac{3}{16\pi}\dfrac{M_{1}}{v_{2}^{2}}m_{3}\,.
\end{align} 
This upper bound on $\epsilon_{1}$ can be translated into a lower bound on the lightest right-handed neutrino mass $M_{N,1}$, i.e. Davidson-Ibarra bound \cite{Davidson:2002qv}:
\begin{align}
M_{N,1}\gtrsim \dfrac{5\times 10^{8}\text{ GeV}}{1+t_{\beta}^{2}}\,,
\end{align} 
which translates into a lower bound for $Y_{N,11}$
\begin{align}
Y_{N,11}\gtrsim \dfrac{7.1\times 10^{8}\text{ GeV}}{v_{S}\left(1+t_{\beta}^{2}\right)}\,.
\label{eq:yn_lower_bound}
\end{align}
By combining equations \eqref{eq:yn_lower_bound} and \eqref{eq:yn_upper_bound} we obtain the range for $Y_{N,11}$:
\begin{align}
\dfrac{7.1\times 10^{8}\text{ GeV}}{v_{S}\left(1+t_{\beta}^{2}\right)}\lesssim Y_{N,11}\lesssim \left(\dfrac{8\pi^{2}\lambda_{S}}{163\log\left(\frac{30 M_{\text{P}}}{m_{s}}\right)}\right)^{1/4}\,.
\label{eq:yn_bounds}
\end{align}
Once $Y_{N,11}$ is constrained by $v_{S}$, $t_{\beta}$ and $\lambda_{S}$ we are left with only nine degrees of freedom, namely $Y_{\nu,ij}$. In order to obtain the left-handed neutrino Yukawa couplings $Y_{\nu}$, we need to calculate \cite{Casas:2001sr}
\begin{align}
Y_{\nu}=\frac{\sqrt{1+t_{\beta}^{2}}}{v}~\mathcal{D}_{\sqrt{M}}O\mathcal{D}_{\sqrt{m_{\nu}}}U_{\nu}^{\dagger}\,,
\end{align}
where 
\begin{align}
&\mathcal{D}_{\sqrt{M}}\equiv\text{diag}\left(\sqrt{M_{N,1}},\sqrt{3M_{N,1}},\sqrt{3M_{N,1}}\right)\,,\\
&\mathcal{D}_{\sqrt{m_{\nu}}}\equiv\text{diag}\left(\sqrt{m_{\nu,1}},\sqrt{m_{\nu,2}},\sqrt{m_{\nu,3}}\right)
\end{align}
denote the diagonalized right-handed Majorana and left-handed Dirac neutrino mass matrix respectively. $U_{\nu}$ is the PMNS-neutrino mixing matrix whose components are given by the best global fits of Ref. \cite{deSalas:2020pgw}. $O$ is a $3\times 3$-orthogonal matrix which is in general complex and consists of three complex angles. However, we follow the same line of reasoning as in Ref. \cite{Ballesteros:2016xej} by taking $O$ to be of unity since we can neglect $\mathcal{O}(1)$ contributions for the RGE stability analysis. Furthermore, the left-handed neutrinos are constrained by neutrino oscillation experiments \cite{deSalas:2020pgw} and cosmological neutrino observations \cite{Planck:2018vyg}. From the PLANCK 2018~\cite{Planck:2018vyg} constraints we obtain the upper bound on the sum of the neutrino masses:
\begin{align}
\sum_{i=1}^{3}m_{\nu,i}<0.12\text{ eV}~~~\left(95\%\text{ C.L. Planck TT,TE,EE+lowE+lensing+BAO}\right)\,.
\end{align}
The experimental best fit constraints of neutrino masses from atmospheric and solar mass splitting \cite{deSalas:2020pgw} for normal ordering (NO) are given by:
\begin{align}
\Delta m_{21}^{2}\left(10^{-5}\text{ eV}^{2}\right)= 7.5\substack{+0.22\\-0.2}~~,~~\Delta m_{31}^{2}\left(10^{-3}\text{ eV}^{2}\right)=2.55\substack{+0.02\\-0.03}~~~\text{(best fit $\pm 1\sigma$, NO)}\,.
\label{eq:neutrino_mass_splitting}
\end{align}
We consider a hierarchical mass ordering $m_{\nu,1}<m_{\nu,2}<m_{\nu,3}$ where we take, as mentioned above, $m_{\nu,1}\equiv m_{1}\simeq 0$. Then, the left-handed neutrino Yukawa matrix is given by:
\begin{align}
Y_{\nu,(i+1)j}\simeq \dfrac{1}{v}\sqrt{\dfrac{3}{\sqrt{2}}Y_{N,11}m_{\nu,(i+1)}\left(1+t_{\beta}^{2}\right)v_{S}}~U_{\nu,j(i+1)}^{\ast}~~~\text{with}~~~Y_{\nu,1j}\simeq 0\,,
\end{align}
where $m_{\nu,2}$ and $m_{\nu,3}$ are given by:
\begin{align}
m_{\nu,2}\simeq \Delta m_{2,1}\simeq 8.66\text{ meV}~~~,~~~m_{\nu,3}\simeq \sqrt{\Delta m_{2,1}^{2}+\Delta m_{3,1}^{2}}\simeq 51.23\text{ meV}\,.
\end{align} 
Since $m_{\nu,3}\gg m_{\nu,2}\gg m_{\nu,1}$ and $U_{\nu,32,33}^{\ast}\gtrsim U_{\nu,ij}^{\ast}$, we only consider $Y_{\nu,32,33}$ for the remainder of this paper, i.e.
\begin{align}
Y_{\nu,32,33}\simeq \dfrac{1}{v}\sqrt{\dfrac{3}{\sqrt{2}}Y_{N,11}m_{\nu,3}\left(1+t_{\beta}^{2}\right)v_{S}}~U_{\nu,32,33}^{\ast}\,.  
\end{align}
The aforementioned nine degrees of freedom are thus constrained by experimental values and by the benchmark relevant parameter $t_{\beta}$. Hence, we can account for BAU by vanilla thermal leptogenesis with hierarchical normal ordering of the light neutrino masses.\\
\\
There are a few caveats to consider regarding the size of the neutrino couplings in the RG-flow of $\lambda_{1S}$. Therefore, we consider the RGE of $\lambda_{1S}$:
\begin{align}
 \mathcal{D} \lambda_{1S} = &\ 
   \lambda_{1S}\left( -\frac32 g_1^2 -\frac92 g_2^2 
 + 4\lambda_{1S} +4\lambda_S +6\lambda_1  \right) + \lambda_{2S}\left(4\lambda_3 + 2\lambda_4\right)+8\lambda_{12S}^2\nonumber\\
 &+ \lambda_{1S}\left( 6Y_b^2 + 2Y_\tau^2 
+ 2\text{Tr}\left(Y_\nu^\dagger Y_\nu\right) 
 + \text{Tr}\left(Y_N^\dagger Y_N\right) \right) 
 - 4\text{Tr}\left(Y_\nu^\dagger Y_\nu Y_N^\dagger Y_N\right)\,. 
\end{align}
As we can see the running is dominated by the size of $\lambda_{1S}$, $\lambda_{2S}$ and $\text{Tr}\left(Y_{\nu}^{\dagger}Y_{\nu}Y_{N}^{\dagger}Y_{N}\right)$ where the latter contributes negatively on the running. For PQTHI-scenarios we do not have to worry about the size of the neutrino Yukawa term since $\lambda_{1S,2S}\gg\text{Tr}\left(Y_{\nu}^{\dagger}Y_{\nu}Y_{N}^{\dagger}Y_{N}\right)$ which is thus negligible. However, for PQI the portal couplings can be of the same size as the neutrino Yukawa term, i.e. $\lambda_{1S,2S}\sim\text{Tr}\left(Y_{\nu}^{\dagger}Y_{\nu}Y_{N}^{\dagger}Y_{N}\right)$, which causes $\lambda_{1S}$ to run negative at higher energies. In order to secure positive portal couplings at the inflationary scale for PQI, we impose a condition on the initial value of $\lambda_{1S}$ to guarantee $\lambda_{1S}(M_{P})>0$:
\begin{align}
\lambda_{1S}\gtrsim \text{Tr}\left(Y_{\nu}^{\dagger}Y_{\nu}Y_{N}^{\dagger}Y_{N}\right)\,.
\label{eq:bau_lepto_l1s_estimate}
\end{align}
For the remainder of the paper, we will only consider the real entries of $Y_{\nu}$, i.e. $\text{Re}(Y_{\nu,ij})$, since the RG-analysis, except for $Y_{\nu}$ itself, is independent of the imaginary part. The RG-running of $Y_{\nu}$ is severly suppressed by multiplicatives of its own value at the electroweak scale (see eq. \eqref{eq:rge_ynu}). Therefore, its value at the electroweak scale can be approximated to be the same at the inflationary scale. The same applies even more so for $Y_N$. We demonstrate this fact in figure \ref{fig:yn_ynu_running} for $Y_{\nu,32,33}$ and $Y_{N,1}$. We can see that the values of the Yukawa couplings are approximately the same at the inflationary scale as they are at the electroweak scale. However, for completeness, we will include the running of the neutrino Yukawas into our RG-analysis but note that their value is almost scale-invariant, as we would expect from looking at the RGEs of eqns \eqref{eq:rge_ynu}-\eqref{eq:rge_yn} of appendix \ref{app:rges}.

\subsection{Interconnection of portal couplings}
\label{subsec:Infl_particle_pheno}
\begin{figure}[ht]
\begin{center}
\includegraphics[width=0.7\textwidth]{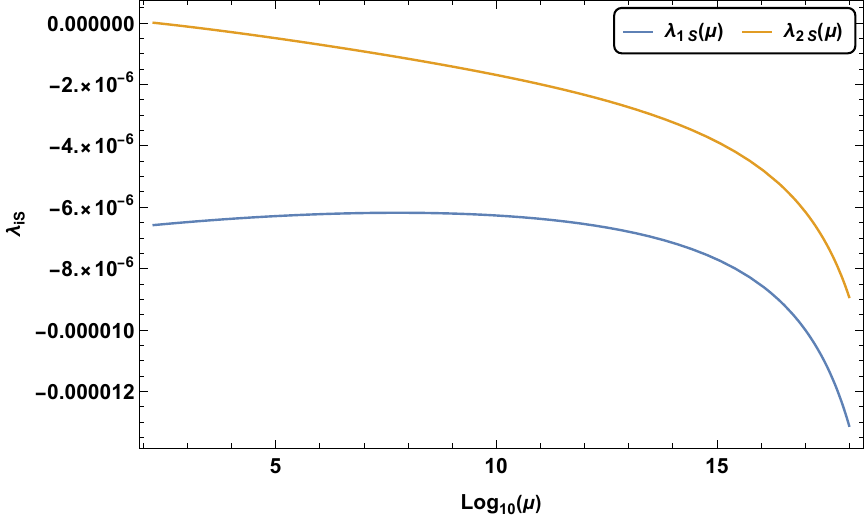}
\end{center}
\caption{RG evolution of portal couplings $\lambda_{1S}$ and $\lambda_{2S}$ 
for $s h_{12}$-inflation using BP-$s h_{12}$ from Table \ref{tab:bps_portal_couplings}.  The top pole mass $m_{t}=\left(172.5\pm 0.7\right)$ GeV given by Ref. \cite{Zyla:2020zbs} is used to set the renormalization scale for the benchmarks to run from low- to high scale.} 
\label{fig:RG_portal_PQTHI}
\end{figure}

In this section we consider the effects of the one-loop RG running of the portal couplings and make the connection between inflation and Higgs phenomenology. We evolve the couplings from the electroweak scale all the way up to the Planck scale and discuss the intricacies of satisfying the constraints for successful inflation (see Table \ref{tab:inflation_cond}) while accommodating a 125 GeV Higgs with additional heavy Higgses at the electroweak/TeV scale. In particular, we will discuss the running of the portal couplings and its impact on the inflationary field direction with regard to inflationary conditions (see Table \ref{tab:inflation_cond} for details). Therefore, we pay attention to the size of the  portal coupling and consider its sign at the Planck scale. We introduce in Table \ref{tab:bps_portal_couplings} benchmark points representing inflation in the $s h_{12}$-, $s h_{2}$- and $s$-direction which satisfy our considerations and convey the statement of our discussion.\\
\begin{table}[ht]
\begin{center}
 \begin{tabular}{|c|c|c|c|}
  \hline
  Parameters &\textbf{BP-$s h_{12}$}&\textbf{BP-$s h_{2}$}&\textbf{BP-$s$}\\
  \hline
  $\lambda_1$&0.07&0.07& 0.07  \\
  $\lambda_2$& 0.263&0.287&0.258 \\
  $\lambda_3$&0.60& 0.54&0.54  \\
  $\lambda_4$&-0.40&-0.14&-0.14 \\
  $\lambda_S$&$6.5\times 10^{-10}$&$4.44\times 10^{-10}$&$10^{-10}$  \\
  $\lambda_{1S}$&$-6.59\times 10^{-6}$&$5.57\times 10^{-6}$&$4.8\times10^{-14}$ \\
  $\lambda_{2S}$&$10^{-15}$&$-4.27\times 10^{-6}$ &$10^{-15}$   \\
  $\lambda_{12S}$&$2.5\times 10^{-16}$& $2.5\times 10^{-16}$&$2.5\times 10^{-16}$ \\
  $\tan \beta$ &5.5&5.5&26   \\
    $Y_{N,1}$&$9\times 10^{-4}$&$9\times 10^{-4}$&$4\times 10^{-5}$ \\
     $Y_{\nu,3}$&$ 5.18\times 10^{-3}$&$5.18\times 10^{-3}$&$1.09\times 10^{-3}$ \\
  $v_S$ &$3\times 10^{10}$ &$3\times 10^{10}$&$3\times 10^{10}$   \\
  \hline
  $m_h\left(\text{GeV}\right)$ &125.2&125.2 &125.1  \\
  $m_H\left(\text{GeV}\right)$ &799.4&798.8&1711.5 \\
  $m_{s}\left(\text{GeV}\right)$&$6.6\times 10^{5}$&$6.3\times 10^{5}$ &$3\times 10^{5}$ \\
  $m_A\left(\text{GeV}\right)$ &799.5&799.5&1711.5 \\
  $m_{H^{\pm}}\left(\text{GeV}\right)$&807.0&802.2&1712.8  \\
  \hline
 \end{tabular}
 \caption{Three benchmarks passing theoretical and experimental constraints representative for inflation in $s h_{12}$-, $s h_{2}$- and $s$-direction. The top pole mass is given by Ref. \cite{Zyla:2020zbs} to be $m_{t}=\left(172.5\pm 0.7\right)$ GeV.}
 \label{tab:bps_portal_couplings}
\end{center}
\end{table}
As discussed in section \ref{sec:mass_spec} we determined that the squared masses of the heavy Higgses strongly depend on $\lambda_{12S} v_{S}^{2}$. In fact, a tiny value for $\lambda_{12S}$ is preferred by high-scale validity analysis \cite{Chakrabarty:2014aya,Oredsson:2018yho} since $\lambda_{12S}v_{S}^{2}$ resembles a soft breaking parameter in a softly broken $U(1)$-symmetric 2HDM\footnote{In section \ref{app:matching_scale} we describe the effective low-energy matching of 2hdSMASH to a softly broken $U(1)$-symmetric 2HDM.} \cite{Clarke:2015bea,Oda:2019njo}. Therefore, we first consider the RG running of $\lambda_{12S}$. We can already see that the running of $\lambda_{12S}$ is proportional to itself:
\begin{align}
\mathcal{D}\lambda_{12S} = &\ 
\lambda_{12S} \left( -\frac32 g_1^2 -\frac92 g_2^2 
 + 2\lambda_3 + 4\lambda_4 + 2\lambda_S + 4(\lambda_{1S} + \lambda_{2S})
 + 3Y_t^2 + 3Y_b^2 + Y_\tau^2 \right)\,.
\label{eq:RGE_lambda_12S}
\end{align}
By considering the value of $\lambda_{12S}$ at the electroweak scale we determine the size of heavy Higgs masses since we fix the PQ-breaking scale at $v_{S}=3\times10^{10}$ GeV. Hence, $\lambda_{12S}$ is chosen to be tiny as argued in section \ref{sec:mass_spec} to acquire a phenomenologically viable model which can be tested at the HL-LHC or future colliders. This smallness is associated with an enhanced Poincaré symmetry which we have discussed in section \ref{app:matching_scale}. The size of $\lambda_{12S}$ will be preserved all the way up to the Planck scale, thus justifying the considerations of Section \ref{sec:inflation}. By analyzing the RG evolution of $\lambda_{1S}$ and $\lambda_{2S}$ we can determine whether we satisfy PQI or PQTHI directions at the Planck scale. The RGEs of $\lambda_{1S}$ and $\lambda_{2S}$ are given by: 
\begin{align}
 \mathcal{D} \lambda_{1S} = &\ 
   \lambda_{1S}\left( -\frac32 g_1^2 -\frac92 g_2^2 
 + 4\lambda_{1S} +4\lambda_S +6\lambda_1  \right) + \lambda_{2S}\left(4\lambda_3 + 2\lambda_4\right)+8\lambda_{12S}^2\nonumber\\
 &+ \lambda_{1S}\left( 6Y_b^2 + 2Y_\tau^2 
+ 2\text{Tr}\left(Y_\nu^\dagger Y_\nu\right) 
 + \text{Tr}\left(Y_N^\dagger Y_N\right) \right) 
 - 4\text{Tr}\left(y_\nu^\dagger Y_\nu Y_N^\dagger Y_N\right) , \\
 \mathcal{D} \lambda_{2S} = &\ 
   \lambda_{2S}\left( -\frac32 g_1^2 -\frac92 g_2^2 
 + 4\lambda_{2S} +4\lambda_S +6\lambda_2  \right) + \lambda_{1S}\left(4\lambda_3 + 2\lambda_4\right)+8\lambda_{12S}^2 \nonumber \\
 &+ \lambda_{2S}\left( 6Y_t^2 
+ \text{Tr}\left(Y_N^\dagger Y_N\right) \right)\,.
\end{align}
Since $\lambda_{12S}$, $Y_{N}$ and $Y_{\nu}$ are very small, we can safely neglect terms involving them. Thus, we can rewrite the RGEs of the portal couplings as follows:
\begin{align}
 \mathcal{D} \lambda_{1S} \simeq &\ 
   \lambda_{1S}\left( -\frac32 g_1^2 -\frac92 g_2^2 
 + 4\lambda_{1S} +4\lambda_S +6\lambda_1 +6y_b^2 + 2y_\tau^2 \right) + \lambda_{2S}\left(4\lambda_3 + 2\lambda_4\right) , \\
 \mathcal{D} \lambda_{2S} \simeq &\ 
   \lambda_{2S}\left( -\frac32 g_1^2 -\frac92 g_2^2 
 + 4\lambda_{2S} +4\lambda_S +6\lambda_2 + 6y_t^2 \right) + \lambda_{1S}\left(4\lambda_3 + 2\lambda_4\right)\,.
\end{align}
There are two terms to consider, namely the first term which is proportional to the evolving portal coupling itself and the second term which is proportional to the other portal coupling. We notice that the combination of both terms will determine the size of the portal coupling at the Planck scale. Unfortunately, we cannot solve these RGEs analytically but we can analyze their contributions and make a statement about the size and the behavior of their evolution so that we end up with the required conditions for PQI or PQTHI at the Planck scale. There are 
three separate scenarios concerning the size of the portal couplings, i.e. 
\begin{align}
\left|\lambda_{1S}\right|<\left|\lambda_{2S}\right|~~,~~\left|\lambda_{1S}\right|>\left|\lambda_{2S}\right|~~,~~\left|\lambda_{1S}\right|\sim\left|\lambda_{2S}\right|\,,
\end{align}
where only two of them will be our main focus. Therefore, we will consider without loss of generality benchmarks BP-$s h_{12}$ and BP-$s h_{2}$ - BP-$s$  of Table \ref{tab:bps_portal_couplings} for the cases $\left|\lambda_{1S}\right|>\left|\lambda_{2S}\right|$ and $\left|\lambda_{1S}\right|\sim\left|\lambda_{2S}\right|$ respectively. 
Starting with the latter will help clarify the various contributions to the RGEs of $\lambda_{1S}$ and $\lambda_{2S}$.\\ 
\\
\textbf{$\left|\lambda_{1S}\right|\sim\left|\lambda_{2S}\right|$:}
Both portal couplings will influence the running of each other. Therefore it is necessary to analyze which contributions will have the dominant influence. Since the top Yukawa coupling is larger than the bottom- and the $\tau$-Yukawa couplings, the influence of $\lambda_{2S}$ will dominate the evolution of $\lambda_{1S}$. Hence, $\lambda_{1S}$ will evolve towards the value of $\lambda_{2S}$ when approaching the Planck scale. This can be circumvented by approximating the RGEs further and modifying the portal couplings accordingly. By comparison, we can neglect $Y_{b,\tau}$, $\lambda_{1,2,S}$ and $g_{1,2}$ since the portal couplings are equal in size and the top Yukawa term dominates the running of $\lambda_{2S}$. This is due to the fact that the top Yukawa dominates in the low-energy regime and becomes in comparison less dominant in the high energy regime. Thus, the running of $\lambda_{2S}$ affects the running of $\lambda_{1S}$ by the coupling term involving $\lambda_{3}$ and $\lambda_{4}$. We therefore assume $\lambda_{3}$ and $\lambda_{4}$ not to grow significantly when approaching the Planck scale\footnote{We require $\lambda_{3}$ and $\lambda_{4}$ to run perturbative to high energies and thus prefer that exponential growth near the Planck scale is suppressed.}. Hence, the RGEs of $\lambda_{1S,2S}$ for $\left|\lambda_{1S}\right|\sim\left|\lambda_{2S}\right|$ can be approximated to:
\begin{align}
\mathcal{D} \lambda_{1S} \approx &\ 
 \lambda_{2S}\left(4\lambda_3 + 2\lambda_4\right) , \\
 \mathcal{D} \lambda_{2S} \approx &\ 
   6\lambda_{2S} y_t^2\,.
\end{align}
The running of $\lambda_{2S}$ will dominate unless we implement a correction to the initial value of $\lambda_{1S}$ so that both portal couplings will preserve their respective sign and size at the Planck scale. Considering the chain rule, we can reformulate the two RGEs in order to obtain a differential equation:
\begin{align}
&\mathcal{D}\lambda_{1S}=\dfrac{\partial \lambda_{1S}}{\partial \lambda_{2S}}\mathcal{D}\lambda_{2S}\approx \lambda_{2S}\left(4\lambda_3 + 2\lambda_4\right)\\
\Rightarrow~& \dfrac{\partial \lambda_{1S}}{\partial \lambda_{2S}}\approx \dfrac{2\lambda_3 + \lambda_4}{3 y_t^2}\,.\label{eq:l1s_cond_eq1}
\end{align}
By integrating both sides, we obtain the necessary correction for $\lambda_{1S}$ to counter the effect of the top Yukawa:
\begin{align}
\delta \lambda_{1S}\approx \dfrac{\lambda_{2S}\left(2\lambda_3 + \lambda_4\right)}{3 y_t^2}\,,
\end{align}
where $\left|\lambda_{1S}\right|\simeq \left|\lambda_{2S}\right|$. Hence, the initial value for $\lambda_{1S}$ is given by: 
\begin{align}
\left|\lambda_{1S}^{\text{corr.}}(\mu_{\textit{EW}})\right|=\left|\lambda_{1S}+\delta\lambda_{1S}\right|_{\mu_{\textit{EW}}}\approx \left|\lambda_{2S}\times\left(1+\dfrac{2\lambda_{3}+\lambda_{4}}{3y_{t}^{2}}\right)\right|_{\mu_{\textit{EW}}}\,,
\label{eq:l1s_initial_cond_equal}
\end{align}
where $\mu_{\text{EW}}$ is the electroweak scale and $\lambda_{1S}^{\text{corr.}}$ is the corrected value of $\lambda_{1S}$. This correction causes both RGEs to be of equal size and preserve their respective sign at the Planck scale. Therefore, this case describes inflation in the $s h_{1}$ or $s h_{2}$ direction where either portal coupling differs by sign change. For illustration we consider the case where $\lambda_{1S}>0$ and $\lambda_{2S}<0$ given by benchmark point BP-$s h_{2}$.  
\begin{figure}[ht]
\begin{center}
\includegraphics[width=0.7\textwidth]{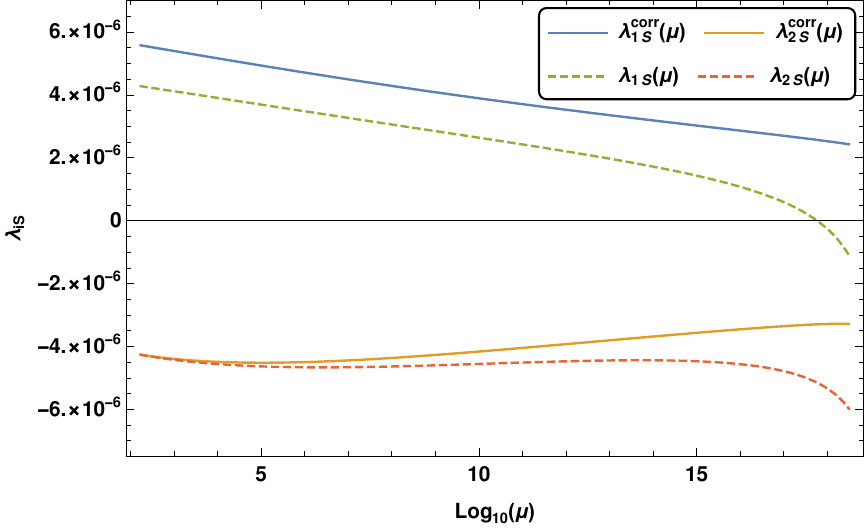}
\end{center}
\caption{RG evolution of $\lambda_{1S}$ and $\lambda_{2S}$ of BP-$s h_{2}$ where $\left|\lambda_{1S}\right|\sim \left|\lambda_{2S}\right|$ with opposite signs. Here, the initial value of $\lambda_{1S}^{\text{corr}}$ is given by eq. \eqref{eq:l1s_initial_cond_equal} which affects the running of $\lambda_{1S}$ positively compared to the running where no correction is applied.} 
\label{fig:RG_portal_PQTHI_opposite_signs_equal}
\end{figure} 
In figure \ref{fig:RG_portal_PQTHI_opposite_signs_equal} we can see the running of  both portal couplings with and without a correction applied to $\lambda_{1S}(\mu)$. As described above, we can see that the correction preserves the sign of both portal couplings all the way up to the Planck scale. The same line of reasoning applies to $s h_{1}$ by switching signs of $\lambda_{1S}$ and $\lambda_{2S}$. For the benchmark point BP-$s$ of Table \ref{tab:bps_portal_couplings} both portal couplings are taken positive. However, the running of $\lambda_{1S}$ and $\lambda_{2S}$ is dominantly determined by the left- and right-handed Yukawa couplings. In section \ref{subsec:bau_thermal_lepto} we discussed the appropriate initial value $\lambda_{1S}$ needs (see eq. \eqref{eq:bau_lepto_l1s_estimate}) in order for both portal couplings to remain positive at the Planck scale.\\
\\
\textbf{$\left|\lambda_{1S}\right|>\left|\lambda_{2S}\right|$:} In general, this case describes either PQI or PQTHI-$sh_{12}$ since both portal couplings will be either positive (PQI) or negative (PQTHI-$sh_{12}$) at the Planck scale, as we will argue now. For PQI, we consider tiny portal couplings which is represented in benchmark BP-$s$. Therefore, we will refer to PQTHI-$sh_{12}$ for a more general discussion where portal couplings are assumed to be sizable compared to PQI. We consider BP-$sh_{12}$ of table \ref{tab:bps_portal_couplings}, where  $\lambda_{1S}<0$ and $\lambda_{2S}>0$ but differ in size, i.e. $\left|\lambda_{1S}\right|>\left|\lambda_{2S}\right|$ at the electroweak scale. Thus, we can approximate the RGEs of $\lambda_{1S}$ and $\lambda_{2S}$ by
\begin{align}
 \mathcal{D} \lambda_{1S} \approx &\ 
   \lambda_{1S}\left( -\frac32 g_1^2 -\frac92 g_2^2 
 + 4\lambda_{1S} +4\lambda_S +6\lambda_1 +6y_b^2 + 2y_\tau^2 \right)  , \\
 \mathcal{D} \lambda_{2S} \approx &\ 
   \lambda_{1S}\left(4\lambda_3 + 2\lambda_4\right)\,, 
\end{align}
where only contributions of $\lambda_{1S}$ are considered since $\lambda_{2S}$ is sub-dominant in comparison. This approximation is illustrated in Figure \ref{fig:RG_portal_PQTHI} where $\lambda_{2S}$ approaches the value of $\lambda_{1S}$ and they become equal at the Planck scale.

\subsection{Stability analysis}
\label{subsec:stability_analysis_of_2hdm}
\begin{figure}[htb!]
  \centering
  \includegraphics[width=1\linewidth]{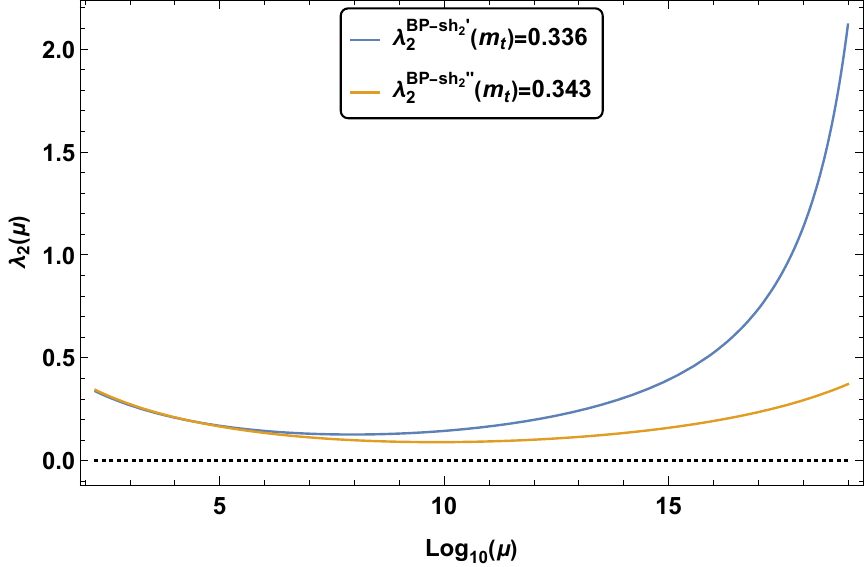}
\caption{Shown are $\lambda_{2}(\mu)$ from benchmark BP-$sh_2 '$ and BP-$sh_2''$ given in table \ref{tab:bps_l2_stab}. The top pole mass $m_{t}=\left(172.5\pm 0.7\right)$ GeV given by Ref. \cite{Zyla:2020zbs} is used to set the renormalization scale for the benchmarks to run from low- to high scale.}
\label{fig:l2_comparison}
\end{figure}

\begin{figure}[htb!]
\centering
\begin{subfigure}{.5\textwidth}
  \centering
  \includegraphics[width=1\linewidth]{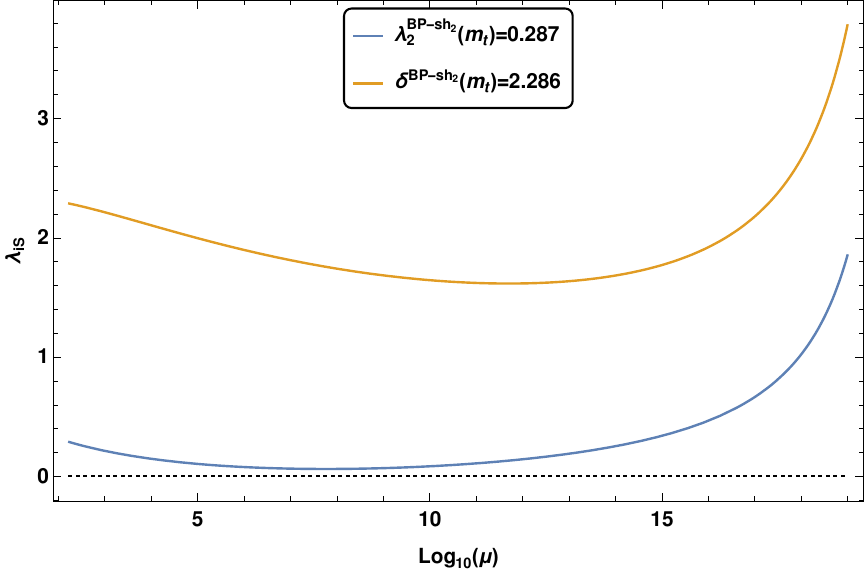}
\end{subfigure}%
\begin{subfigure}{.5\textwidth}
  \centering
  \includegraphics[width=1\linewidth]{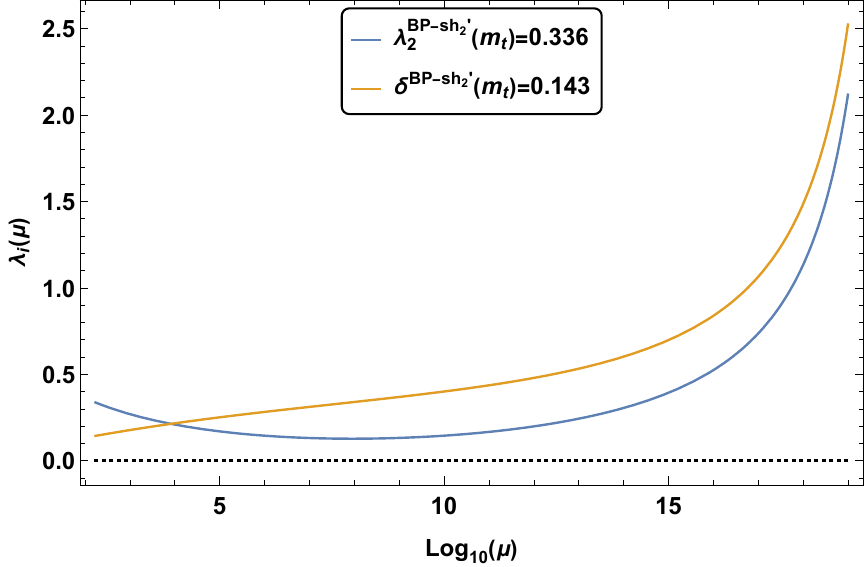}
\end{subfigure}
\caption{Shown are the RG-running of $\lambda_{2}$ and $\delta$ given by benchmark BP-$sh_2$ (left) and the RG-running of $\lambda_2$ and $\delta$ given by benchmark BP-$sh_{2}'$ of table \ref{tab:bps_l2_stab}. The renormalization scale is set at the top pole mass $m_{t}=\left(172.5\pm 0.7\right)$ GeV, cf. Ref. \cite{Zyla:2020zbs}, to run from low- to high scale.}
\label{fig:RG_tl_l2}
\end{figure}

\begin{figure}[htb!]
  \centering
  \includegraphics[width=1\linewidth]{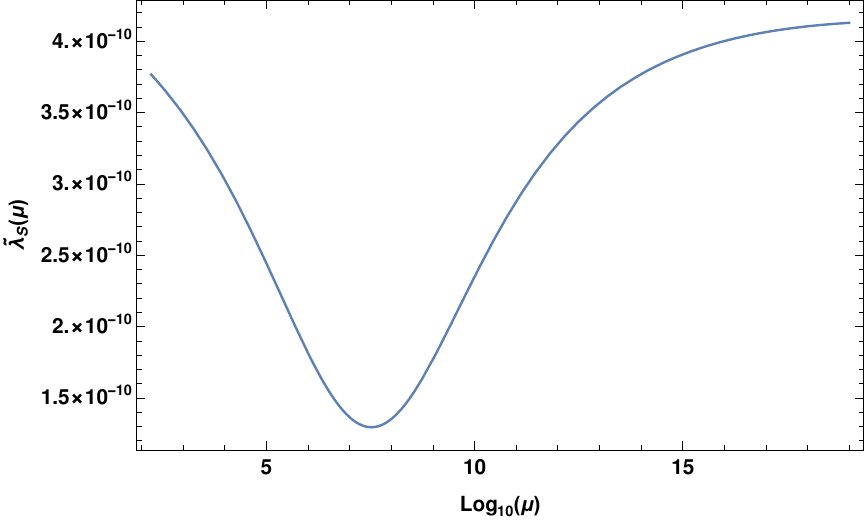}
\caption{RG-evolution of $\tilde{\lambda}_{S}(\mu)\equiv \lambda_S(\mu)-\frac{\lambda_{2S}^{2}(\mu)}{\lambda_{2}(\mu)}$ from benchmark BP-$sh_2$. The renormalization scale is set at the top pole mass $m_{t}=\left(172.5\pm 0.7\right)$ GeV, cf. Ref. \cite{Zyla:2020zbs}, to run from low- to high scale.}
\label{fig:lsl2s}
\end{figure}

\begin{figure}[htb!]
\begin{center}
\includegraphics[width=0.7\textwidth]{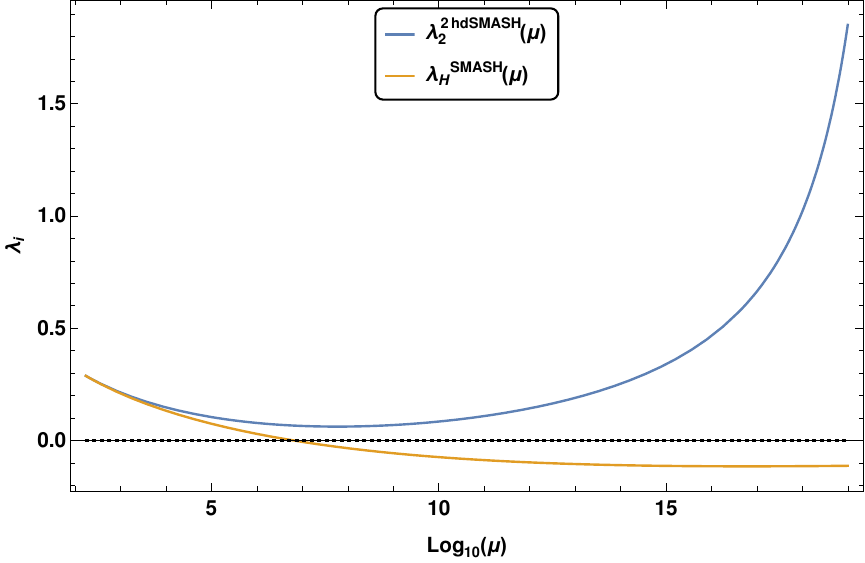}
\end{center}
\caption{Depicted are the RG evolutions of $\lambda_{2}(\mu)$ for 2hdSMASH (blue solid curve) and SMASH (orange solid curve) for PQTHI using BP-$sh_{2}$ of Table \ref{tab:bps_portal_couplings}. The renormalization scale is set at the top pole mass $m_{t}=\left(172.5\pm 0.7\right)$ GeV, cf. Ref. \cite{Zyla:2020zbs}, to run from low- to high scale.} 
\label{fig:l2rgethresholdPQTHI}
\end{figure}

In this section we analyze vacuum stability, perturbative unitarity and high-scale validity of 2hdSMASH. Vacuum stability is mostly endangered by $\lambda_{2}$ running negative. For a Type-II 2HDM, the Higgs doublet $\Phi_{2}$ couples to the top-quark which can force $\lambda_2$ into negative values. The same phenomenon has been observed in the SM, which is regarded as the top-quark instability. We show, that we can circumvent this problem by stabilizing the running of $\lambda_{2}$ with the mixing parameter $\lambda_{34}$. For that reason, we consider the necessary BfB condition of the 2HDM
\begin{align}
\lambda_{34} \geq -\sqrt{\lambda_1 \lambda_2}\,,
\label{eq:nec_bfb_stab}
\end{align}
which we can translate into the following form
\begin{align}
\dfrac{\lambda_{34}^{2}}{\lambda_{1}}\geq -\lambda_{2}\,.
\label{eq:nec_vac_stab}
\end{align}
We identify in eq. \eqref{eq:nec_vac_stab} the parameter $\delta$ given by
\begin{align}
\delta\equiv \dfrac{\lambda_{34}^{2}}{\lambda_1}\,,
\end{align}
which we will use to stabilize the running effects of $\lambda_2$ as we will argue in the following.\\
\\
The instability can be understood by considering the one-loop RGE of $\lambda_{2}$:
\begin{align}
\mathcal{D} \lambda_2 &=  
\frac34 g_1^4 + \frac32 g_1^2g_2^2 + \frac94 g_2^4 -\frac{9}{5} g_{1}^{2} \lambda_2 -9 g_{2}^{2} \lambda_2 +12 \lambda_{2}^{2} +2 \lambda_{2s}^{2} \\ 
 &+4 \lambda_{3}^{2} +4 \lambda_3 \lambda_4 +2 \lambda_{4}^{2}+12 \lambda_2 Y_{t}^{2} -12 Y_{t}^{4}\,,\notag
\end{align} 
where the dominant negative top Yukawa contribution $\propto Y_{t}^{4}$ is competing with other dominant quartic coupling terms. It proves useful to consider $\lambda_{3}$ and $\lambda_{4}$ for stabilization since it contributes positively to its RG-running and $\lambda_{1}$ is absent. However, enhancing $\lambda_{1}$  would only contribute indirectly to the RG-running of $\lambda_2$ by its mediator couplings $\lambda_{3}$ and $\lambda_{4}$. Hence, we fix $\lambda_{1}$ and vary $\lambda_{3}$ and $\lambda_{4}$ for stabilization. By considering the RGE of $\lambda_{1}$, we recognize that the same $\lambda_{3}$ and $\lambda_{4}$ contributions affect its running 
\begin{align}
\mathcal{D} \lambda_1 &=  
   \frac34 g_1^4 + \frac32 g_1^2g_2^2 + \frac94 g_2^4 -\lambda_1\left( 3 g_1^2 +9g_2^2 \right)
 + 12\lambda_1^2 + 4\lambda_3\lambda_4 + 4\lambda_3^2+ 2\lambda_4^2
 + 2\lambda_{1S}^2\nonumber \\
 &+ 12\lambda_1Y_b^2 - 12Y_b^4 
  +4\lambda_1 Y_\tau^2 - 4Y_\tau^4
  + 4\lambda_2\text{Tr}\left(Y_\nu^\dagger Y_\nu\right) -4\text{Tr}\left(Y_\nu^\dagger Y_\nu Y_\nu^\dagger Y_\nu\right) \,,
\end{align}
where negative terms, such as the down-type Yukawa contribution is sub-dominant in comparison. Therefore, any enhancement of $\lambda_{3}$ and $\lambda_{4}$ will enhance $\lambda_{1}$ by going to higher energies and thus endangering perturbative unitarity. This can be avoided by applying the perturbative unitarity bounds for $\lambda_{1}$ and $\lambda_{34}$, such that
\begin{align}
\left|\delta\right|\equiv \left|\dfrac{\lambda_{34}^{2}(\mu)}{\lambda_{1}(\mu)}\right|<8\pi~~,~~\forall \mu 
\end{align}
is bounded from above. Furthermore, we have to ensure that we maintain a $125$ GeV light Higgs. Its mass is approximately
\begin{align}
m_{h}^{2}\approx  \frac{v_{S}^{2}}{\left(1+t_{\beta}^2\right)^2} \left[ {\lambda_{1}+ t_{\beta}^4\, \lambda_{2}+2\, t_{\beta}^2\, \lambda_{34}}{}
\right]\,,
\end{align}
where portal terms are neglected for now. We can see that for $t_{\beta}>1$, as considered in our case, the $\lambda_2$ term is $t_{\beta}^{2}$ times larger than the one with $\lambda_{34}$. Therefore, any enhancement in $\lambda_{34}$ is met with a slight decrease in $\lambda_{2}$ since its decrease will counterbalance the increase of $\lambda_{34}$ by a factor of $t_{\beta}^{2}$. Thus, $\lambda_{34}$ is capable to influence the RG-running of $\lambda_2$ without violating Higgs phenomenology.\\
\\
There are two intertwining effects which stabilize the RG-running of $\lambda_2$, namely
\begin{itemize}
\item[1)] initial values given to $\lambda_2$, $\lambda_3$ and $\lambda_4$ and
\item[2)] an RG-running effect caused by $\lambda_3$ and $\lambda_4$.
\end{itemize}
Both effects, namely 1) and 2), contribute to the RG stabilization of $\lambda_2$ and are interconnected. Here, the emphasis is on the RG-running effects since they are responsible to uplift $\lambda_2$ from its negative top Yukawa running term at energy scales $\mu\gtrsim m_t$. In order to gain some analytical understanding, we consider the Coleman-Weinberg potential, cf. Ref. \cite{Weinberg:1973am}, which we will utilize to acquire $\lambda_{2}(\mu)$, i.e. the integrated version\footnote{This can be verified by using the \textit{Callan-Symanzik} equation.} of $\mathcal{D}\lambda_2$, to identify the dominant RG-running contributions. The full expression of the Coleman-Weinberg potential is given in appendix \ref{app:coleman_weinberg}, whose general analytical form is given by
\begin{align}
V_{\text{CW}}(\phi_i)&=\dfrac{1}{64\pi^{2}}\\
&\times\left(\sum_{b} g_{b}m_{b}^{4}(\phi_i)\left[\log\left(\frac{m_{b}^{2}(\phi_i)}{\Lambda^{2}}\right)-\frac{3}{2}\right]-\sum_{f} g_{f}m_{f}^{4}(\phi_i)\left[\log\left(\frac{m_{f}^{2}(\phi_i)}{\Lambda^{2}}\right)-\frac{5}{2}\right]\right)\,,\notag
\end{align}
where $g_{f/b}$ are the degrees of freedom and $m_{f/b}$ are the masses of fermions/bosons. By performing a matching to an effective tree-level potential in the $h_2$-direction
\begin{align}
V_{\text{eff}}\simeq V_{0}+V_{\text{CW}}\simeq \dfrac{\lambda_{2}^{\text{eff}} h_{2}^{4}}{8}
\end{align}
with 
\begin{align}
V_{0}\simeq \dfrac{\lambda_{2} h_{2}^{4}}{8}\,,
\end{align}
where $h_1\simeq s\simeq 0$, we acquire an RG-improved potential for $h_2$. As a result we obtain the effective coupling $\lambda_{2}^{\text{eff}}$ by computing
\begin{align}
\lambda_{2}^{\text{eff}}=\dfrac{8}{h_{2}^{4}}V_{\text{eff}}(h_2)\,,
\end{align}
which is given by
\begin{align}
&\lambda_{2}^{\text{eff}}\simeq \lambda_2(m_t)+\frac{\lambda_{2S}^2 \left(\log \left(\frac{h_{2}^2 \lambda_{2S}}{2 m_{t}^2}\right)-\frac{3}{2}\right)+\lambda_{3}^2 \left(\log \left(\frac{h_{2}^2 \lambda_{3}}{2 m_{t}^2}\right)-\frac{3}{2}\right)}{16\pi^{2}}\label{eq:full_eff_l20}\\
&+\frac{(\lambda_{3}+\lambda_{4})^2 \left(\log \left(\frac{h_{2}^2 (\lambda_{3}+\lambda_{4})}{2 m_{t}^2}\right)-\frac{3}{2}\right)+6 \left(\lambda_{2}^2 \left(\log \left(\frac{27 h_{2}^2 \lambda_{2}}{16
   m_{t}^2}\right)-\frac{3}{2}\right)-Y_{t}^4 \left(\log \left(\frac{h_{2}^2 Y_{t}^2}{2 m_{t}^2}\right)-\frac{3}{2}\right)\right)}{16 \pi ^2}\,,\notag
\end{align}
\normalsize
where we neglected the sub-dominant gauge couplings $g_{1,2,3}$. Furthermore, we can neglect the portal coupling term $\propto\lambda_{2S}^{2}$ since its contribution is small in comparison\footnote{We list in table \ref{tab:bps_l2_stab} three benchmarks, which represent the most extreme case where portal couplings are sizable, i.e. benchmarks for PQTHI-$sh_2$. The contribution of the shown portal couplings is sub-dominant to eq. \eqref{eq:full_eff_l20}.}. Thus, we find
\begin{align}
\lambda_{2}^{\text{eff}}&\approx\lambda_{2}(m_t)+\frac{\lambda_{3}^2 \left(\log \left(\frac{h_{2}^2 \lambda_{3}}{2 m_{t}^2}\right)-\frac{3}{2}\right)+(\lambda_{3}+\lambda_{4})^2 \left(\log \left(\frac{h_{2}^2 (\lambda_{3}+\lambda_{4})}{2
   m_{t}^2}\right)-\frac{3}{2}\right)}{16\pi^{2}}\label{eq:full_eff_l2} \\
&+\frac{6 \left(\lambda_{2}^2 \left(\log \left(\frac{27 h_{2}^2 \lambda_{2}}{16 m_{t}^2}\right)-\frac{3}{2}\right)-Y_{t}^4 \left(\log \left(\frac{h_{2}^2 Y_{t}^2}{2
   m_{t}^2}\right)-\frac{3}{2}\right)\right)}{16 \pi ^2}\notag
\end{align}
\normalsize
with running scale $\mu\equiv h_2$. We can identify four stabilizing contributions, namely $\lambda_2(m_t)$, $\lambda_2^{2}$, $\lambda_{34}^{2}$ and $\lambda_{3}^{2}$. The former two, i.e. $\lambda_2(m_t)$ and $\lambda_2^{2}$, are capable of stabilizing its running when large initial conditions are applied. Moreover, we can see that contributions proportional to $\lambda_{34}^{2}$ and $\lambda_{3}^{2}$, enter quadratically as well and provide a further source of stabilization. In the case where $\lambda_4$ is negative, as we consider in our benchmarks, the most important parameter for RG-running stability becomes $\lambda_3$ which enters quadratically in two terms of eq. \eqref{eq:full_eff_l2}. This phenomenon is illustrated in figure \ref{fig:l2_comparison} where $\lambda_2(\mu)$ is depicted for benchmark points BP-$sh_2 '$ and BP-$sh_2 ''$ of table \ref{tab:bps_l2_stab}.
Naively, we would expect that benchmark BP-$sh_2 '$ has a sub-dominant effect on the running of $\lambda_2$ compared to BP-$sh_2 ''$. Despite the fact, that the initial value of $\lambda_2$ is smaller in comparison, the dominance of $\lambda_3$ causes $\lambda_2(\mu)$ to run higher. This leaves BP-$sh_2 '$ with the biggest impact.
\begin{table}[htb!]
\begin{center}
 \begin{tabular}{|c|c|c|c|}
  \hline
  Parameters &\textbf{BP-$s h_{2}$}&\textbf{BP-$sh_{2}'$}&\textbf{BP-$sh_{2}''$}\\
  \hline
  $\lambda_1$&0.07& 0.07 & 0.07  \\
  $\lambda_2$&0.287&0.336 & 0.343\\
  $\lambda_3$&0.54&0.54 &0.44 \\
  $\lambda_4$&-0.14&-0.44&-0.44 \\
  $\lambda_S$&$4.44\times 10^{-10}$&$4.44\times 10^{-10}$&$4.44\times 10^{-10}$  \\
  $\lambda_{1S}$&$5.57\times 10^{-6}$&$5.57\times 10^{-6}$&$5.57\times 10^{-6}$ \\
  $\lambda_{2S}$&$-4.27\times 10^{-6}$ &$-4.27\times 10^{-6}$&$-4.27\times 10^{-6}$   \\
  $\lambda_{12S}$& $2.5\times 10^{-16}$&$2.5\times 10^{-16}$&$2.5\times 10^{-16}$ \\
  $\tan \beta$ &5.5&5.5&5.5   \\
    $Y_{N,1}$&$9\times 10^{-4}$&$9\times 10^{-4}$&$9\times 10^{-4}$ \\
     $Y_{\nu,3}$&$ 5.18\times 10^{-3}$&$5.18\times 10^{-3}$&$5.18\times 10^{-3}$ \\
  $v_S$ &$3\times 10^{10}$ &$3\times 10^{10}$  &$3\times 10^{10}$\\
  \hline
  $m_h\left(\text{GeV}\right)$ &125.2&125.1&125.2  \\
  $m_H\left(\text{GeV}\right)$ &798.7&799.5&799.7 \\
  $m_{s}\left(\text{GeV}\right)$&$6.3\times 10^{5}$ &$6.3\times 10^{5}$&$6.3\times 10^{5}$ \\
  $m_A\left(\text{GeV}\right)$ &799.5&799.5&799.5 \\
  $m_{H^{\pm}}\left(\text{GeV}\right)$&802.1&807.7&807.7  \\
  \hline
 \end{tabular}
 \caption{List of three benchmarks representing the most extreme case of portal coupling configuration which pass theoretical and experimental constraints and differ by $\lambda_2$, $\lambda_3$ and $\lambda_4$.}
 \label{tab:bps_l2_stab}
\end{center}
\end{table}
Moreover, we can associate the tree-level stability condition of eq. \eqref{eq:nec_vac_stab} to an RG-running stability condition
\begin{align}
\delta(\mu)+\lambda_2(\mu)\geq 0~~~~,~~~~\forall \mu\,,
\end{align}
where negative $\lambda_2(\mu)$ can be counterbalanced by $\delta(\mu)$. This effect can be seen in figure \ref{fig:RG_tl_l2} by considering benchmarks BP-$sh_2$ and BP-$sh_2'$ of table \ref{tab:bps_l2_stab}. The tree-level effect is given by BP-$sh_2$ and depicted in figure \ref{fig:RG_tl_l2} (left) with $\delta^{\text{BP-}sh_{2}}>\lambda_{2}^{\text{BP-}sh_{2}}$. By comparison, we can see in figure \ref{fig:RG_tl_l2} (right) the RG-running effect of benchmark BP-$sh_2 '$ which is mainly caused by the term $\sim\lambda_{3}^{2}$ of eq. \eqref{eq:full_eff_l2}.
In our analysis we neglected terms $\propto\lambda_{2S}^{2}$ in the RG-running of $\lambda_2$. This is due to the fact, that $\lambda_{2S}$ is constrained by another stability condition, namely
\begin{align}
\tilde{\lambda}_{S}\equiv\lambda_{S}-\dfrac{\lambda_{2S}^{2}}{\lambda_2}\geq 0\,.
\end{align}
This stability condition would be violated if we chose $\lambda_{2S}^{2}\sim\lambda_{2}$, since $\lambda_{S}\sim 10^{-10}$. Thus, the portal coupling is constrained to be $\lambda_{2S}\lesssim \sqrt{\lambda_{S}}$ (see sec. \ref{subsec:stability_analysis_of_ls}). We show in figure \ref{fig:lsl2s} the RG-evolution of $\tilde{\lambda}_{S}$ and its scale-invariance.
We note, that $\lambda_{3}$ and $\lambda_{34}$ are mainly responsible for stabilizing the RG-running of $\lambda_2$. By erasing $\lambda_1$, $\lambda_3$, $\lambda_4$, $\lambda_{1S}$ and $\lambda_{12S}$ from our model, we would recover the SMASH model. In particular, in the absence of $\lambda_3$ and $\lambda_4$, the RG-running of $\lambda_2$ in 2hdSMASH becomes the RGE of $\lambda_{H}$ in SMASH. By comparison, 2hdSMASH provides further parameters for stabilization, i.e. $\lambda_{3}$ and $\lambda_{4}$, which is depicted in figure \ref{fig:l2rgethresholdPQTHI}.
We conclude, the mixing parameters $\lambda_3$ and $\lambda_4$ assist stabilizing the RG-running of $\lambda_2$. Furthermore, the two intertwining effects, i.e. tree-level initial conditions and RG-effects, provide an analytic understanding how $\lambda_2(\mu)$ behaves. The analysis of this section is representative of the analysis done for all benchmarks acquired in section \ref{subsec:benchmark_points}. We chose benchmarks of type PQTHI-$sh_2$ in table \ref{tab:bps_l2_stab} as an extreme case where portal couplings become sizable and thus provide a more general discussion. We were able to see that portal couplings are constrained and have a sub-dominant effect. This can be seen on tree-level by the following relation
\begin{align}
\lambda_2 + \frac{1}{2}\left(\delta-\frac{\lambda_{2S}^{2}}{\lambda_S}\right)\geq 0\,,
\end{align}
where $\delta$ is typically larger. Moreover, we demand that perturbative unitarity constraints \eqref{pert1}-\eqref{pert6} must be obeyed along the RG evolution, thus ensuring high-scale validity by avoiding Landau-Poles, cf. Ref. \cite{Oda:2019njo}. In section \ref{subsec:benchmark_points}, we give benchmark points which satisfy all theoretical constraints at tree- and one-loop level. The analysis of this section can be easily adopted to scenarios where we consider PQI and other PQTHI directions.
\subsection{Benchmark points}
\label{subsec:benchmark_points}

The discussion of inflation in section \ref{sec:PQ-and mixed inflation} and its connection to electroweak physics by means of RG-analysis in sections \ref{subsec:stability_analysis_of_ls}-\ref{subsec:stability_analysis_of_2hdm}, culminates to the discussion of benchmark points which satisfy theoretical and experimental constraints. We present benchmarks which are consistent with BfB- and perturbative unitarity conditions from electroweak- all the way up to the Planck scale, thus providing high-scale validity and vacuum stability. Furthermore, the benchmarks are tested against experimental constraints from B-physics and collider experiments on the Higgs sector, thus ensuring 2hdSMASH to be phenomenologically viable.\\
For this part of the analysis, 2hdSMASH was implemented in \texttt{SARAH-v4.14.4}~\cite{Staub:2013tta}, \texttt{SPheno-v4.0.4}~\cite{Porod:2003um}, \texttt{HiggsSignals-v2}~\cite{Bechtle:2013xfa} and \texttt{HiggsBounds-v5}~\cite{Bechtle:2020pkv}.\\ With
\texttt{SARAH-v4.14.4}, the necessary equations were provided to use in \texttt{SPheno-v4.0.4}, such as the RGEs given in appendix \ref{app:rges}, the scalar- and fermionic masses and its decays. For a given benchmark, this was computed with \texttt{SPheno-v4.0.4} and checked with\\ \texttt{HiggsSignals-v2} to account for a low-energy signal-strength measurement of the $125$ GeV SM-like Higgs scalar at the LHC. Furthermore, the same \texttt{SPheno} output was cross-checked with \texttt{HiggsBounds-v5} against constraints from heavy Higgs searches at LEP, Tevatron and LHC.
   \begin{table}[ht]
\begin{center}
 \begin{tabular}{|c|c|c|c|c|c|}
  \hline
  Parameters &\textbf{BP1}&\textbf{BP2}&\textbf{BP3} &\textbf{BP4} & \textbf{BP5} \\
  \hline
  $\lambda_1$&0.07&0.07 &0.07 & 0.07&0.07  \\
  $\lambda_2$&0.287& 0.263&0.257&0.258& 0.257 \\
  $\lambda_3$&0.54&0.60 &0.24 & 0.54 & 0.24  \\
  $\lambda_4$&-0.14&-0.4&0.27&-0.14 &-0.28 \\
  $\lambda_S$&4.44$\times 10^{-10}$&6.5$\times 10^{-10}$&1.0$\times 10^{-10}$ &1.0$\times 10^{-10}$& 1.0$\times 10^{-10}$   \\
  $\lambda_{1S}$&5.57$\times 10^{-6}$&-6.59$\times 10^{-6}$&4.8$\times 10^{-14}$&4.8$\times 10^{-14}$ &3.6$\times 10^{-13}$ \\
  $\lambda_{2S}$&$-4.27\times 10^{-6}$&1.0$\times 10^{-15}$&1.0$\times 10^{-15}$&1.0$\times 10^{-15}$ &1.0$\times 10^{-15}$   \\
  $\lambda_{12S}$&2.5$\times 10^{-16}$&2.5$\times 10^{-16}$&2.5$\times 10^{-16}$& 2.5$\times 10^{-16}$&2.5$\times 10^{-16}$  \\
  $\tan \beta$ &5.5&5.5&26&26 &18   \\
$Y_{N,1}$&$9\times 10^{-4}$&$9\times 10^{-4}$&$4\times 10^{-5}$&$4\times 10^{-5}$ & $10^{-4}$\\
    $Y_{\nu,3}$&$ 5.175\times 10^{-3}$&$5.175\times 10^{-3}$&$1.09\times 10^{-3}$ &$1.09\times 10^{-3}$&$1.2\times 10^{-3}$\\ 
  $v_S$ &3.0$\times 10^{10}$&3.0$\times 10^{10}$&3.0$\times 10^{10}$&3.0$\times 10^{10}$ &3.0$\times 10^{10}$   \\
  \hline
  $m_h\left(\text{GeV}\right)$ &125.2&125.3&125.0& 124.9&125.4  \\
  $m_H\left(\text{GeV}\right)$ &798.7&799.4&1711.5&1711.5 &1425.2 \\
  $m_{s}\left(\text{GeV}\right)$&6.3$\times 10^{5}$ &6.7$\times 10^{5}$&3.0$\times 10^{5}$&3.0$\times 10^{5}$ &3.0$\times 10^{5}$ \\
  $m_A\left(\text{GeV}\right)$ &799.5&799.5&1711.5&1711.5 &1425.2 \\
  $m_{H^{\pm}}\left(\text{GeV}\right)$ &802.1&807.0&1709.1&1712.8 &1422.2 \\ 
  \hline
 \end{tabular}
 \caption{List of benchmarks passing the theoretical constraints and experimental constraints as discussed in the text with a pole top mass of $m_{t}=\left(172.5\pm 0.7\right)$ GeV according to Ref. \cite{Zyla:2020zbs}.}
 \label{tab:bp}
\end{center}
\end{table}
With the benchmarks \textbf{BP1-BP5} listed in table \ref{tab:bp} we cover all interesting parameter configurations which also allow for PQI and PQTHI. The 2HDM couplings $\lambda_1$-$\lambda_4$ are of $\mathcal{O}(1)$ such that the SM-like Higgs mass $m_h=(125.10\pm 0.14)$ GeV \cite{Zyla:2020zbs} is obtained. Furthermore, we consider $\tan\beta\gtrsim 5.5$ to accommodate the constraint on the axion decay constant $f_a$ as discussed in appendix \ref{app:axion_bounds}. There is a stringent constraint from $B\rightarrow X_s \gamma$ \cite{BaBar:2012fqh} on the charged Higgs sector of the Type-II 2HDM which provides a lower bound on the charged Higgs masses 
\begin{align}
m_{H^{\pm}}>650 \, \text{GeV}\,.
\end{align}
This lower bound was recently corrected to be higher in Type-II 2HDM models by computation of $(B \rightarrow X_s \gamma)$ at NNLO QCD level \cite{Misiak:2020vlo}
\begin{align}
m_{H^{\pm}}>800 \, \text{GeV}\,.
\end{align}
For choosing $\lambda_{12S}\simeq \mathcal{O}(\frac{v}{v_S})^2$, as discussed in section \ref{sec:mass_spec} (cf. Ref. \cite{Espriu:2015mfa}), the heavy Higgs sector becomes nearly degenerate with $m_H\approx m_A\approx m_{H^{\pm}}$. The constraints from electroweak precision tests, namely $S,T,U$ variables, are satisfied for the nearly degenerate mass spectrum of the neutral heavy Higgs sector. We note that the charged Higgs mass has a small mass splitting compared to the other heavy\\ Higgses which is conceived by considering the change in value for $\lambda_4$ of table \ref{tab:bp}. The portal couplings $\lambda_{1S}$ and $\lambda_{2S}$ do not have a sizable effect on the scalar masses. However, they are considered in the RG-analysis and connect inflation with particle phenomenology, as discussed in section \ref{subsec:Infl_particle_pheno}. The portal couplings are chosen such that successful inflation is guaranteed and constraints from neutrino physics for thermal leptogenesis and BAU are considered for RG-evolution. The neutrino Yukawa couplings given in table \ref{tab:bp} satisfy the constraints from neutrino oscillation experiments and BAU as discussed in sec. \ref{subsec:bau_thermal_lepto}. Moreover, the portal couplings are naturally small which protects the SM-like Higgs mass from high-scale radiative corrections. The benchmarks given in table \ref{tab:bp} are in the TeV range which is established by $\lambda_{12S}\simeq \mathcal{O}(\frac{v}{v_S})^2$. Due to this relation, $\lambda_{12S}v_S^{2}$ is of order $\mathcal{O}(v^{2})$ and resembles a soft breaking parameter of the effective 2HDM low-energy theory. This provides an additional parameter to control the mass spectrum without endangering Higgs phenomenology and its RG evolution all the way up to the Planck scale, cf. Ref.~\cite{Oredsson:2018yho}.\\
Therefore, the benchmarks provided in table \ref{tab:bp} facilitate a mass spectrum at the TeV scale which are allowed by theoretical and experimental constraints. The smallness of the portal couplings imply a considerable suppression of Higgs to axion decays. Hence, 2hdSMASH is indistinguishable from other extended Higgs sectors without axion at upcoming collider experiments based on Higgs decays. However, signals from axion dark matter searches may serve as detection probe of 2hdSMASH.

\section{Allowed Parameter Space}
\label{sec:allowedhiggs}
\begin{figure}[ht!]
    \centering
     \includegraphics[scale=0.46]{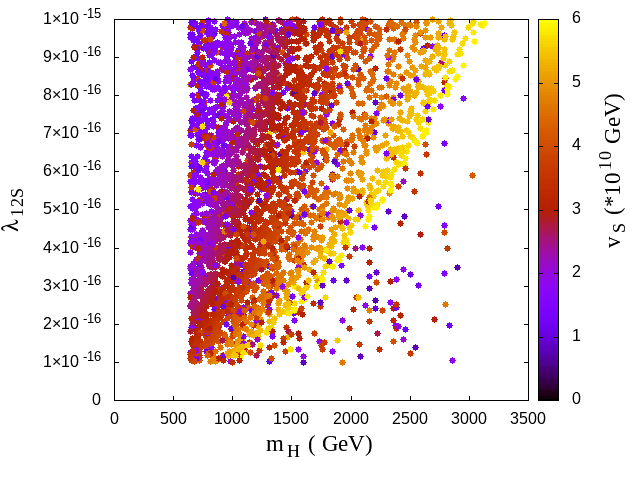} \hfill \hfill
    \includegraphics[scale=0.46]{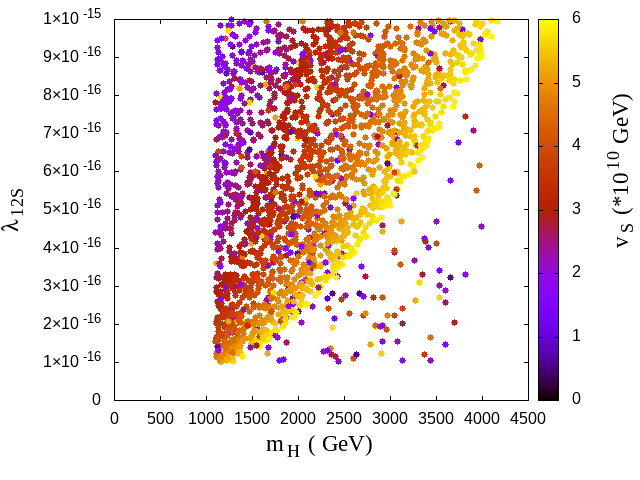}
    \caption{Variation of $\lambda_{12S}$ against  $m_H$ with $v_s$ on the coloured palette for $\tan \beta=5.5$ (left) and 10 (right).}
    \label{fig:l12svs}
\end{figure}

\begin{figure}[ht!]
    \centering
     \includegraphics[scale=0.46]{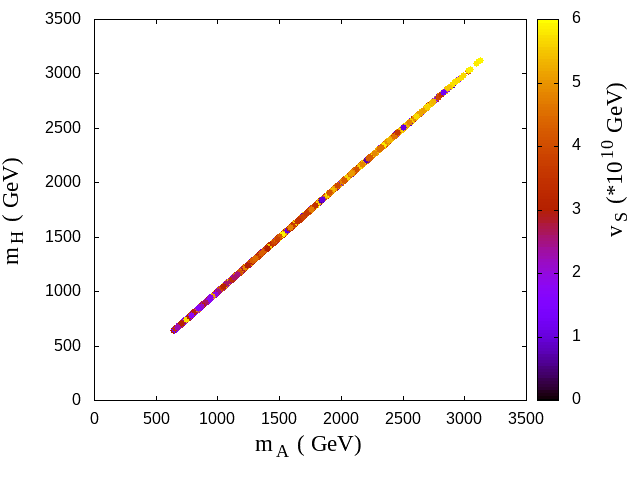} \hfill \hfill
      \includegraphics[scale=0.46]{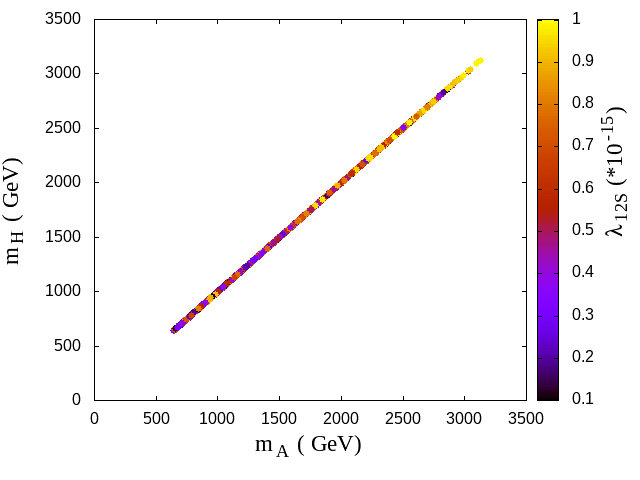}\hfill \hfill
          \includegraphics[scale=0.46]{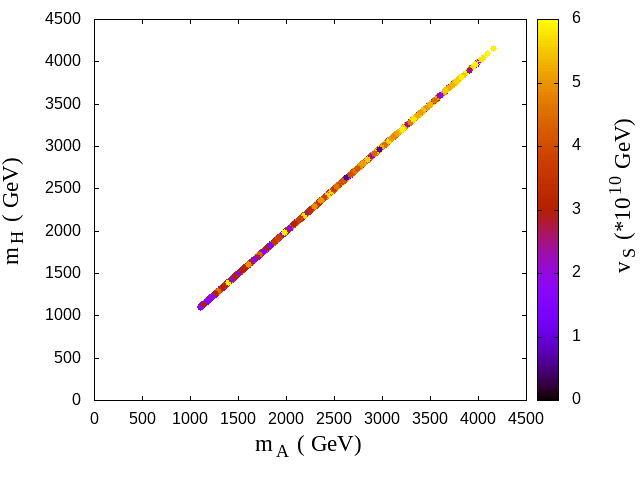}\hfill \hfill
     \includegraphics[scale=0.46]{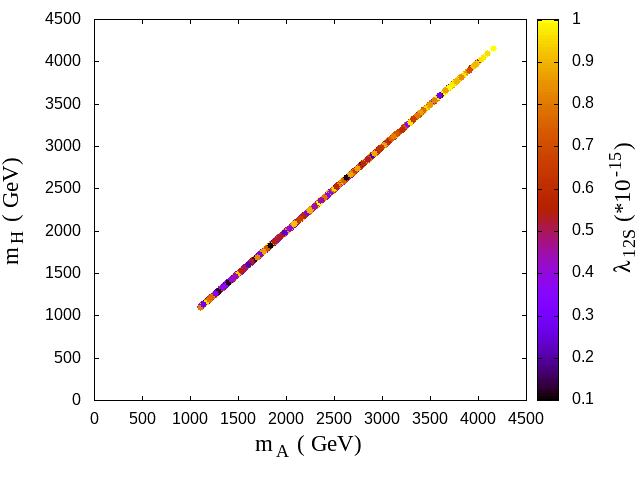}  \\          

                 \caption{ The  mass planes of the light CP-even  and CP-odd Higgs against $v_S$ (left) and $\lambda_{12S}$ (right) for $\tan \beta=5.5$ (top) and 10 (bottom).}   
   \label{fig:masses1}
 \end{figure}

\begin{figure}[ht!]
              \includegraphics[scale=0.46]{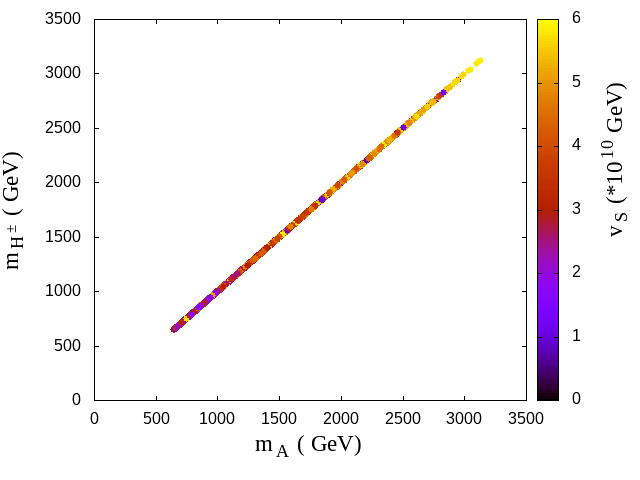} \hfill
            \includegraphics[scale=0.46]{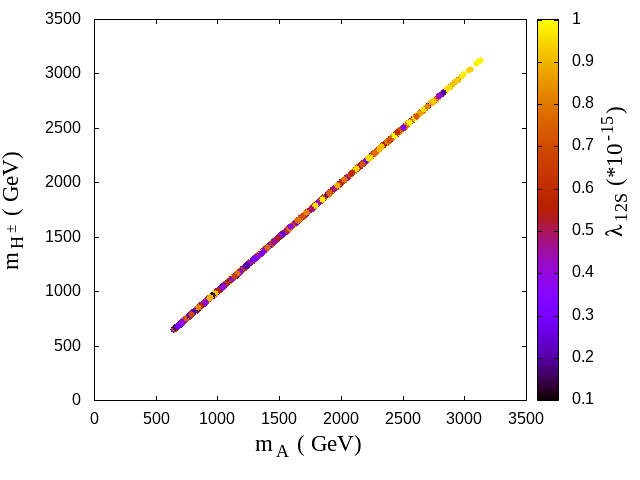} \\
             \includegraphics[scale=0.46]{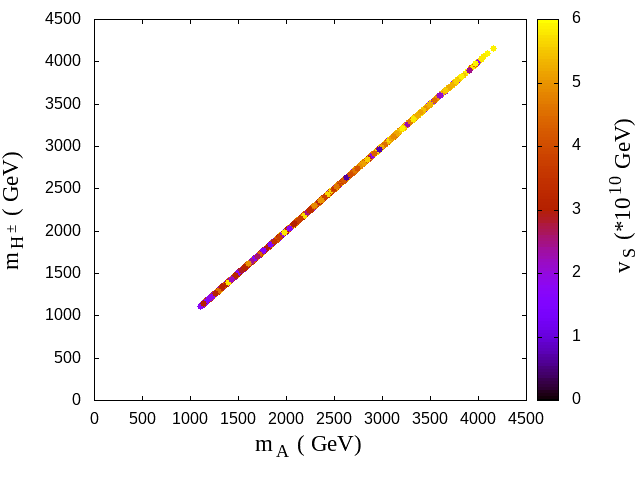} \hfill
            \includegraphics[scale=0.46]{tanb10/mHpm_mAvs.png}              
                 \caption{The mass planes of the  charged Higgs and CP-odd Higgs against $v_S$ (left) and $\lambda_{12S}$ (right) for $\tan \beta=5.5$ (top) and 10 (bottom).}
    \label{fig:masses11}
 \end{figure}

\begin{figure}[ht!]
 \centering
                 \includegraphics[scale=0.46]{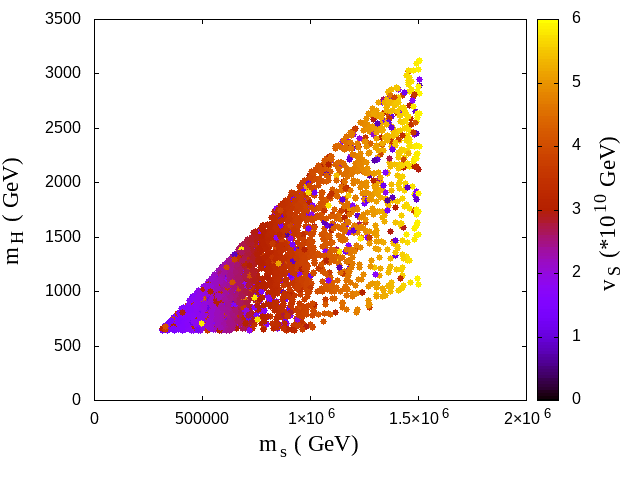} \hfill  \hfill
                 \includegraphics[scale=0.46]{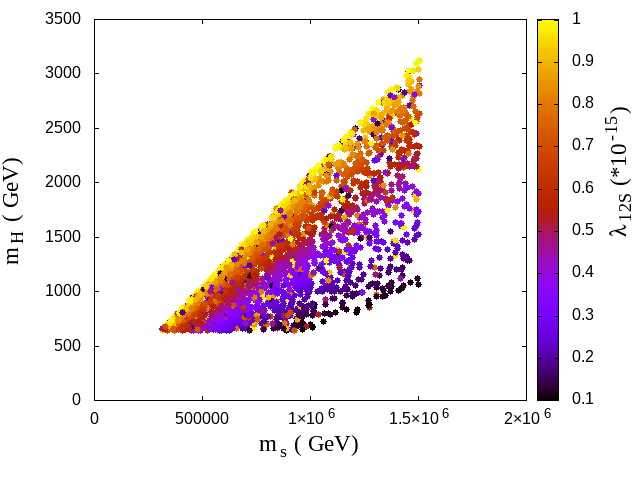} \\
     \includegraphics[scale=0.46]{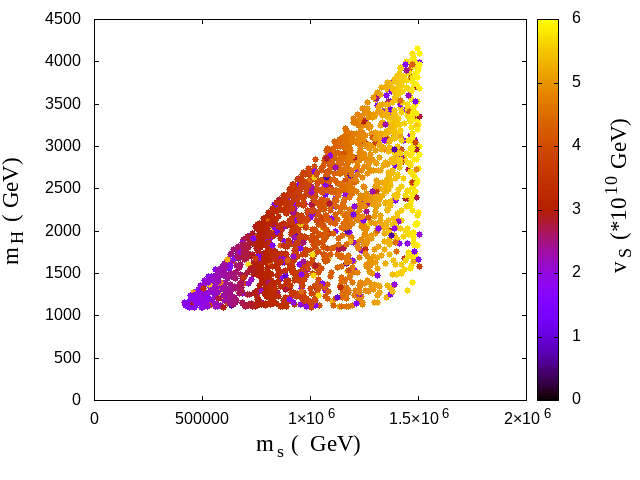} \hfill \hfill
      \includegraphics[scale=0.46]{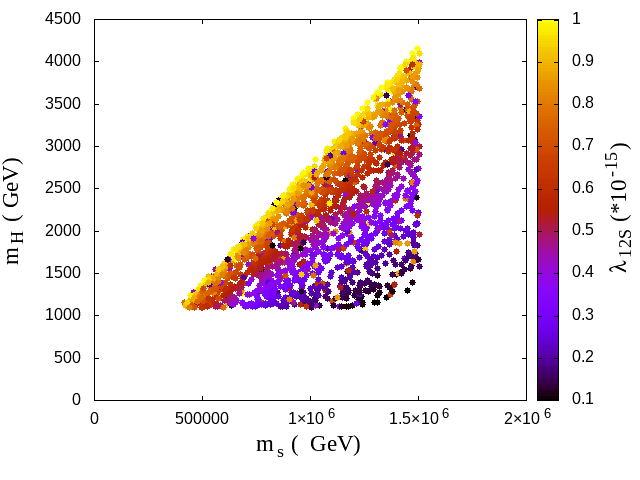} \\
         \caption{Different mass planes of the heavy CP-even scalar against $v_s$ (left) and $\lambda_{12s}$ (right) for $\tan \beta=5.5$  (top) and $10$ (bottom).}
    \label{fig:masses2}
\end{figure}

In this section, we discuss the constraints from the Higgs sector on the allowed parameter space and scan over the parameter space. We choose \textbf{BP2} and vary  $v_S$ and $\lambda_{12S}$  for  $\tan\beta$ = 5.5 and 10    to illustrate the different aspects of the mass spectra over the allowed range of $v_S$ (eq.~\ref{eq:preferred_v_S range}) 
and $\lambda_{12S} \simeq 10^{-16}$-$10^{-15}$.  We ensure the points are allowed by \texttt{HiggsBounds-v5}~\cite{Bechtle:2020pkv} and \texttt{HiggsSignals-v2}~\cite{Bechtle:2013xfa} and satisfy collider bounds on heavy Higgs masses from Tevatron, LEP and LHC as well as signal strength measurements for the 125 GeV SM-like Higgs as can be seen in Figs.~\ref{fig:l12svs}-\ref{fig:masses2} for both $\tan \beta=5.5$ and 10.  

As discussed in sec.~\ref{sec:mass_spec}, the Higgs masses, at the tree-level, are determined by $\tan \beta, v_S$ and $\lambda_{12S}$. Low values of $\lambda_{12S}$ ($\sim \mathcal{O}(\frac{v}{v_S})^2$) are favoured from inflationary context as discussed in sec.~\ref{sec:inflation}  as well as provide testable benchmark scenarios~\cite{Espriu:2015mfa} for future colliders. Figure~\ref{fig:l12svs} (left) shows the variation of $\lambda_{12S}$ against  $m_H$, the mass of the heavy CP-even Higgs  with $v_S$ on the coloured palette for $\tan \beta = 5.5$. We observe for a fixed $v_S$, $m_H$ decreases as $\lambda_{12S}$ decreases, while, for a fixed $\lambda_{12S}$, $m_H$ increases as $v_S$ increases.  Fig~\ref{fig:l12svs} (right) shows the $\lambda_{12S}-m_H$ plane for a higher $\tan \beta = 10$. For high tan $\beta$ the mass scale of the heavy Higgses allowed from Higgs constraints increases with an increase in $\tan \beta$. 
In Fig.~\ref{fig:masses1} and ~\ref{fig:masses11}, we observe that the charged Higgs, the pseudoscalar and the heavy CP-even Higgs are mass degenerate and  accessible at upcoming colliders while being allowed from current experimental constraints  for two values of $\tan \beta = 5.5$ and 10. 
Fig.~\ref{fig:masses2}  shows the variation of the allowed mass ranges for $h, H, A$ and $\rho$ with varying $v_S$ and $\lambda_{12S}$ for fixed $\tan \beta = 5.5$ and 10. While $m_{\rho}$ is typically at the PQ scale, lighter scalars and pseudoscalar are still achievable for low values of $\lambda_{12S}$ over the entire range of $v_S$ and a large fraction of the parameter space is within the scope of detection at the upcoming HL-LHC~\cite{Bahl:2020kwe}.
However, for such low values of the portal couplings, the Higgs couple primarily to the SM particles and the model is an effective U(1) symmetric 2HDM at low energy scales as discussed in 
appendix~\ref{app:matching_scale}. One must look forward to the upcoming axion experiments in order to constrain the mass of the axion and consequently, $v_S$. The mass spectrum would then be dictated by $\tan \beta$ and $\lambda_{12S}$ and hence considerably difficult to distinguish between 2HDM and 2hdSMASH at low energy experiments, probing the Higgs sector. Therefore, experimental observations at PLANCK and BICEP/KECK could provide conclusive evidence  of 2hdSMASH while precision measurements of the Higgs sector at upcoming lepton colliders in correlation with studies of axion dark matter could provide complimentary evidence to observe or constrain the model in the upcoming decade.

\section{Summary and Conclusions}
\label{sec:conclusions}
We have investigated the 2hdSMASH model  consisting of two SU(2) Higgs doublets, a SM complex scalar singlet and three SM right-handed neutrinos addressing dark matter, inflation and neutrino masses. In this paper, we focus on the interconnection of inflation and low energy phenomenology and present representative benchmarks consistent with theoretical and experimental constraints.\\
We have realized effective single inflation with stable attractor solutions thereby neglecting isocurvature perturbation which we have analyzed in section \ref{sec:PQ-and mixed inflation}. The orthogonal directions are verified by analysis to be heavy~\cite{Matlis:2022iwr}. Therefore 2hdSMASH accommodates
four effective single field directions, namely the direction of the PQ-scalar $s$ termed PQ-Inflation (PQI) and three effective single field mixed directions $s h_{1}$, $s h_{2}$ and $s h_{12}$  termed PQ-Two-Higgs-Inflation (PQTHI). These field directions are acquired by establishing a hierarchy of non-minimal couplings to gravity $\xi_{S}\gg\xi_{1,2}\simeq 0$, thereby avoiding unpredictive inflation in 2HDM-directions. Both, PQI and PQTHI-$s h_{ij}$ require $\xi_S \lesssim 1$ to ensure predictive inflation without breaking unitarity. This sets stringent constraints on the effective PQ-scalar self coupling $\tilde{\lambda}_{S}$ \eqref{eq:effective_lambda_s} , namely $\tilde{\lambda}_{S}\sim 10^{-10}\xi_{S}^{2}$, which are required by the PLANCK 2018~\cite{Planck:2018jri,Planck:2018vyg} constraint on the scalar perturbation amplitude $A_{S}\sim 10^{-10}$, i.e. $\tilde{\lambda}_{S}\sim 10^{-10}$. Accounting for post-inflationary expansion of the universe and considering the PLANCK 2018~\cite{Planck:2018jri,Planck:2018vyg} and BICEP/KECK XIII \cite{BICEP:2021xfz} measurements on the spectral tilt $n_s$ and the tensor-to-scalar-ratio $r$, we obtain the following constraints on $\tilde{\lambda}_{S}$, $\xi_{S}$ and the e-fold number $N_\ast$, i.e. $$9\times 10^{-10}\gtrsim\tilde{\lambda}_{S}\gtrsim 2.2\times 10^{-12} ~~,~~ 1.35\times 10^{-2}\lesssim\xi_{S}\lesssim 1 ~~,~~ 59.3 \lesssim N_{\ast}\lesssim 60.7\,.$$ For $\xi_{S}\sim 1$ we obtain the following range for $n_s$ and $r$: $$0.966 \lesssim n_{s}\lesssim 0.967 \text{ and } 0.0037\gtrsim r\gtrsim 0.0036$$ satisfying the PLANCK 2018~\cite{Planck:2018jri,Planck:2018vyg} and BICEP/KECK XIII \cite{BICEP:2021xfz} measurements as can be seen in figure \ref{fig:ns_r}. The two variants of inflation realizing these predictions, namely PQI and PQTHI, can be distinguished via the portal couplings. While all portal couplings are required to fulfill $\left|\lambda_{1S,2S,12S}\right|\ll\sqrt{\tilde{\lambda}_{S}}$ to ensure stability of $\lambda_{S}$ all the way up to the Planck scale, there are two distinguishable scenarios of portal coupling regimes, namely
\begin{align}
&\lambda_{12S}\ll\lambda_{S}\lesssim \lambda_{1S,2S}\ll \lambda_{1,2,34}~~~\text{with}~~~\lambda_{iS}^{2}/\lambda_S\ll \lambda_{1,2,34} &&\text{(I)}\notag\\
& \lambda_{12S}\lesssim \lambda_{1S}\sim \lambda_{2S}\ll \lambda_{1,2,34,S}\,, &&\text{(II)}\notag
\end{align}
describing PQTHI and PQI respectively. These two scenarios are associated with an enhanced Poincaré symmetry $\mathcal{G}_{P}^{\text{PQ}}\times \mathcal{G}_{P}^{\text{2HDM}}$ where the 2HDM-sector effectively decouples from the PQ-scalar sector as it was shown in Ref.~\cite{Clarke:2015bea}. 
This accommodates a scale-invariant theory where radiative corrections are negligible due to the smallness of the portal couplings. In particular, a tiny value for $\lambda_{12S}$ ensures high-scale validity since $\lambda_{12S}v_S^2$ resembles a low energy soft breaking parameter of a softly broken $U(1)$-symmetric 2HDM. Moreover, the portal couplings are required to preserve their sign at the Planck scale following the inflationary conditions
\begin{align}
&\lambda_{1S,2S}>0\,,\notag\\
&  \left(\lambda_{iS}>0 \wedge \lambda_{jS}<0\right)\vee \left(\lambda_{1S,2S}<0\right) \,, \notag
\end{align}
for PQI and PQTHI-$sh_{i,j,ij}$ respectively. 

These constraints were considered in an extensive analysis on the one-loop RG-running of $\lambda_{1S}$ and $\lambda_{2S}$, thereby accounting for low energy effects from thermal leptogenesis, BAU, Higgs phenomenology as well as vacuum stability and perturbative unitarity. For PQI, we obtained constraints from thermal leptogenesis and BAU in the one-loop RG-analysis on the portal couplings
\begin{align*}
    \lambda_{1S}\gtrsim \text{Tr}\left(Y_{\nu}^{\dagger}Y_{\nu}Y_{N}^{\dagger}Y_{N}\right)\,.
\end{align*}
ensuring $\lambda_{1S}$ to remain positive all the way up to the Planck scale. For PQTHI-scenarios, we considered two cases, namely PQTHI-$sh_{i}$ represented by $\left|\lambda_{1S}\right|_{\mu_{\text{ew}}}\simeq \left|\lambda_{2S}\right|_{\mu_{\text{ew}}}$ and PQTHI-$sh_{12}$ represented by $\left|\lambda_{1S}\right|_{\mu_{\text{ew}}}\lessgtr \left|\lambda_{2S}\right|_{\mu_{\text{ew}}}$. Considering PQTHI-$sh_{i}$, we obtained a correction to the initial value of $\lambda_{1S}$ which counterbalances the negative top Yukawa contribution of $\lambda_{2S}$,
\begin{align}
\left|\lambda_{1S}^{\text{corr.}}(\mu_{\textit{EW}})\right|=\left|\lambda_{1S}+\delta\lambda_{1S}\right|_{\mu_{\textit{EW}}}\approx \left|\lambda_{2S}\times\left(1+\dfrac{2\lambda_{3}+\lambda_{4}}{3y_{t}^{2}}\right)\right|_{\mu_{\textit{EW}}}\,,
\end{align}
where $\lambda_{1S}^{\text{corr.}}$ is the corrected initial value of $\lambda_{1S}$. In the case of PQTHI-$sh_{12}$, we found the responsible couplings, i.e. $\lambda_3$ and $\lambda_4$, which force $\lambda_{2S}$ to run towards $\lambda_{1S}$. Furthermore, we obtained an analytic understanding of RG-running in the 2HDM, where the couplings $\lambda_3$ and $\lambda_4$ provide RG-running stability. This is ensured by the relation
\begin{align}
\delta(\mu)+\lambda_2(\mu)\equiv \dfrac{\lambda_{34}^{2}(\mu)}{\lambda_{1}(\mu)}+\lambda_2(\mu)\geq 0~~~~,~~~~\forall \mu
\end{align}
which prevents $\lambda_{2}(\mu)$ to run negative. We identified two intertwining effects which stabilize $\lambda_2 (\mu)$ by 1) the initial value and 2) by RG-effects caused by $\lambda_3^{2}$ and $\lambda_{34}^{2}$. Moreover, we included perturbative unitarity conditions in our RG-analysis which causes the couplings to be bounded from above. Amongst others, we considered the most prominent perturbative unitarity conditions on the couplings which we used for stabilization, i.e.
\begin{align}
\left|\lambda_{1,3}(\mu)\right|<8\pi~~,~~\left|\lambda_{3}(\mu)\pm \lambda_{4}(\mu)\right|<8\pi~~,~~\left|\delta (\mu)\right|<8\pi\,,
\end{align} 
in order to avoid Landau poles. From this elaborate RG-stability analysis we acquired viable benchmark points which account for all theoretical and experimental constraints, i.e. vacuum stability, perturbative unitarity, inflationary cosmology, thermal leptogenesis, BAU and Higgs phenomenology. The latter was confirmed by testing the benchmarks with \texttt{HiggsSignals-v2} and \texttt{HiggsBounds-v5} which are in accordance with current LEP, Tevatron and LHC data.
\\

In summary, the salient points of our work are,
\begin{itemize}
   
    \item Realising inflation in the singlet scalar directions set a stringent constraint on the effective self-coupling $\tilde{\lambda}_S \sim 10^{-10}$. This sets stringent constraints on the portal couplings $\lambda_{1S}$ and $\lambda_{2S}$. On the other hand, BfB conditions also set an upper bound on $\lambda_{12S}$ depending on $\lambda_{1S}$ and $\lambda_{2S}$. 
    
     \item We observed that the favoured range of portal couplings from inflation, especially $\lambda_{12S}$ along with $v_S$ and $\tan \beta$ have a strong bearing on the mass of the heavy Higgs spectra.  For axionic dark matter, the present constraints on $f_a$, and thermal relic abundance restricts the singlet \textit{vev} $v_S$ to be $\mathcal{O}(10^9$-$10^{10})$ GeV. The large hierarchy between $v$ and $v_S$ leads to a naturally \textit{compressed} heavy Higgs spectrum, at the tree-level. 
     Ref.~\cite{Espriu:2015mfa} previously discussed that $\mathcal{O}$(TeV) scale Higgs spectra are obtainable for portal couplings of the order $\frac{v^2}{v^2_S}$. Additionally, we observe that this region remains favoured from the inflationary point of view which also favour such low portal couplings, thereby  predicting  $\mathcal{O}$(TeV) scale BSM Higgses accessible at HL-LHC and future colliders.
    
    \item We obtained representative benchmarks satisfying both theoretical conditions (BfB, high scale validity, perturbative unitarity) and experimental  constraints from the Higgs sector consistent with inflationary conditions at the high scale and within the projected reach of the HL-LHC.  
    
    \item Lastly, we note that although a light TeV scale Higgs sector may be possible, it would be  difficult to distinguish this model from 2HDM at LHC and its future upgrades. Such searches  coupled with axion dark matter searches could be good complementary channels with observations at axion detection experiments, while observations by PLANCK would be more conclusive in discovering or excluding this model.  
    
\end{itemize}

We have omitted the discussion on preheating and reheating which is crucial to understand whether each inflationary scenario is able to reheat the universe efficiently enough to provide the correct thermal history of our universe. Subsequently, this will affect the number of e-folds needed to sufficiently inflate the universe and its complete thermal predictions regarding PQ-symmetry restoration, leptogenesis and BAU. The predictions of this paper, do however provide an answer to whether inflation in one or the other direction is preferred while accommodating phenomenological constraints. We leave the remaining investigation of the thermal history in 2hdSMASH for a future study.

During the final stages of our study, we came across Ref.~\cite{Sopov:2022bog} which addresses a similar problem as in this paper. However, our work is different from theirs in several respects including the model which has been studied. While both studies deal with the $\nu$DFSZ model, Ref.~\cite{Sopov:2022bog} works with a trilinear U(1) breaking term, discussing the inflationary directions in a top specific $\nu$DFSZ model termed as VISH$\nu$ which serves as a complementary model for our study. In our work, we deal with a quadratic U(1) breaking term and determine the inflationary conditions subject to all theoretical conditions including unitarity, boundedness-from-below and high-scale validity to arrive at phenomenologically relevant benchmarks for future collider searches. 

\section*{Acknowledgements}

 The authors acknowledge support by the Deutsche Forschungsgemeinschaft (DFG, German Research Foundation) under Germany's Excellence Strategy EXC 2121 "Quantum Universe"- 390833306. This work has been partially funded by the Deutsche Forschungsgemeinschaft (DFG, German Research Foundation) - 491245950. JD acknowledges support from the  HEP  Dodge Family Endowment Fellowship at the Homer L.Dodge Department of Physics $\&$ Astronomy at the University of Oklahoma. The authors thank A.H. Sopov and C.~Tamarit for helpful discussions. 
 
\section*{Appendix}
\appendix 
\section{Derivation of CP-even/odd Scalar Masses}
\label{app:deriv_scalar_masses}
We derive the charged, neutral CP-odd and CP-even scalar masses by diagonalizing the corresponding squared mass matrices respectively. We will see that computing the charged and CP-odd scalar masses is simpler than computing the CP-even scalar masses. In fact, we will have to use an expansion in order of $v/v_S$ for the latter, while the computation of the former is exact. In the following we will start with the derivation of the charged and CP-odd scalar masses, i.e. $H^{\pm}$ and $A$, before we move on to the neutral CP-even scalar masses.\\
\\
For the charged CP-even and the neutral CP-odd masses, we expand the Higgs doublets $\Phi_{i}$ and the PQ-scalar singlet $S$ about their vacuum:
\begin{align}
 \Phi_1 =& (h_1^{+} \,,\, \frac{1}{\sqrt{2}}(v_1+h_1+ia_1))^T\,, \\
 \Phi_2 =& (h_2^{+} \,,\, \frac{1}{\sqrt{2}}(v_2+h_2+ia_2))^T\,, \\
  S =& \frac{1}{\sqrt{2}}(v_S+ s + ia_S) \, .    
\end{align}
in order to derive the squared mass matrices:
\begin{align}
\mathcal{M}_{H^{\pm}}^{2}&=
\begin{pmatrix}
\frac{v_{2} \left(v_{S}^2 \lambda_{12S} -\lambda_{4} v_{1} v_{2}\right)}{2 v_{1}} & \frac{1}{2} \left(\lambda_{4} v_{1} v_{2}-v_{S}^2 \lambda_{12S} \right) \\
 \frac{1}{2} \left(\lambda_{4} v_{1} v_{2}-v_{S}^2 \lambda_{12S} \right) & \frac{v_{1} \left(v_{S}^2 \lambda_{12S} -\lambda_{4} v_{1} v_{2}\right)}{2 v_{2}}
\end{pmatrix}  \,,\\
\notag\\
\mathcal{M}_{A}^{2}&=
\begin{pmatrix}
\frac{v_{2} v_{S}^2 \lambda_{12S} }{2 v_{1}} & -\frac{v_{S}^2 \lambda_{12S} }{2} & -v_{2} v_{S} \lambda_{12S}  \\
 -\frac{v_{S}^2 \lambda_{12S} }{2} & \frac{v_{1} v_{S}^2 \lambda_{12S} }{2 v_{2}} & v_{1} v_{S} \lambda_{12S}  \\
 -v_{2} v_{S} \lambda_{12S}  & v_{1} v_{S} \lambda_{12S}  & 2 v_{1} v_{2} \lambda_{12S}
\end{pmatrix}\,,
\end{align}
which can be diagonalized by considering the characteristic polynomial in an eigenvalue equation:
\begin{align}
\det\left(\mathcal{M}_{i}^{2}- \mathbb{1}\lambda\right)=0\,.
\end{align}
The obtained eigenvalues correspond to the charged CP-even and neutral CP-odd masses. In the case of the charged CP-even eigenvalues, we acquire a charged Goldstone boson which is "eaten" by the $W^{\pm}$-bosons and a charged Higgs with a squared mass given by:
\begin{align}
m_{H^{\pm}}^{2}=\frac{1}{2} v_{S}^2 \left(\frac{\left(t_{\beta}^2+1\right) \lambda_{12S} }{t_{\beta}}-\frac{\lambda_{4} v^2}{v_{S}^2}\right)\,.
\end{align}
For the CP-odd eigenvalues, we get two pseudo-Nambu-Goldstone bosons and one pseudoscalar Higgs. One of the two pseudo-Nambu-Goldstone bosons is "eaten" by the $Z$-Boson while the other one corresponds to the axion acquiring a mass from the mixing with the neutral pion. The squared pseudoscalar Higgs mass is given by:
\begin{align}
m_{A}^{2}=\frac{2\lambda_{12S} v_{S}^{2}}{1+t_{\beta}^{2}}  \left(\frac{v^2}{v_{S}^{2}} t_{\beta}+\frac{\left(1+t_{\beta}^2\right)^{2}}{4 t_{\beta}}\right)\,.
\end{align}
We proceed with the CP-even scalar masses and derive the squared-mass matrix in the right basis by making an ansatz for the unitary rotation matrix $R$. The procedure is the following:
\begin{itemize}
\item[1.)] Calculate the squared-mass matrix in the correct basis, i.e. the basis-vectors need to be specified with respect to the physical field basis given by $\left(h,\beta,s\right)$:
 \begin{align}
 \mathcal{M}_{0^{+}}'^{2}
=\begin{pmatrix}
\left.\dfrac{\partial^{2} V}{\partial h \partial h}\right|_{v,v_{S}}&\left.\dfrac{1}{h}\dfrac{\partial^{2} V}{\partial h \partial \beta}\right|_{v, v_{S}}&\left.\dfrac{\partial^{2} V}{\partial h \partial s}\right|_{v, v_{S}}\\
\left.\dfrac{1}{h}\dfrac{\partial^{2} V}{\partial \beta \partial h}\right|_{v, v_{S}}&\left.\dfrac{1}{h^{2}}\dfrac{\partial^{2} V}{\partial \beta \partial \beta}\right|_{v, v_{S}}&\left.\dfrac{1}{h}\dfrac{\partial^{2} V}{\partial \beta \partial s}\right|_{v, v_{S}}\\
\left.\dfrac{\partial^{2} V}{\partial s \partial h}\right|_{v, v_{S}}&\left.\dfrac{1}{h}\dfrac{\partial^{2} V}{\partial s \partial \beta}\right|_{v, v_{S}}&\left.\dfrac{\partial^{2} V}{\partial s \partial s}\right|_{v, v_{S}}
\end{pmatrix}\,.
 \end{align}
\item[2.)] Make use of the fact that $v/v_{S}\ll 1$ by factoring out $v_{S}$ from $\mathcal{M}_{0^{+}}'^{2}$:
\scriptsize{\begin{align}
&\mathcal{M}_{0^{+}}'^{2}=\label{eq:sq_mass_matrix}\\
&v_{S}^{2}\begin{pmatrix}
 \left(\frac{v}{v_{S}}\right)^{2}\frac{ \left(\lambda_{2} t_{\beta}^4+2 \lambda_{34} t_{\beta}^2+\lambda_{1}\right)}{\left(t_{\beta}^2+1\right)^2} & \left(\frac{v}{v_{S}}\right)^{2}\frac{t_{\beta} \left((\lambda_{2}-\lambda_{34}) t_{\beta}^2-\lambda_{1}+\lambda_{34}\right)}{\left(t_{\beta}^2+1\right)^2} & \left(\frac{v}{v_{S}}\right)\frac{\left(\lambda_{2S} t_{\beta}^2-2 \lambda_{12S}  t_{\beta}+\lambda_{1S}\right)}{t_{\beta}^2+1} \\
 \left(\frac{v}{v_{S}}\right)^{2} \frac{t_{\beta} \left((\lambda_{2}-\lambda_{34}) t_{\beta}^2-\lambda_{1}+\lambda_{34}\right)}{\left(t_{\beta}^2+1\right)^2} & \left(\frac{v}{v_{S}}\right)^{2}\frac{t_{\beta}^2 (\lambda_{1}+\lambda_{2}-2 \lambda_{34})}{\left(t_{\beta}^2+1\right)^2}+\frac{\left(t_{\beta}^2+1\right) \lambda_{12S} }{2 t_{\beta}} & \left(\frac{v}{v_{S}}\right)\frac{\left(\left(t_{\beta}^2-1\right) \lambda_{12S} +t_{\beta} (\lambda_{2S}-\lambda_{1S})\right)}{t_{\beta}^2+1} \\
 \left(\frac{v}{v_{S}}\right)\frac{ \left(\lambda_{2S} t_{\beta}^2-2 \lambda_{12S}  t_{\beta}+\lambda_{1S}\right)}{t_{\beta}^2+1} & \left(\frac{v}{v_{S}}\right)\frac{ \left(\left(t_{\beta}^2-1\right) \lambda_{12S} +t_{\beta} (\lambda_{2S}-\lambda_{1S})\right)}{t_{\beta}^2+1} & \lambda_{S}
\end{pmatrix}\,.\notag
\end{align}}
\normalsize
\item[3.)] Choose the unitary matrix, cf. Ref. \cite{Lombardi:2018wed}\footnote{This ansatz was introduced in Ref. \cite{Lombardi:2018wed} in order to diagonalize the neutral CP even squared mass matrix of the DFSZ model with a cubic term $\propto c\, \Phi_2^{\dagger}\Phi_1 S +h.c.$ instead of a quartic term.}
\begin{align}
R=\exp\left\lbrace A\left(\frac{v}{v_{S}}\right)+B\left(\frac{v}{v_{S}}\right)^{2}\right\rbrace\,,
\end{align} 
where $A,B\in \mathbb{R}^{3\times 3}$ with $A^{T}=-A$ and $B^{T}=-B$, to diagonalize the squared-mass matrix $\mathcal{M}_{0^{+}}'^{2}$.
\item[4.)] Match $A\left(\frac{v}{v_{S}}\right)+B\left(\frac{v}{v_{S}}\right)^{2}$ in powers of $v/v_{S}$ with $\mathcal{M}_{0^{+}}'^{2}$:
\begin{align}
& A\left(\dfrac{v}{v_{S}}\right)+B\left(\dfrac{v}{v_{S}}\right)^{2}=\begin{pmatrix}
0&B_{12}\left(\dfrac{v}{v_{S}}\right)^{2}&A_{13}\left(\dfrac{v}{v_{S}}\right)\\
-B_{12}\left(\dfrac{v}{v_{S}}\right)^{2}&0&A_{23}\left(\dfrac{v}{v_{S}}\right)\\
-A_{13}\left(\dfrac{v}{v_{S}}\right)&-A_{23}\left(\dfrac{v}{v_{S}}\right)&0
\end{pmatrix}\\
\nonumber\\
&\text{with }A_{12}=B_{13}=B_{23}=0\,.\nonumber
\end{align}
\item[5.)]Diagonalize the squared-mass matrix by calculating:
\begin{align}
\mathcal{D}=R^{T}\mathcal{M}_{0^{+}}'^{2}R
\end{align} 
and expanding $\mathcal{D}$ up to second order in $v/v_{S}$, cf. Refs. \cite{Espriu:2015mfa,Lombardi:2018wed}
\begin{align}
\mathcal{D}&=\left.\sum_{n=0}^{\infty}\left.\dfrac{\partial^{n}}{\partial x^{n}}\left(R^{T}\mathcal{M}_{0^{+}}^{2}R\right)\right|_{x=0}\dfrac{x^{n}}{n!}\right|_{x=\left(\frac{v}{v_{S}}\right)}\simeq \tilde{\mathcal{D}}+\mathcal{O}\left(\left(\dfrac{v}{v_{S}}\right)^{3}\right)\\
\Rightarrow ~~\tilde{\mathcal{D}}&=\begin{pmatrix}
m_{h}^{2}/v_{S}^{2}&0&0\\
0&m_{H}^{2}/v_{S}^{2}&0\\
0&0&m_{s}^{2}/v_{S}^{2}
\end{pmatrix}
=
\begin{pmatrix}
\tilde{\mathcal{D}}_{11}&\tilde{\mathcal{D}}_{12}&\tilde{\mathcal{D}}_{13}\\
\tilde{\mathcal{D}}_{12}&\tilde{\mathcal{D}}_{22}&\tilde{\mathcal{D}}_{23}\\
\tilde{\mathcal{D}}_{13}&\tilde{\mathcal{D}}_{23}&\tilde{\mathcal{D}}_{33}
\end{pmatrix}\,.
\end{align}
\item[6.)] Utilize the fact that we have three equations $\left(\tilde{\mathcal{D}}_{12,13,23}\overset{!}{=}0\right)$ for three parameters $\left(B_{12}, A_{13,23}\right)$ and solve these equations for these parameters:
\begin{align}
\tilde{\mathcal{D}}_{12,13,23}\overset{!}{=}0~ \Longrightarrow ~B_{12}~,~A_{13,23} \,.
\end{align}
\newpage
\item[7.)] Use the results of $B_{12}$, $A_{13,23}$ to obtain $\tilde{\mathcal{D}}_{11,22,33}$ in order to calculate the masses: 
\begin{align}
\frac{m_{h}^{2}}{v_S^2}&=  \frac{\left(\frac{v}{v_S}\right)^2}{\left(1+t_{\beta}^2\right)^2} \left[ {\lambda_{1}+ t_{\beta}^4\, \lambda_{2}+2\, t_{\beta}^2\, \lambda_{34}}{}
-\dfrac{\left(\lambda_{1S}+t_{\beta}^2\,\lambda_{2S} -2 t_{\beta} \lambda_{12S} \right)^2}{\lambda_{S}}\right]
\\
&+ {\mathcal O}\left( \left( \frac{v}{v_S} \right)^4\right)\,,\notag
\\
\frac{m_{H}^{2}}{v_S^2}&= \dfrac{\left(1+t_{\beta}^2\right)\lambda_{12S}}{2t_{\beta}}
+ \frac{t_{\beta}}{\left(1+t_{\beta}^{2}\right)^{2}}\,\Biggl[ 
 \dfrac{2\left(\left(\lambda_{1S}-\lambda_{2S}\right)t_{\beta}+\lambda_{12S}\left(1-t_{\beta}^{2}\right)\right)^{2}}{\lambda_{12S}\left(1+t_{\beta}^2\right)-2 t_{\beta}\lambda_{S} }
\\
&+\left(\lambda_{1}+\lambda_{2}-2\lambda_{34}\right)t_{\beta}\Biggr]
\left(\frac{v}{v_S}\right)^2
+ {\mathcal O}\left( \left( \frac{v}{v_S} \right)^4\right) \,,\notag\\
\frac{m_{s}^{2}}{v_S^2}&= \lambda_{S}  + 
\frac{t_{\beta}}{\left(1+t_{\beta}^{2}\right)^{2}}\,\Biggl[\dfrac{\left(\lambda_{1S}+\lambda_{2S}t_{\beta}^{2}-2t_{\beta}\lambda_{12S}\right)^{2}}{\lambda_{S}}
\\
&- \dfrac{2 t_{\beta}^{2}\left(\left(\lambda_{1S}-\lambda_{2S}\right)t_{\beta}+\lambda_{12S}\left(1-t_{\beta}^{2}\right)\right)^{2}}{\lambda_{12S}\left(1+t_{\beta}^2\right)-2 t_{\beta}\lambda_{S} }
\Biggr]\left(\frac{v}{v_S}\right)^2 + {\mathcal O}\left( \left( \frac{v}{v_S} \right)^4\right)\,.
\notag
\end{align}
These analytically derived masses are valid for \textbf{BP1}-\textbf{BP4} of section \ref{sec:mass_spec}.
\item[8.)] The limit of small portal couplings $\left(\lambda_{1S,2S,12S}\ll 1\right)$ represents the decoupling of the 2HDM from the PQ-scalar singlet. We choose the following ansatz for the unitary matrix $R$:
\begin{align}
R=\exp\left\lbrace M\left(\dfrac{v}{v_{s}}\right)^{2}\right\rbrace\,,
\end{align}
which makes the components $\left(\mathcal{M}_{0^{+}}'^{2}\right)_{12,13,23}$ marginally small since they are weighted up to second order in $v/v_{s}$. Thus, we obtain 
\begin{align}
&m_{h}^{2} \approx  \frac{v^{2}}{\left(1+t_{\beta}^{2}\right)^{2}} \left(\lambda_{1}+\lambda_{2}t_{\beta}^{4}+2\lambda_{34}t_{\beta}^{2}\right)\,,\\
&m_{H}^{2} \approx \dfrac{v_{S}^2 \lambda_{12S}\left(1+t_{\beta}^{2}\right)}{2 t_{\beta}^{2}}+ \dfrac{t_{\beta}^{2}\left(\lambda_{1}+\lambda_{2}-2\lambda_{34}\right)v^{2}}{\left(1+t_{\beta}^{2}\right)^{2}}\,,\\
&m_{s}^{2}\approx \lambda_{S}v_{S}^{2}\,,
\end{align}
\vspace*{-0.9cm}
which are represented by \textbf{BP4} of section \ref{sec:mass_spec}.
\end{itemize}
\newpage
\section{Constraints on the Axion Decay Constant}
\label{app:axion_bounds}
There are astrophysical and cosmological constraints on the 
axion decay constant $f_a$. 
In fact, the measured duration of the neutrino signal of the supernova SN 1987A provides an upper bound on the emission rate of axions~\cite{Raffelt:2006cw} that can be translated into a lower bound on the decay constant. Most recent improved determinations  
of the axion emission rate result in~\cite{Carenza:2019pxu,Carenza:2020cis}
\begin{equation}
f_a \gtrsim 
\left\{
\begin{array}{l}
5.2\times 10^8\,{\rm GeV}\,,\      {\rm for}\ \tan\beta \lesssim 0.5\,,\\
9.8\times 10^8\,{\rm GeV}\,,\     {\rm for}\ \tan\beta \gtrsim 5\,,
\end{array}
\right.
\end{equation}
where $\tan \beta \equiv v_2/v_1$. 

Axion dark matter considerations bring further restrictions. We will argue that preheating in 2hdSMASH is quite similar to preheating in SMASH~\cite{Ballesteros:2016euj,Ballesteros:2016xej}. In particular, the PQ symmetry is restored in the preheating phase and later broken during the radiation dominated era of the hot big bang phase. Correspondingly, axion dark matter is produced not only by 
the misalignment mechanism~\cite{Preskill:1982cy,Abbott:1982af,Dine:1982ah}, but also by the decay of topological defects~\cite{Davis:1986xc,Lyth:1991bb}: strings --  vortex-like defects, which are formed at the PQ phase transition -- and
domain walls -- surface-like defects by which the strings are attached when the temperature of the Universe becomes comparable to 
critical temperature of the QCD crossover. The structure of the domain walls is determined by the domain wall number $N$. For $N>1$, as in 2hdSMASH,  domain walls are stable, as long as the $U(1)_{\rm PQ}$  is an exact global symmetry. In this case, they would overclose the Universe, in conflict with standard cosmology~\cite{Zeldovich:1974uw,Sikivie:1982qv}. 
However, there are very good arguments that the PQ symmetry -- like any global symmetry -- is not protected from explicit symmetry breaking effects by PLANCK-scale-suppressed operators appearing in the low-energy effective Lagrangian~\cite{Holman:1992us,Kamionkowski:1992mf,Barr:1992qq,Ghigna:1992iv}. 
However, these operators modify also the axion potential, eventually shifting its minimum away from zero, thereby destroying the solution of the strong CP problem, namely, the axion quality problem.
Crucially, this drawback is absent in models where the Peccei-Quinn symmetry is not ad hoc but instead an automatic or accidental symmetry of an exact discrete $Z_{\mathcal N}$ symmetry~\cite{Georgi:1981pu}. In fact, for $\mathcal N \geq 9$, the discrete symmetry can protect the axion against semiclassical gravity effects~\cite{Dias:2002hz,Dias:2002gg,Dias:2003zt,Carpenter:2009zs,Harigaya:2013vja,Dias:2014osa,Ringwald:2015dsf}, while it can at the same time provide for the small explicit symmetry breaking term needed to make the 
$N >1$ models cosmologically viable. Assuming such a $Z_{\mathcal N}$ symmetry at action in 2hdSMASH, the 
required value of $f_a$ to explain all the cold dark matter in the universe by axions is~\cite{Ringwald:2015dsf,IAXO:2019mpb}\footnote{Recently, there has been an investigation~\cite{Beyer:2022ywc} that argued   that for domain wall number $N=6$ in DFSZ models $f_{a}$ $\lesssim 5.4 \times 10^8$ GeV.} 
\begin{eqnarray}
4.4\times 10^7\,{\rm GeV}\lesssim f_a\lesssim 9.9\times 10^9\,{\rm GeV},\   {\rm for}\ {\mathcal N}=9, \\
1.3\times 10^9\,{\rm GeV}\lesssim f_a\lesssim 9.9\times 10^9\,{\rm GeV},\  {\rm for}\ {\mathcal N}=10.
\end{eqnarray}
Clearly, part of the parameter range required for ${\mathcal N}=9$ is already excluded by the SN 1987A constraint. 
We conclude, that the preferred PQ scale range of 2hdSMASH is  
\begin{eqnarray}
3.1\times 10^9\,{\rm GeV}\lesssim v_S\lesssim 5.9\times 10^{10}\,{\rm GeV}\,,\      {\rm for}\ \tan\beta \lesssim 0.5\,,\\
5.9\times 10^9\,{\rm GeV}\lesssim v_S\lesssim 5.9\times 10^{10}\,{\rm GeV}\,,\      {\rm for}\ \tan\beta \gtrsim 5\,.
\end{eqnarray}

\section{Theoretical Constraints}
\label{app:theoretical_constraints}
We give the derivation of the two theoretical constraints, namely the Bounded from Below conditions, and the perturbative unitarity conditions for the 2hdSMASH model.
\subsection{Bounded from below conditions}
\label{app:bfb}
We derive the BfB conditions in 2hdSMASH by imposing copositivity (conditionally positive conditions) where the biquadratic form of the quartic scalar potential is positive on non-negative vectors. This is realized by applying Sylvester's criterion. We quote the copositivity condition from Refs. \cite{Kannike:2012pe,Kannike:2016fmd} and Sylvester's criterion from Ref. \cite{Kannike:2012pe}  in the following: \\
\\
\underline{\textbf{Copositivity}:}\\
\textit{``A symmetric matrix $A$ is copositive if the quadratic form $x^{T} A x \geqslant 0$ for all vectors $x \geqslant 0$ in the non-negative orthant $\mathbb{R}_{+}^{n}$. (The notation $x \geqslant 0$ means that $x_i \geqslant 0$ for each $i = 0, \ldots, n$.)  A symmetric matrix $A$ is strictly copositive if the quadratic form $x^{T} A x > 0$ for all vectors $x > 0$ in the non-negative orthant $\mathbb{R}_{+}^{n}$."}, cf. Refs. \cite{Kannike:2012pe,Kannike:2016fmd}.\\
\\
\underline{\textbf{Sylvester's criterion}:}\\
\textit{``\dots for a symmetric matrix $A$ to be positive semidefinite, the principal minors of $A$ have to be non-negative. (The principal minors are determinants of the principal submatrices. The principal submatrices of $A$ are obtained by deleting rows and columns of $A$ in a symmetric way, i.e. if the $i_{1}, \ldots, i_{k}$ rows are deleted, then the $i_{1}, \ldots, i_{k}$ columns are deleted as well. The largest principal submatrix of $A$ is $A$ itself.) Thus if the matrix $A$ is positive, all of its submatrices, in particular the diagonal elements $a_{ii}$ have to be non-negative."}, cf. Ref. \cite{Kannike:2012pe}.\\
\\
Note that Sylvester's criterion is a necessary requirement for a matrix $A$ to be copositive and by extension $V_{4}>0$. In order to utilize copositivity and thus Sylvester's criterion, we write the quartic part of the scalar potential $V_{4}$:
\begin{align}
V_{4}&=\frac{1}{8}\left(\lambda_{1}h_{1}^{4}+\lambda_{2}h_{2}^{4}+\lambda_{S}s^{4}\right)\\
&+\frac{1}{4}\left(h_{1}^{2}h_{2}^{2}(\lambda_{3}+\zeta_{4}\lambda_{4})+\lambda_{1S}h_{1}^{2}s^{2}+\lambda_{2S}h_{2}^{2}s^{2}-2\lambda_{12S}\zeta_{12S} h_{1}h_{2}s^{2}\right)>0\,,\notag
\end{align}
where
\begin{align}
&\Phi_{1}^{\dagger}\Phi_{2}=\zeta_{4}h_{1}h_{2}~~,~~\zeta_{4}=\dfrac{\left|\Phi_{1}^{\dagger}\Phi_{2}\right|^{2}}{\left|\Phi_{1}\right|^{2}\left|\Phi_{2}\right|^{2}}\,,\\
&\Phi_{1}^{\dagger}\Phi_{2}S^{2}=\zeta_{12S}h_{1}h_{2}s^{2}~~,~~\zeta_{12S}=\dfrac{\text{Re}\left(\Phi_{1}^{\dagger}\Phi_{2}S^{2}\right)}{\left|S\right|^{2}\left|\Phi_{1}\right|\left|\Phi_{2}\right|}
\end{align}
with $\zeta_{4}\in \left[0,1\right]$ and $\zeta_{12S}\in \left[-1,1\right]$. As a first step, we consider the biquadratic form by imposing $\lambda_{12S}=0$:
\begin{align}
&V'_{4}\equiv \left. V_{4}\right|_{\lambda_{12S}=0}=\begin{pmatrix}
h_{1}^{2} & h_{2}^{2} & s^{2}\end{pmatrix}
\begin{pmatrix}
\lambda_{1} & \lambda_{3}+\zeta_{4}\lambda_{4} & \lambda_{1S}\\
\lambda_{3}+\zeta_{4}\lambda_{4} & \lambda_{2} & \lambda_{2S}\\
\lambda_{1S} & \lambda_{2S} & \lambda_{S}
\end{pmatrix}
\begin{pmatrix}
h_{1}^{2} \\ h_{2}^{2} \\ s^{2}\end{pmatrix}
>0
\end{align}
to obtain the \underline{necessary} BFB conditions by applying Sylvester's criterion:
\begin{align}
& \lambda_{1}>0~~,~~\lambda_{2}>0~~,~~\lambda_{3}+\min\left\lbrace0,\lambda_{4}\right\rbrace>-\sqrt{\lambda_{1}\lambda_{2}}\,,\notag\\
\notag\\
&\lambda_{S}>0~~,~~\sqrt{\lambda_{1}\lambda_{S}}>\lambda_{1S}>-\sqrt{\lambda_{1}\lambda_{S}}~~,~~\sqrt{\lambda_{2}\lambda_{S}}>\lambda_{2S}>-\sqrt{\lambda_{2}\lambda_{S}}
\,,\notag\\
&\sqrt{\lambda_{1}\lambda_{2}\lambda_{S}}+\lambda_{2S}\sqrt{\lambda_{1}}+\lambda_{1S}\sqrt{\lambda_{2}}+\left(\lambda_{3}+\min\left\lbrace0,\lambda_{4}\right\rbrace\right)\sqrt{\lambda_{S}}\notag\\
+&\sqrt{2\left(\left(\lambda_{3}+\min\left\lbrace0,\lambda_{4}\right\rbrace\right)+\sqrt{\lambda_{1}\lambda_{2}}\right)\left(\lambda_{1S}+\sqrt{\lambda_{1}\lambda_{S}}\right)\left(\lambda_{2S}+\sqrt{\lambda_{2}\lambda_{S}}\right)}>0\,,\notag
\end{align}
where the last condition is given by the combination of the following two conditions\footnote{Either of the two conditions of eqns. \eqref{eq:nec_cond_for_bfb_mixed_1}-\eqref{eq:nec_cond_for_bfb_mixed_2} has to hold for $V'_{4}>0$} \cite{Kannike:2012pe,Kannike:2016fmd}:
\begin{align}
&\det A >0\,,\label{eq:nec_cond_for_bfb_mixed_1}\\
&\sqrt{a_{11}a_{22}a_{33}}+a_{12}\sqrt{a_{33}}+a_{13}\sqrt{a_{22}}+a_{23}\sqrt{a_{11}}>0\label{eq:nec_cond_for_bfb_mixed_2}
\end{align} 
with $a_{ij}$ as the components of 
$A\in \mathbb{R}^{3\times 3}$ where 
\begin{equation*}
    A=\begin{pmatrix}
\lambda_{1} & \lambda_{3}+\zeta_{4}\lambda_{4} & \lambda_{1S}\\
\lambda_{3}+\zeta_{4}\lambda_{4} & \lambda_{2} & \lambda_{2S}\\
\lambda_{1S} & \lambda_{2S} & \lambda_{S}
\end{pmatrix}.
\end{equation*} By considering the $s$-dependent portal terms of the quartic scalar potential with $\zeta_{12S}=\pm 1$ we get:
\begin{align}
V_{4}^{\text{Portal}}&=\lambda_{1S}h_{1}^{2}s^{2}+\lambda_{2S}h_{2}^{2}s^{2}-2\left(\pm\lambda_{12S} h_{1}h_{2}s^{2}\right)\\
&=\begin{pmatrix}
h_{1}s & h_{2}s
\end{pmatrix}
\begin{pmatrix}
\lambda_{1S} & \mp\lambda_{12S}\\
\mp\lambda_{12S} & \lambda_{2S} 
\end{pmatrix}
\begin{pmatrix}
h_{1}s \\ h_{2}s
\end{pmatrix}\notag
\end{align} 
to obtain the \underline{sufficient} BFB conditions:
\begin{align}
\lambda_{1S}>0~~,~~\lambda_{2S}>0~~,~~\lambda_{1S}\lambda_{2S}-\left|\lambda_{12S}\right|^{2} >0\,.
\end{align}
It's important to note that the sufficient BfB conditions can be overly strict and therefore exclude parameter points which are BfB but do not pass these conditions. Sufficient BfB conditions ensure that $V_{4}>0$ but they are not necessary. For that reason we chose to break these sufficient BfB conditions and checked numerically whether individual parameter points are BfB. We have used the \texttt{Mathematica} package \texttt{BFB} by Ref. \cite{Ivanov:2018jmz}.
\subsection{Perturbative unitarity bounds}
\label{app:perturbative_unitarity}

The full tree-level perturbative unitarity constraints are calculated by requiring that the unique eigenvalues of the $2\to 2$ scalar scattering matrix (scalar S-matrix) $\mathcal{M}_{2\to 2}$ are below the upper bound of $8\pi$ (suggested in Refs. \cite{Muhlleitner:2016mzt} and \cite{Horejsi:2005da}). The following calculations are done by using the perturbative unitarity \texttt{Mathematica} package included in the directory of \texttt{ScannerS} which is described in Ref. \cite{Muhlleitner:2020wwk}. Therefore we follow the calculations on perturbative unitarity of Ref. \cite{Muhlleitner:2020wwk}.\\
\\
First, we need to derive the $2\to 2$ scalar S-matrix which is given by:
\begin{align}
\mathcal{M}_{AB\to CD}=\bra{AB}\mathcal{M}\ket{CD}=\dfrac{1}{\sqrt{\left(1+\delta_{AB}\right)\left(1+\delta_{CD}\right)}}\dfrac{\partial^{4} V_{4}}{\partial A\partial B\partial C\partial D}\,,
\end{align}
where the $\delta_{ij}$- functions are necessary symmetry factors. Here, we have used the quartic scalar potential $V_{4}$ in the gauge eigenbasis $\left(h_{1,2},s,h_{1,2}^{\pm},a_{1,2,S}\right)$ which is given by the following field definition:
\begin{align}
\Phi_{1}&=\begin{pmatrix}
h_{1}^{+}\\
\dfrac{1}{\sqrt{2}}\cdot\left(h_{1}+v_{1}+i\cdot a_{1}\right)
\end{pmatrix}\,,\\
\Phi_{2}&=\begin{pmatrix}
h_{2}^{+}\\
\dfrac{1}{\sqrt{2}}\cdot\left(h_{2}+v_{2}+i\cdot a_{2}\right)
\end{pmatrix}\,,\\
S&=\dfrac{1}{\sqrt{2}}\cdot\left(s+v_{S}+i\cdot a_{S}\right)\,.
\end{align}
The resulting $2\to 2$ scalar S-matrix $\mathcal{M}_{2\to 2}$ is block diagonal and its eigenvalues $\mathcal{M}_{2\to 2}^{i}$ for $i=\left\lbrace 1,\dots,n\right\rbrace$ are bounded by $8\pi$:
\begin{align}
\left|\mathcal{M}_{2\to 2}^{i}\right|< 8\pi\,.
\end{align}
It's important to note that the eigenvalues are basis independent since unitary transformations are basis independent (see Ref.\cite{Muhlleitner:2020wwk,Kanemura:1993hm} for details). Part of the eigenvalues reproduce the same eigenvalues for a soft $U(1)$-symmetric 2HDM in the absence of the PQ-scalar $s$:
\begin{align}
&e_{1,2,3}=\lambda_{1,2,3}\,,\\
&e_{4}=\lambda_{3}-\lambda_{4}\,,\\
&e_{5}=\lambda_{3}+\lambda_{4}\,,\\
&e_{\pm}=\frac{1}{2}\left(\lambda_{1}+\lambda_{2}\pm\sqrt{\left(\lambda_{1}-\lambda_{2}\right)^{2}+4\lambda_{4}^{2}}\right)\,.
\end{align}
By including the PQ-scalar $s$ we get the following additional eigenvalues:
\begin{align}
&s_{1,2}=\lambda_{1S,2S}\,,\\
&s_{3\pm}=\frac{1}{2}\left(\lambda_{1S}+\lambda_{2S}\pm \sqrt{16\lambda_{12S}^{2}+\left(\lambda_{1S}-\lambda_{2S}\right)^{2}}\right)\,,\\
&s_{4\pm}=\frac{1}{2}\left(\lambda_{3}+2\lambda_{4}+\lambda_{S}\pm \sqrt{16\lambda_{12S}^{2}+\left(\lambda_{3}+2\lambda_{4}-\lambda_{S}\right)^{2}}\right)\,,\\
&\frac{1}{2}k_{1,2,3}\,,
\end{align}
where the eigenvalues $\frac{1}{2}|k_{1,2,3}|$ are the real roots of the following cubic potential:
\begin{align}
&\frac{1}{2} \left(48 \lambda_{1} \lambda_{2S}^2+48 \lambda_{1S}^2 \lambda_{2}-144 \lambda_{1} \lambda_{2} \lambda_{S}-64 \lambda_{1S} \lambda_{2S} \lambda_{3}-32 \lambda_{1S} \lambda_{2S} \lambda_{4}+64 \lambda_{3}^2 \lambda_{S}\right)\\
+&\frac{1}{2} \left(64 \lambda_{3} \lambda_{4} \lambda_{S}+16 \lambda_{4}^2 \lambda_{S}+x \left(36 \lambda_{1} \lambda_{2}+24 \lambda_{1} \lambda_{S}-8 \lambda_{1S}^2+24\lambda_{2} \lambda_{S}-8 \lambda_{2S}^2\right)\right)\nonumber\\
&-\frac{x}{2} \left(\lambda_{3}^2-16 \lambda_{3} \lambda_{4}-4 \lambda_{4}^2\right)+\frac{1}{2}\left(x^2 (-6 \lambda_{1}-6 \lambda_{2}-4 \lambda_{S})+x^3\right)\,.\nonumber
\end{align}
Taking the absolute value of all eigenvalues, we obtain the full tree-level perturbative unitarity conditions for 2hdSMASH:
\begin{align}
&\left|\lambda_{1,2,3,1S,2S}\right|<8\pi\,,\\
&\left|\lambda_{3}\pm\lambda_{4}\right|<8\pi\,,\\
&\left|\frac{1}{2}\left(\lambda_{1}+\lambda_{2}\pm\sqrt{\left(\lambda_{1}-\lambda_{2}\right)^{2}+4\lambda_{4}^{2}}\right)\right|<8\pi\,,\\
&\left|\frac{1}{2}\left(\lambda_{1S}+\lambda_{2S}\pm \sqrt{16\lambda_{12S}^{2}+\left(\lambda_{1S}-\lambda_{2S}\right)^{2}}\right)\right|<8\pi\,,\\
&\left|\frac{1}{2}\left(\lambda_{3}+2\lambda_{4}+\lambda_{S}\pm \sqrt{16\lambda_{12S}^{2}+\left(\lambda_{3}+2\lambda_{4}-\lambda_{S}\right)^{2}}\right)\right|<8\pi\,,\\
&\frac{1}{2}\left|k_{1,2,3}\right|<8\pi\,.
\end{align}

 
\section{\boldmath Renormalisation Group Equations}
\label{app:rges}

Following is the full list of one-loop RGEs in the 
2hdSMASH model\footnote{Ref.~\cite{Clarke:2015bea} has also given the one-loop RGEs of 2hdSMASH, but neglected terms quadratic in  $\lambda_{12S}$ systematically. We have restored them here using  \texttt{SARAH} and \texttt{PYR@TE}.}:
\begin{align}
 \mathcal{D} M_{11}^2 = &\
   M_{11}^2 \left( 6\lambda_1 -\frac32 g_1^2-\frac92 g_2^2 +6Y_t^2
\right)
 + M_{22}^2 \left( 4\lambda_3 + 2\lambda_4  \right)
 + M_{SS}^2  2\lambda_{1S} , \\
 \mathcal{D} M_{22}^2 = &\
   M_{22}^2 \left( 6\lambda_2 -\frac32 g_1^2-\frac92 g_2^2 +6Y_b^2 +2Y_\tau^2
   +2\text{Tr}\left(Y_\nu^\dagger Y_\nu\right)\right) 
 + M_{11}^2 \left( 4\lambda_3 + 2\lambda_4  \right)
 + M_{SS}^2  2\lambda_{2S},  \\
\mathcal{D} M_{SS}^2 = &\ 
   M_{SS}^2 \left( 4\lambda_S + \text{Tr}\left(Y_N^\dagger Y_N\right)\right)
 + M_{11}^2  4\lambda_{1S} 
 + M_{22}^2  4\lambda_{2S} , \\
\mathcal{D} \langle S \rangle^2 = &\ 
  -\text{Tr}\left( Y_N^\dagger Y_N \right) \langle S \rangle^2 
  \;\;\; \text{[i.e. the wave function renormalisation]}, \\
\mathcal{D} g_{\{1,2,3\}} = &\ \{7,-3,-7\} g_{\{1,2,3\}}^3 , \\
 \mathcal{D} \lambda_1 = &\ 
   \frac34 g_1^4 + \frac32 g_1^2g_2^2 + \frac94 g_2^4 -\lambda_1\left( 3 g_1^2 +9g_2^2 \right)
 + 12\lambda_1^2 + 4\lambda_3\lambda_4 + 4\lambda_3^2+ 2\lambda_4^2
 + 2\lambda_{1S}^2\nonumber \\
 &+ 12\lambda_1Y_b^2 - 12Y_b^4 
  +4\lambda_1 Y_\tau^2 - 4Y_\tau^4
  + 4\lambda_2\text{Tr}\left(Y_\nu^\dagger Y_\nu\right) -4\text{Tr}\left(Y_\nu^\dagger Y_\nu Y_\nu^\dagger Y_\nu\right), \\
\mathcal{D} \lambda_2 = &\ 
   \frac34 g_1^4 + \frac32 g_1^2g_2^2 + \frac94 g_2^4 -\lambda_2\left( 3 g_1^2 +9g_2^2 \right)
 + 12\lambda_2^2 + 4\lambda_3\lambda_4 + 4\lambda_3^2+ 2\lambda_4^2
 + 2\lambda_{2S}^2 \nonumber\\
 &+ 12\lambda_2Y_t^2 - 12Y_t^4 
, \\
 \mathcal{D} \lambda_3 = &\ 
   \frac34 g_1^4 - \frac32 g_1^2g_2^2 + \frac94 g_2^4 -\lambda_3\left( 3 g_1^2 +9g_2^2 \right)
 + \left(6\lambda_3+2\lambda_4\right)\left(\lambda_1+\lambda_2\right) +4\lambda_3^2+2\lambda_4^2
 + 2\lambda_{1S}\lambda_{2S} \nonumber \\
 &+ \lambda_3\left( 6Y_t^2+6Y_b^2+2Y_\tau^2+2\text{Tr}\left(Y_\nu^\dagger Y_\nu\right) \right)
 - 12Y_t^2Y_b^2  
, \\
 \mathcal{D} \lambda_4 = &\ 
   3 g_1^2g_2^2 -\lambda_4\left( 3g_1^2 + 9g_2^2 \right)
 + 2\lambda_4\left(\lambda_1+\lambda_2\right) + 8\lambda_4\lambda_3 + 4\lambda_4^2 +4\lambda_{12S}^2 \nonumber \\
 &+ \lambda_4\left( 6Y_t^2+6Y_b^2+2Y_\tau^2+2\text{Tr}\left(Y_\nu^\dagger Y_\nu\right) \right)
 + 12Y_t^2Y_b^2 
, \\
\mathcal{D} \lambda_S = &\ 
   10\lambda_S^2 +2\lambda_S \text{Tr}\left( Y_N^\dagger Y_N\right)
   + 4\lambda_{1S}^2 + 4\lambda_{2S}^2 +8\lambda_{12S}^2 -2\text{Tr}\left( Y_N^\dagger Y_N Y_N^\dagger Y_N \right), \\
 \mathcal{D} \lambda_{1S} = &\ 
   \lambda_{1S}\left( -\frac32 g_1^2 -\frac92 g_2^2 
 + 4\lambda_{1S} +4\lambda_S +6\lambda_2  \right) + \lambda_{2S}\left(4\lambda_3 + 2\lambda_4\right)+8\lambda_{12S}^2\nonumber\\
 &+ \lambda_{1S}\left( 6Y_b^2 + 2Y_\tau^2 
+ 2\text{Tr}\left(Y_\nu^\dagger Y_\nu\right) 
 + \text{Tr}\left(Y_N^\dagger Y_N\right) \right) 
 - 4\text{Tr}\left(Y_\nu^\dagger Y_\nu Y_N^\dagger Y_N\right) , \\
 \mathcal{D} \lambda_{2S} = &\ 
   \lambda_{2S}\left( -\frac32 g_1^2 -\frac92 g_2^2 
 + 4\lambda_{2S} +4\lambda_S +6\lambda_2  \right) + \lambda_{1S}\left(4\lambda_3 + 2\lambda_4\right)+8\lambda_{12S}^2 \nonumber \\
 &+ \lambda_{2S}\left( 6Y_t^2 
+ \text{Tr}\left(Y_N^\dagger Y_N\right) \right) , \\
 \mathcal{D}\lambda_{12S} = &\ 
\lambda_{12S} \left( -\frac32 g_1^2 -\frac92 g_2^2 
 + 2\lambda_3 + 4\lambda_4 + 2\lambda_S + 4\lambda_{1S} + 4\lambda_{2S}
 + 3Y_t^2 + 3Y_b^2 + Y_\tau^2\right)\notag\\
 &+ \lambda_{12S}\left(\text{Tr}\left(Y_\nu^\dagger Y_\nu \right) + \text{Tr}\left(Y_N^\dagger Y_N \right) \right), \\
 \mathcal{D} Y_t = &\ 
   Y_t\left( -\frac{17}{12}g_1^2-\frac94 g_2^2-8g_3^2
   +\frac92 Y_t^2+\frac12 Y_b^2 
\right) 
, \\
 \mathcal{D} Y_b = &\ 
   Y_b\left( -\frac{5}{12}g_1^2-\frac94 g_2^2-8g_3^2
   +\frac92 Y_b^2+\frac12 Y_t^2 + Y_\tau^2
\right) , \\
 \mathcal{D} Y_\tau = &\ 
   Y_\tau\left( -\frac{15}{4}g_1^2-\frac94g_2^2
   +\frac52 Y_\tau^2+ 3Y_b^2
\right) , \\
\mathcal{D}Y_\nu = &\ 
   Y_\nu \left( -\frac34 g_1^2 -\frac94 g_2^2 + 3Y_b^2 + \text{Tr}\left(Y_\nu^\dagger Y_\nu\right) \right)
   + Y_\nu Y_\tau^2 -\frac32 \text{Diag}\left(0,0,Y_\tau^2\right)Y_\nu \notag\\
   &+ \frac32 Y_\nu Y_\nu^\dagger Y_\nu + \frac12 Y_\nu Y_N^\dagger Y_N\label{eq:rge_ynu}, \\
\mathcal{D}Y_N = &\ 
   \frac12 \text{Tr}\left( Y_N^\dagger Y_N\right) Y_N + Y_N Y_N^\dagger Y_N
   +Y_N Y_\nu^\dagger Y_\nu + Y_\nu^T Y_\nu^\ast Y_N \label{eq:rge_yn},
\end{align}
where $\mathcal{D}\equiv (4\pi)^2\frac{d}{d\ln\mu_R}$. 
\section{2HDM inflation in 2hdSMASH}
\label{app:2hdm inflation}
\begin{figure}
\centering
\begin{subfigure}{.5\textwidth}
  \centering
  \includegraphics[width=1.\linewidth]{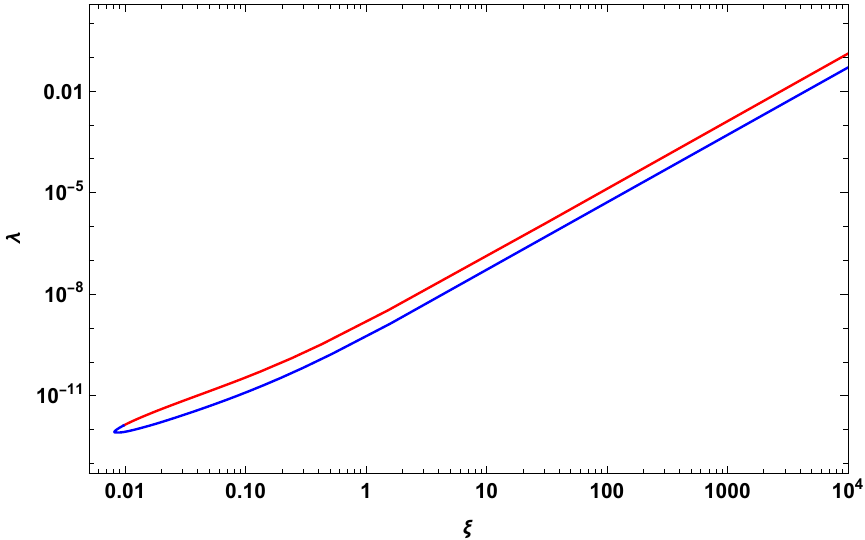}
\end{subfigure}%
\begin{subfigure}{.5\textwidth}
  \centering
  \includegraphics[width=1.\linewidth]{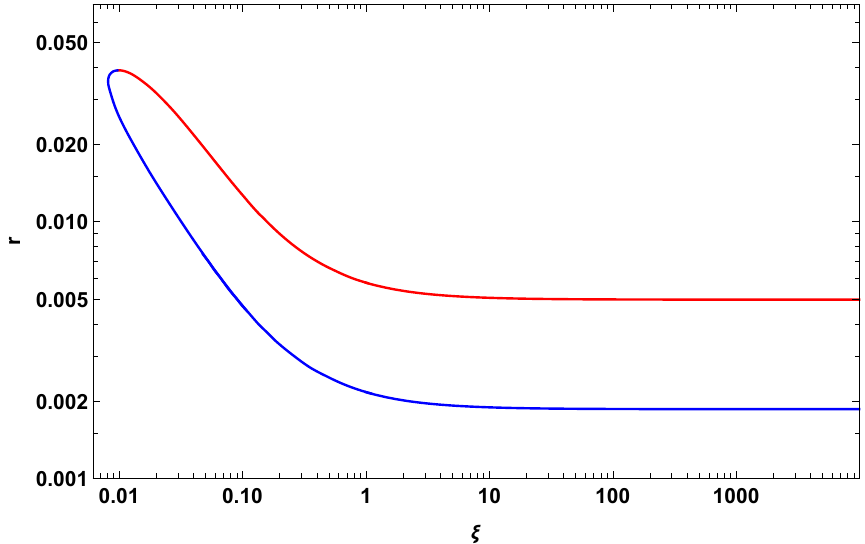}
\end{subfigure}
\caption{Shown are the 95\% C.L. contours of $\lambda$ vs. $\xi$ (left) and $r$ vs. $\xi$ (right). The red and blue curves indicate the redder and bluer spectrum of $n_{s}$.}
\label{fig:r_vs_xi_and_lambda_vs_xi_2hdm}
\end{figure}

\begin{figure}[htb]
\centering
\includegraphics[width=1\linewidth]{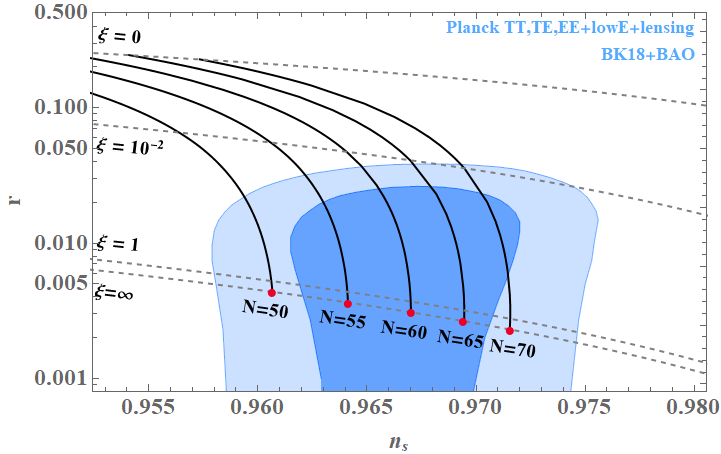}
\caption{Shown are the 95\% and 68\% C.L. contours of $r$ vs. $n_{s}$ (blue) and the inflationary predictions by 2hdSMASH for 2HDM-inflation (solid black). The dashed curves correspond to the isocontours of constant $\xi$. The main inflationary prediction for the 2HDM are given by $\xi=\infty$ and $N\in\left[50,70\right]$ (red dot).}
\label{fig:ns_r_2hdm}
\end{figure}

In this section we will introduce THI by deriving its field directions, its inflationary conditions and its slow-roll potential. Inflation in the 2HDM model has already been discussed in numerous Refs. \cite{Gong:2012ri,Nakayama:2015pba,Choubey:2017hsq,Modak:2020fij}. 
In order to decouple the 2HDM-field directions from the PQ-direction we implement a hierarchy of non-minimal couplings, i.e. $\xi_{1,2}\gg\xi_{S}$. The effective scalar potential is thus given by 
\begin{align}
\label{VEpot_quartic_angles2}
&\tilde{V}_{\text{quartic}}(\vartheta,\gamma)\simeq \\ &M_{p}^{4}~\dfrac{t_{\gamma}^4 \left(\lambda_{1}+\lambda_{2} t_{\vartheta}^4+2 \lambda_{34} t_{\vartheta}^2\right)
+2 t_{\gamma}^2 \left(t_{\vartheta}^2+1\right) \left(\lambda_{1S}+\lambda_{2S} t_{\vartheta}^2\right)
+\lambda_{S} \left(t_{\vartheta}^2+1\right)^2}{8 t_{\gamma}^{4}\left(\xi_{1}+\xi_{2}t_{\vartheta}^{2}\right)^2} \,.
\nonumber
\end{align}
The Jacobian from eq. \eqref{eq:jacobian_Vpot_angles} for the 2HDM-field directions is determined by $\gamma_{\text{THI}}$:
\begin{align}
J\left(\vartheta,\gamma_{\text{THI}}\right)&=
\left.\begin{pmatrix}
\dfrac{\partial \tilde{V}_{\text{quartic}}(\vartheta,\gamma)}{\partial \vartheta} & \dfrac{\partial \tilde{V}_{\text{quartic}}(\vartheta,\gamma)}{\partial \gamma}
\end{pmatrix}^{T}\right|_{\gamma=\gamma_{\text{THI}}}\\
&=
\begin{pmatrix}
0 & \dfrac{t_{\vartheta} \left(t_{\vartheta}^2+1\right) \left(-\lambda_{1} \xi_{2}+\lambda_{34} \xi_{1}+t_{\vartheta}^2 (\lambda_{2} \xi_{1}-\lambda_{34} \xi_{2})\right)}{2 \left(\xi_{1}+\xi_{2} t_{\vartheta}^2\right)^3}
\end{pmatrix}^{T}\,,\notag
\end{align} 
for which we acquire via $J_{2}\left(\vartheta,\gamma_{\text{THI}}\right)=0$ the extrema $\vartheta_{\text{THI}}$:
\begin{align}
\vartheta_{\text{THI}}=
\begin{cases}
\vartheta_{h_{1}}=0 \\
\vartheta_{h_{2}}=\dfrac{\pi}{2}\\
\vartheta_{h_{12}}=\arctan\left(\sqrt{\dfrac{\lambda_{34}\xi_{1}-\lambda_{1}\xi_{2}}{\lambda_{34}\xi_{2}-\lambda_{2}\xi_{1}}}\right)
\end{cases}\,.\label{eq:theta_THI}
\end{align} 
The corresponding inflationary vacuum energies are thus given by \cite{Gong:2012ri}
\begin{align}
\tilde{V}_{0}^{\text{THI}}=
\begin{cases}
\tilde{V}_{0}^{h_{1}}\simeq \frac{\lambda_{1}}{8\xi_{1}^{2}}\\
\tilde{V}_{0}^{h_{2}}\simeq \frac{\lambda_{2}}{8\xi_{2}^{2}}\\
\tilde{V}_{0}^{h_{12}}\simeq \frac{\lambda_{1}\lambda_{2}-\lambda_{34}^{2}}{8\left(\lambda_{1}\xi_{2}^{2}+\lambda_{2}\xi_{1}^{2}-2\lambda_{34}\xi_{1}\xi_{2}\right)}
\end{cases}
\end{align}
with quartic couplings $\lambda_{i}$
\begin{align}
\text{2HDM-}h_{1}:\, \lambda_{1}~~~,~~~\text{2HDM-}h_{2}:\, \lambda_{2}~~~,~~~\text{2HDM-}h_{12}:\, \lambda_{12}\equiv \lambda_{1}\lambda_{2}-\lambda_{34}^{2}\,.
\end{align}
The critical points $\left\lbrace\gamma_{\text{THI}},\vartheta_{\text{THI}}\right\rbrace$ of eq. \eqref{eq:theta_THI} correspond to the three field space directions, i.e. $h_{1}$, $h_{2}$ and $h_{12}$. The hessian matrix in two dimensional space is required
\begin{align}
H\left(\vartheta_{\text{THI}},\gamma_{\text{THI}}\right)=\left.
\begin{pmatrix}
\dfrac{\partial^{2}\tilde{V}_{\text{quartic}}(\vartheta,\gamma) }{\partial \gamma^{2}} & \dfrac{\partial^{2}\tilde{V}_{\text{quartic}}(\vartheta,\gamma) }{\partial \gamma\partial \vartheta} \\ \dfrac{\partial^{2}\tilde{V}_{\text{quartic}}(\vartheta,\gamma) }{\partial \gamma \partial \vartheta} & \dfrac{\partial^{2}\tilde{V}_{\text{quartic}}(\vartheta,\gamma) }{\partial \vartheta^{2}}
\end{pmatrix}
\right|_{\substack{\gamma=\gamma_{\text{THI}}\\ \vartheta=\vartheta_{\text{THI}}}}
\end{align}
in order to obtain the sufficient minimum conditions
\begin{align}
&\det H\left(\vartheta_{\text{THI}},\gamma_{\text{THI}}\right)\geq 0\,,\label{eq:minima_conditions}\\
&H_{1,1}\left(\vartheta_{\text{THI}},\gamma_{\text{THI}}\right)\geq 0 ~~\text{or}~~ H_{2,2}\left(\vartheta_{\text{THI}},\gamma_{\text{THI}}\right)\geq 0\notag \,.
\end{align}
These are calculated to be:
\begin{align*}
&\textbf{2HDM-$\left(h_{1}\right)$:}\\
&\kappa_{1}\equiv\lambda_{34}\xi_{1}-\lambda_{1}\xi_{2}>0 \,,\\
&\kappa_{1s}\equiv \lambda_{1S}>0\,.\\
&\\
&\textbf{2HDM-$\left(h_{2}\right)$:}\\
&\kappa_{2}\equiv\lambda_{34}\xi_{2}-\lambda_{2}\xi_{1}>0\,,\\
&\kappa_{2s}\equiv \lambda_{2S}>0\,.\\
&\\
&\textbf{2HDM-$\left(h_{12}\right)$:}\\
&-\kappa_{1}\kappa_{2}\left(\kappa_{1}\xi_{2}+\kappa_{2}\xi_{1}\right)>0 \,,\\
&\kappa_{1} \kappa_{2} \left(\kappa_{1}+\kappa_{2}\right) \left(\kappa_{1} \xi_{2}+\kappa_{2} \xi_{1}\right) \left(\kappa_{1} \lambda_{2s}+\kappa_{2} \lambda_{1S}\right)>0\,.
\end{align*}
We note that these conditions simplify drastically for tiny portal couplings and reproduce the known results for pure 2HDM inflation \cite{Gong:2012ri} where $\kappa_{1s}=\kappa_{2s}\simeq 0$. As discussed in section \ref{sec:inflation} this is regarded as the limit from case (I) to case (II). For the remainder, we keep the discussion on the minimum conditions for 2HDM-inflation as general as possible and consider case (I) in order to quantify the results in comparison to PQI and PQTHI.\\
\\
These minimum conditions, however, just describe a part of the inflationary conditions. Since inflation proceeds along a valley (minimum) of a given field space direction it requires that orthogonal field space directions are ridges (maxima). Therefore, the minimum conditions need to be supplemented by the maximum conditions of the other field space directions in order to fully describe the inflationary conditions. This applies, in particular to the 2HDM-$h_{1}$ and 2HDM-$h_{2}$ direction. For the 2HDM-$h_{12}$ direction this is already an automatic feature since this field direction is described by a mixture of both field directions and thus requires each individual field direction to be a ridge. However, the minimum conditions for the 2HDM-$h_{12}$ direction needs more refining. Effectively, the first one of the two conditions states
\begin{align}
\kappa_{1}\leq 0~~~,~~~\kappa_{2}\leq 0 \,.
\end{align}
Therefore, the second condition for $\xi_{1,2}>0$ reads
\begin{align}
\kappa_{1} \lambda_{2s}+\kappa_{2} \lambda_{1S}>0\,,
\label{eq:2hdm_h12_last_factor_condition}
\end{align}
which is only true for 
\begin{align}
\kappa_{1s}\equiv\lambda_{1S}<0~~~,~~~\kappa_{2s}\equiv\lambda_{2S}<0\,.
\end{align}
Moreover, there are kinetic mixing terms for the 2HDM-$h_{12}$ direction. These are quantified by
\begin{align}
K=\frac{6\xi_{1} \xi_{2}}{\Omega^{2}} \frac{h_{1} h_{2}}{M_{p}^2}\,,
\end{align}
where we avoid in the field space metric by assuming a hierarchy for the non-minimal couplings, i.e. $\xi_i\gg\xi_j$. The inflaton field $\phi$ proceeds along the mixed direction composed by $h_1$ and $h_2$ but is parametrically close to either $h_1$ or $h_2$. Both options are equivalent since either self-coupling, i.e. $\lambda_1$ and $\lambda_2$, is of $\mathcal{O}(1)$ and allows for the same predictions. We therefore choose without loss of generality to make the 2HDM-$h_{12}$-direction parametrically close to the $h_2$ direction\footnote{The same discussion applies to the 2HDM-$h_{12}$ direction being parametrically close to $h_1$.}. Hence, we assume a hierarchy for the non-minimal couplings, i.e. $\xi_2\gg\xi_1$. Given this hierarchy, the inflationary conditions for the 2HDM-$h_{12}$ direction simplify
\begin{align}
\kappa_{1s,2s,2}\simeq\lambda_{1S,2S,34}\leq 0~~\wedge~~\kappa_{1}\simeq\lambda_{1}\geq 0\,.
\end{align}
The field space metric of \eqref{eq:field_space} reduces for all 2HDM-directions to a two dimensional field space and becomes quite simple under the above assumptions for the 2HDM-$h_{12}$ direction
\begin{align}
\mathcal{G}_{ij}^{h_{12}}\simeq
\dfrac{b}{\Omega_{12}^{2}}
\begin{pmatrix}
 1  & 0   \\
  0 & \frac{\Omega_{12}^{2}+6\xi_{2}^{2}\frac{\phi^{2}}{M_{p}^{2}}}{\Omega_{12}^{2}}
\end{pmatrix}\,,
\end{align}
where $b$ is the mixing parameter given by
\begin{align}
b\equiv 1+\left|\frac{\lambda_{34}}{\lambda_{1}}\right|\,,
\end{align}
which is determined via  $\cos^{2}\vartheta_{h_{12}}=1-b^{-1}$. The frame function $\Omega_{12}^{2}$ for the mixed direction becomes
\begin{align}
\Omega_{12}^{2}=b+\xi_{2}\dfrac{\phi^{2}}{M_{p}^{2}}\,,
\end{align}
which reduces to the frame function for the 2HDM-$h_{2}$ direction when $\lambda_{34}\to 0$. This would be the limit where the 2HDM becomes an effective SM. However, we demand $\lambda_{34}\neq 0$ since it will be useful for RG-analysis (see section \ref{sec:infl_implications_portal_couplings}). In order to justify our approach for describing 2HDM-$h_{12}$-inflation in an effective single field regime, we need to compute the instantaneous  masses of $h_1$ and $h_2$
\begin{align}
\left. m_{h_{i}}^{2}\right|_{\substack{ \vartheta=\vartheta_{h_{12}}\\ \gamma=\gamma_{\text{THI}}}}&\simeq \left. \mathcal{G}^{h_i h_i}\partial_{h_i}^{2}\tilde{V}_{\text{quartic}}\right|_{\substack{ \vartheta=\vartheta_{h_{12}}\\ \gamma=\gamma_{\text{THI}}}}
\simeq
\begin{cases}
-\frac{\lambda_{34}}{\xi_{2}} & \left(h_{1}\right)\\
~\\
\frac{\lambda_{34}^{2}}{6\lambda_{1}\xi_{2}^{2}} & \left(h_{2}\right)
\end{cases}\,.
\end{align}
To estimate whether the orthogonal directions are heavy, i.e. stabilized, or light, i.e. dynamical, we relate those masses to the Hubble rate $\mathcal{H}^{2}\approx \tilde{V}/3M_{p}^{2}$ 
\begin{align}
\dfrac{m_{h_{i}}^{2}}{\mathcal{H}^{2}}\simeq
\begin{cases}
-\frac{24 \lambda_{1} \lambda_{34} \xi_{2}}{\lambda_{1} \lambda_{2}-\lambda_{34}^2}\gtrsim 1 & \left(h_{1}\right)\\
~\\
\frac{4 \lambda_{34}^2}{\lambda_{1} \lambda_{2}-\lambda_{34}^2}\lesssim 1 & \left(h_2\right)
\end{cases}\,,
\end{align}
which shows that $h_1$ is stabilized for $\xi_{2}\gg 1$ and $h_2$ is dynamical for sufficiently small $\lambda_{34}$, i.e. $\lambda_{1}\lambda_{2}\geq 5\lambda_{34}^{2}$.\\
\begin{figure}[htb!]
\centering
\begin{subfigure}{.5\textwidth}
  \centering
  \includegraphics[width=.9\linewidth]{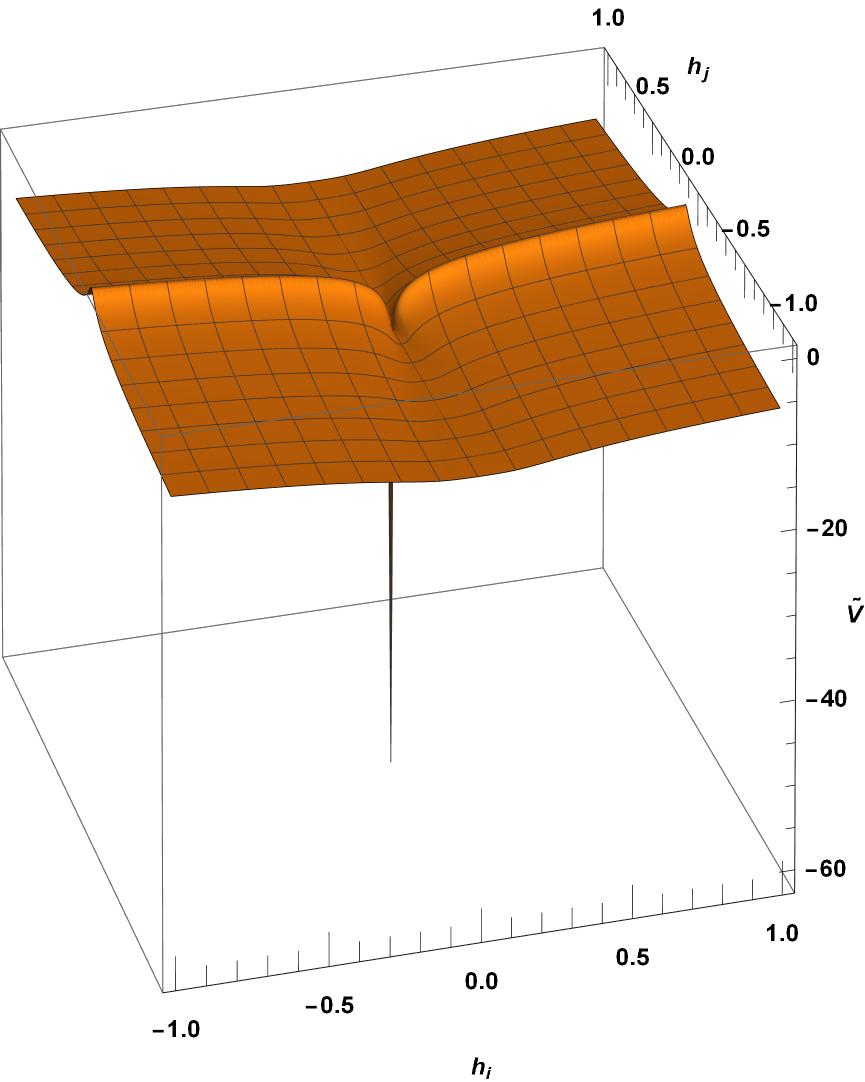}
\end{subfigure}%
\begin{subfigure}{.5\textwidth}
  \centering
  \includegraphics[width=.9\linewidth]{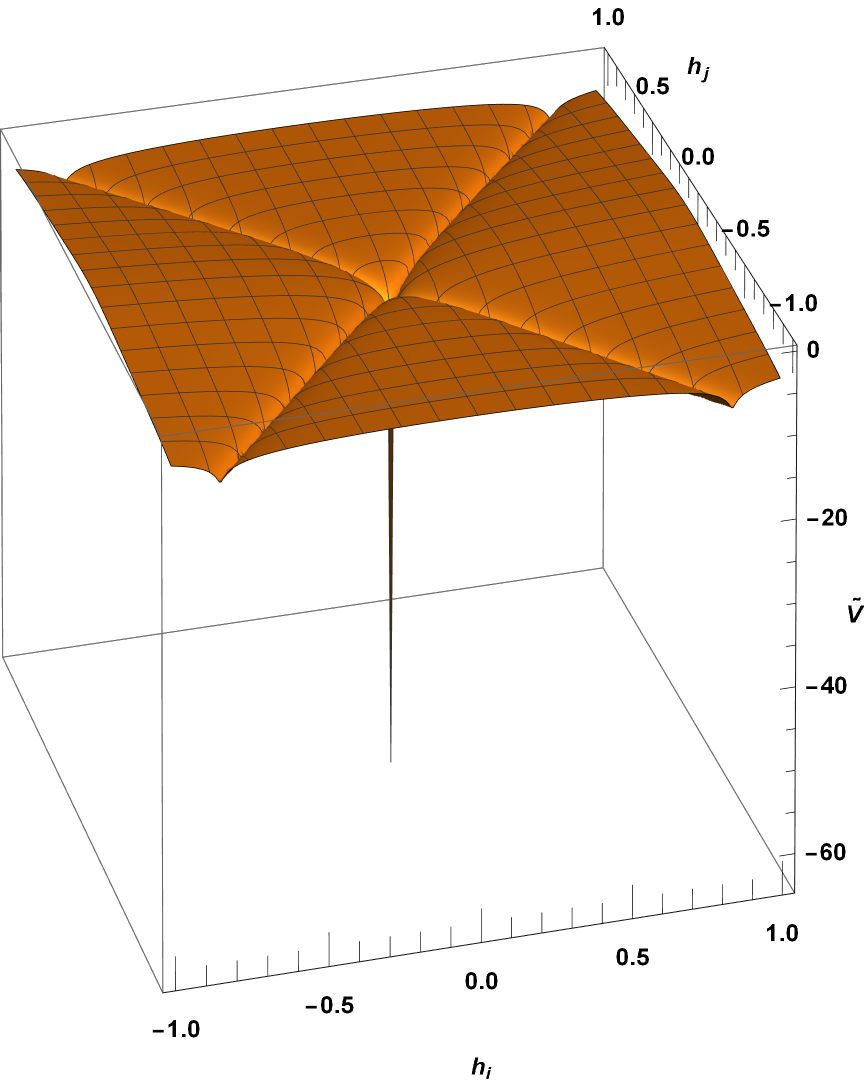}
\end{subfigure}
\caption{Decadic log Einstein frame scalar potential depicting the 2HDM-field directions in 2hdSMASH shown in units of $M_p$ as a function of the two Higgs fields $h_1$ and $h_2$ for pure single field directions (left) and mixed single field direction (right).}
\label{fig:2hdm_pots}
\end{figure}
\\
There are three inflationary trajectories where effective single field inflation  proceeds with the inflaton $\phi$ which can be seen in figure \ref{fig:2hdm_pots}. Since $\phi$ is a non-canonical field we introduce a canonical normalization in the $h_{1}$- and $h_{2}$-field directions via
\begin{align}
&\Omega^{2}\dfrac{d\chi_{i}}{d\phi_i}=\sqrt{\Omega^{2}+6\xi_{i}^{2}\dfrac{\phi^{2}}{M_{p}^{2}}}~~~\text{for }i=1,2\,.
\label{eq:2hdm_chi_i}
\end{align}
The canonically normalized 2HDM-$h_{12}$-field direction is determined by a mixture of $h_1$ and $h_2$ which is controlled by the mixing parameter $b$ and is given by
\begin{align}
&\Omega_{12}^{2}\dfrac{d\chi_{12}}{d\phi}=\sqrt{b\left(\Omega_{12}^{2}+6\xi_{2}^{2}\dfrac{\phi^{2}}{M_{\text{P}}^{2}}\right)}\,.
\label{eq:2hdm_chi_ij}
\end{align}
This allows us to determine the effective single-field slow-roll potential with canonically normalized inflaton field $\chi_{j}$, for $j=1,2,12$:
\begin{align}
\tilde{V}_{j}^{\text{THI}}(\phi)=\dfrac{\lambda^{\text{THI}}_{j}~\phi^{4}(\chi_{j})}{8\left(1+\xi_{j}\frac{\phi^{2}(\chi_{j})}{M_{P}^{2}}\right)^{2}}\,.
\end{align}
The conditions on the coupling parameters for inflation in each 2HDM field direction along with the corresponding effective single field slow-roll potential are summarized in Table \ref{tab:inflation_cond_pure_Higgs_inflation}\\
\\
\begin{table}[htb!]
\begin{center}
\scalebox{0.70}{%
\begin{tabular}{ |c||c|c|c| }
 \hline
 inflation along & Potential \eqref{VEpot_quartic_angles2} minimized at & Inflationary conditions & Einstein frame slow roll potential  \\
 \hline \hline
$h_{1}$& \begin{tabular}{@{}c@{}}$\gamma_{0} =\dfrac{\pi}{2}$\\~\\ $\vartheta_{0}=0$ \end{tabular}&  \begin{tabular}{@{}c@{}}\\ $\kappa_{1}\geq 0$~,~ $\kappa_{1s}\geq 0$\\~\\
$\kappa_{2}\leq 0$~,~ $\kappa_{2s}\leq 0$\\~\end{tabular}
& $\dfrac{\lambda_{1}}{8}\phi^4\left( 1 + \xi_1\frac{\phi^2}{M_P^2}\right)^{-2}$ \\ 
\hline 
$h_{2}$& \begin{tabular}{@{}c@{}}$\gamma_{0} =\dfrac{\pi}{2}$\\~\\ $\vartheta_{0}=\dfrac{\pi}{2}$ \end{tabular}
&  \begin{tabular}{@{}c@{}}\\
$\kappa_{2}\geq 0$~,~ $\kappa_{2s}\geq 0$\\~\\
$\kappa_{1}\leq 0$~,~ $\kappa_{1s}\leq 0$\\~\end{tabular}& $\dfrac{\lambda_{2}}{8}\phi^4\left( 1 + \xi_2\frac{\phi^2}{M_P^2}\right)^{-2}$ \\ 
\hline 
$h_{12}$&\begin{tabular}{@{}c@{}}$\gamma_{0} =\dfrac{\pi}{2}$\\~\\ $\vartheta_{0}=\arctan\left(\sqrt{\frac{\kappa_{1}}{\kappa_{2}}}\right)$ \end{tabular}
&  \begin{tabular}{@{}c@{}}\\ $\kappa_{1}\leq 0$~,~$\kappa_{2}\leq 0$\\~\\$\kappa_{1s}\leq 0$~,~$\kappa_{2s}\leq 0$\\~
   \end{tabular}
& \begin{tabular}{@{}c@{}}\\
 $\dfrac{\lambda_{12}}{8}\phi^4\left( 1 +\xi_{2}\frac{\phi^2}{M_P^2}\right)^{-2}$ \\~
\end{tabular}\\
\hline   
\end{tabular}}
\end{center}
\captionsetup{justification=centering}
\caption{Conditions \& characteristics for $h_1$-, $h_2$- and $h_{12}$-inflation in 2hdSMASH.}
\label{tab:inflation_cond_pure_Higgs_inflation}
\end{table}
Inflation occurs when the Hubble radius decreases at which point the potential energy density dominates the universe. This process is quantified by the slow-roll parameters demanding that the inflaton rolls down the potential well slowly. In slow-roll approximation, these parameters are expressed as
\begin{align}
\epsilon_V= \dfrac{M_{p}^{2}}{2}\left(\dfrac{\tilde{V}'(\chi)}{\tilde{V}(\chi)}\right)^{2}~~,~~\eta_V= M_{p}^{2} \dfrac{\tilde{V}''(\chi)}{\tilde{V}(\chi)}
\end{align}
where primes denote derivatives w.r.t. $\chi$ and are used to calculate the inflationary observables created during slow-roll inflation. Due to this prolonged stage of slow-roll inflation, the inflaton produces quantum fluctuations along the field trajectory. These quantum fluctuations are transferred to scalar metric perturbations in form of density waves $P_{s}(k)$ and tensor metric perturbations in form of gravitational waves $P_{t}(k)$. We give $P_{s}(k)$ and $P_{t}(k)$ as (see Ref.\cite{Matlis:2022iwr})
\begin{align}
P_{s}(k)=A_{s}\left(\dfrac{k}{k_{\ast}}\right)^{n_{s}-1+\dots}~~,~~P_{t}(k)=A_{t}\left(\dfrac{k}{k_{\ast}}\right)^{n_{t}+\dots}
\end{align}
with $n_{s}$ as the spectral scalar index and $n_{t}$ as the tensor spectral index, quantifying the deviation from scale-invariance of the power spectra. The scalar- and tensor metric perturbations are reduced to the scalar perturbation amplitude $A_s$ and to the tensor perturbation amplitude $A_t$ for $k=k_{\ast}$ when the modes exit the horizon at the pivot scale $k_{\ast}$. Moreover, the gravitational amplitude $A_t$ is normalized by $A_s$ to acquire the tensor-to-scalar ratio $r$ which relates gravitational waves to CMB measurements. Additionally, the spectral tensor index $n_t$ is related to the tensor-scalar ratio $r$ for single field inflation via
\begin{align}
n_{t}=-\dfrac{r}{8}\,,
\end{align}
which leaves us with the relevant inflationary observables $A_s$, $n_s$ and $r$ in slow-roll approximation for single field inflation:
\begin{align}
A_{s}=\dfrac{1}{24\pi^{2}M_{Pl}^{4}}\dfrac{\tilde{V}}{\epsilon_V}~~,~~n_{s}=1-6\epsilon_V+2\eta_{V}~~,~~r=16\epsilon_V\,.
\end{align}
As discussed in Refs. \cite{Gong:2012ri,Nakayama:2015pba,Choubey:2017hsq,Modak:2020fij}, we consider large non-minimal couplings in order to satisfy the scalar density perturbations of the CMB given by PLANCK \cite{Planck:2018vyg,Planck:2018jri}. We will show that this is required for the 2HDM-field directions by calculating the expression for $A_s$ and constraining the non-minimal couplings $\xi_{1,2}$ with CMB measurements by PLANCK \cite{Planck:2018vyg,Planck:2018jri}. Moreover, we will calculate $n_s$ and $r_s$ in the large non-minimal coupling limit and relate their expressions to the number of e-folds and make the well-know predictions for Higgs inflation \cite{Bezrukov:2007ep}.\\
\\
In the large field limit, i.e. $\xi_{1,2}\gg 1$, we can approximate $\phi(\chi)$ and its inverse $\chi(\phi)$ by solving equations \eqref{eq:2hdm_chi_i} and \eqref{eq:2hdm_chi_ij} 
\begin{align}
&\phi_i(\chi_i)\simeq \dfrac{M_{p}}{\sqrt{\xi_i}}\sqrt{\exp\left(\sqrt{\dfrac{2}{3}}\dfrac{\chi_i}{M_{p}}\right)-b}\,,\\
&\chi(\phi_i)\simeq \sqrt{\dfrac{3}{2}}M_{p}\log\left(b~M_{p}^{2}+\xi_i\frac{\phi_i^{2}}{M_{p}^2}\right)\,,
\label{phi_chi_large_xi}
\end{align}
from which we acquire with $\phi(\chi)$ the following slow-roll potential
\begin{align}
\tilde{V}(\chi_i)\simeq\dfrac{\lambda}{8\xi^{2}}M^{4}_{p}\left(1-b\exp\left(-\sqrt{\dfrac{2}{3}}\dfrac{\chi}{M_{p}}\right) \right)^{2}
\end{align}
with mixing parameter $b$ which quantifies whether inflation proceeds in a mixed direction ($b\neq 1$) or in a non-mixed direction ($b=1$). By taking the first and second partial derivatives of $\tilde{V}$ allows us to compute the slow-roll parameters and thus the inflationary observables $A_s$, $n_s$ and $r$ in the large $\xi$ limit \cite{Ballesteros:2016xej}
\begin{align}
&A_{s}\simeq \dfrac{\lambda_i}{128\pi^{2}\xi_i^{2}~b}\dfrac{(1-b~x_i)^{4}}{x_i^2}~~,~~n_{s}\simeq  1-\dfrac{8~b}{3}\dfrac{b+x_i}{\left(b-x_i\right)^{2}}~~,~~\simeq \dfrac{64~b^{2}}{3\left(x_i-b\right)^{2}}\label{eq:As_ns_r_large_xi2}
\end{align}
with $x_i\equiv \exp\left(\sqrt{2/3}\chi_i/M_p\right)$. This reduces for $b=1$ to 
\begin{align}
&A_{s}\simeq \dfrac{3 \lambda  \sinh^4\left(\frac{\chi }{\sqrt{6} M_{p}}\right)}{2 \xi ^2}~~,~~r\simeq \dfrac{64~b^{2}}{3\left(x_i-b\right)^{2}}\,,\\
&n_{s}\simeq  \frac{4}{3} \left(\coth \left(\frac{\chi }{\sqrt{6} M_{p}}\right)- \text{csch}^2\left(\frac{\chi}{\sqrt{6} M_{p}}\right)-1\right)-\frac{1}{3}\,.
\end{align}
These quantities can be constrained by CMB measurements. Current constraints on $r$, $A_{s}$ and $n_{s}$ are given by PLANCK 2018 data \cite{Planck:2018vyg},\cite{Planck:2018jri} which were recently updated by Ref. \cite{BICEP:2021xfz}:
\begin{align}
&r_{0.002}<0.0387985 ~~\text{  ($95\%$C.L., TT,TE,EE+lowE+lensing+BK18+BAO)}\,,\\
&A_{s}=(2.105\pm 0.030)\times 10^{-9} ~~\text{  ($68\%$C.L., TT,TE,EE+lowE+lensing+BAO)}\,,\\
&n_{s}=0.9665\pm 0.0038 ~~\text{  ($68\%$C.L., TT,TE,EE+lowE+lensing+BAO)}\,.
\end{align}
The scalar perturbation amplitude $A_s$ can be used to relate the non-minimal coupling $\xi_i$ to the self-coupling $\lambda_i$. From eq. \eqref{eq:As_ns_r_large_xi2} we can approximate $A_s\propto \lambda_i/\xi_i^{2}$ for which we obtain an approximate relation
\begin{align}
&A_s\propto\dfrac{\lambda_i}{\xi_i^{2}}\propto 10^{-9}\\
\Rightarrow~~&\xi_i\propto \sqrt{\lambda_i}10^{-5}\,.
\label{eq:lambda_xi_relation}
\end{align}
We show this $\lambda$-$\xi$ relation in figure \ref{fig:r_vs_xi_and_lambda_vs_xi_2hdm} (left), which we acquired numerically. This relation is furthermore constrained by $n_s$ and $r$: the non-minimal coupling $\xi$ is constrained by $r$ as shown in figure \ref{fig:r_vs_xi_and_lambda_vs_xi_2hdm} (right) and $r$ is constrained by $n_s$ with PLANCK 2018 \cite{Planck:2018vyg},\cite{Planck:2018jri} and BICEP 2021 \cite{BICEP:2021xfz} data. The blue and red curves correspond to a bluer and redder spectrum of $n_s$, respectively.
The 2HDM quartic couplings are of order $\lambda\sim \mathcal{O}(1)$ at the Planck scale. This is required by RG running and Higgs phenomenology to accommodate a $125$ GeV Higgs mass at low energies. Later, we will see that the size of 2HDM quartic couplings cannot be changed by considering different field space directions since the RGEs are strongly coupled. From figure \ref{fig:r_vs_xi_and_lambda_vs_xi_2hdm} (left) and from eq. \eqref{eq:lambda_xi_relation} we can see that the non-minimal coupling is of the order $\xi\sim 10^{4}$. \\
\\ 
In order to determine the inflationary observables $A_s$, $n_s$ and $r$ in the large $\xi$ limit from equation \eqref{eq:As_ns_r_large_xi2} properly, we require $\chi_I$ and $\chi_E$ at horizon crossing and at the end of inflation, respectively. Therefore, we need to compute $\chi(N)$ to determine the number of e-folds required for successful inflation. The number of e-folds during inflation between some initial time $t_I$ and the time at the end of inflation is given by
\begin{align}
N=\int_{t_{I}}^{t_{\text{end}}}\mathcal{H}(t) dt\simeq \int_{\chi_{I}}^{\chi_{\text{end}}}\dfrac{d\chi}{\sqrt{2\epsilon}}
\label{eq:N_int_form}
\end{align}
and is approximated to 
\begin{align}
N\simeq \dfrac{1+6\xi_i}{8M^{2}_{p}}\left(\phi_{I}^{2}-\phi_{E}^{2}\right)-\dfrac{3b}{4}\log\left(\dfrac{b~M^{2}_{p}+\xi_{i}\phi^{2}_{I}}{b~M^{2}_{p}+\xi_{i}\phi^{2}_{E}}\right)\,.
\end{align}
However, the exact number of e-folds can be obtained by solving the Klein-Gordon equation for the canonically normalized field $\chi$. The Klein-Gordon equation for $\chi$ with regard to $N$ can be derived to be
\begin{align}
\dfrac{d^{2}\chi}{dN^{2}}+3\dfrac{d\chi}{dN}-\dfrac{1}{2 M_{p}^{2}}\left(\dfrac{d\chi}{dN}\right)^{3}+\sqrt{2\epsilon}\left(3M_{p}-\dfrac{1}{2M_{p}}\left(\dfrac{d\chi}{dN}\right)^{2}\right)=0 \,,
\end{align}
which is evaluated until the condition where $\epsilon=1$ marks the end of inflation. This differential equation, however, can only be computed numerically and requires appropriate initial conditions. These initial conditions can be estimated by determining $\chi(\Delta N)$
\begin{align}
&\chi(\Delta N)\simeq\sqrt{\dfrac{3}{2}} M_{p} \log \left(\dfrac{b^2(1+ 6\xi)+8 \Delta N \xi }{b (1+6 \xi)}\right)\approx 5 M_{p}\,,\\
&\chi'(\Delta N)\simeq M_{p} \dfrac{4 \sqrt{6} M_{p}}{\Delta N \left(b^2(1+6\xi)+8 \Delta N \xi \right)}\approx 0.02 M_{p} \,.
\end{align}
Furthermore, we utilize the e-fold dependencies of $A_s$, $n_s$ and $r$. This gives us a better understanding of the parameter scale. In the large $\xi$ limit, these quantities are given by
\begin{align}
n_{s}\left(\Delta N\right)\simeq 1-\dfrac{ 2 }{\Delta N }~~,~~r\simeq \dfrac{12 b }{\Delta N^{2} }~~,~~A_{s}(\Delta N)\simeq \dfrac{\lambda \Delta N^2 }{144 \pi^2 b \xi^2 }\,,
\end{align}
where in the non-mixed direction they are given by
\begin{align}
n_{s}\left(\Delta N\right)\simeq 1-\dfrac{2}{\Delta N}~~,~~ r\simeq \dfrac{12}{\Delta N^{2} }~~,~~A_{s}(\Delta N)\simeq \dfrac{\Delta N^2}{144 \pi^2}\dfrac{\lambda}{\xi^2}\,.
\end{align} 
These inflationary predictions resemble the results of Starobinsky-type inflation \cite{Starobinsky:1980te,Mukhanov:1981xt,Starobinsky:1982ee,Gorbunov:2012ns} and Higgs inflation \cite{Bezrukov:2007ep}. As in Higgs inflation and Starobinsky's $R^2$-inflation, the spectral index $n_s$ and the tensor-to-scalar ratio $r$ are solely dependent on $\Delta N$ and independent of $\xi$. In figure \ref{fig:ns_r_2hdm} we plot the $r$ vs. $n_s$ predictions for 2HDM inflation in the large $\xi$ limit. The general $n_s$ and $r$ predictions are shown as black solid lines which go from low to large $\xi$ where the red dots at $\xi=\infty$ mark our $r$ vs. $n_s$ predictions. Each solid black line corresponds to a given number of e-fold which is in the range of $N\in \left[50,70\right]$. Furthermore, we show the isocontours of constant $\xi$ as gray dashed lines.
In fact, the $n_{s}$ and $r$ predictions are true for any chaotic inflationary model with a single field attractor. The predictions are constrained by the 95\% and 68\% C.L. contours of PLANCK 2018 data \cite{Planck:2018jri,Planck:2018vyg} and the BICEP update \cite{BICEP:2021xfz}. Furthermore, we note the remarkable $r$ vs $n_s$ prediction for $N=60$, which we will also encounter for PQI and PQTHI. 

As mentioned before, we require for THI non-minimal couplings of the order of $\xi\sim 10^{4}$. Various authors have discussed the negative impact of such a large non-minimal coupling on perturbative unitarity (e.g. Refs. \cite{Barbon:2009ya,Burgess:2010zq}). The claim is that the unitarity scale is below the inflationary scale, i.e.  $\Lambda_{U}=M_{P}/\xi_{1,2} \lesssim M_{P}/\sqrt{\xi_{1,2}}=\Lambda_{\text{Inf}}$ for $\xi_{1,2}\gg 1$, which puts predictiveness under scrutiny. Recently, this issue was addressed by Ref. \cite{2022JHEP...06..132K} in which they claim that the unitarity scale during inflation is actually higher than previously calculated and thus resolving this issue. However, the results of Ref. \cite{2022JHEP...06..132K} are beyond the scope of this work and we leave it to future explorations to investigate whether this applies to Higgs inflation in our model. For our considerations, large non-minimal couplings are unnatural since there isn't an explanation on how these large couplings were generated. The approach by the authors of Refs. \cite{Giudice:2010ka,Barbon:2015fla,Ballesteros:2016xej} alleviate the above concerns on predictive power by non-minimal couplings of the size $\xi_{i}\lesssim 1$ which implies $\lambda\lesssim 10^{-10}$. We adopt these considerations of Refs. \cite{Giudice:2010ka,Barbon:2015fla,Ballesteros:2016xej} which is according to our philosophy more natural. Non-minimal couplings with the size of $\xi_{i}\lesssim 1$ can be radiatively generated in the very early universe. Therefore, we require the inflaton's quartic self-coupling to be of order $\lambda\lesssim 10^{-10}$. However, the RGEs for $\lambda_{1}$ and $\lambda_{2}$ are dominated by couplings which are of $\mathcal{O}(1)$ as can be seen by eqs. \eqref{eq:l1_rge}-\eqref{eq:l2_rge}
\begin{align}
 \mathcal{D} \lambda_1 = &\ 
   \frac34 g_1^4 + \frac32 g_1^2g_2^2 + \frac94 g_2^4 -\lambda_1\left( 3 g_1^2 +9g_2^2 \right)
 + 12\lambda_1^2  \label{eq:l1_rge}\\
 &+ 4\lambda_3\lambda_4 + 4\lambda_3^2+ 2\lambda_4^2
 + 2\lambda_{1S}^2+ 12\lambda_1Y_b^2 - 12Y_b^4\notag\\ 
  &+4\lambda_1 Y_\tau^2 - 4Y_\tau^4
  + 4\lambda_2\text{Tr}\left(Y_\nu^\dagger Y_\nu\right) -4\text{Tr}\left(Y_\nu^\dagger Y_\nu Y_\nu^\dagger Y_\nu\right),\notag \\
\mathcal{D} \lambda_2 = &\ 
   \frac34 g_1^4 + \frac32 g_1^2g_2^2 + \frac94 g_2^4 -\lambda_2\left( 3 g_1^2 +9g_2^2 \right)
 + 12\lambda_2^2 \label{eq:l2_rge}\\
 &+ 4\lambda_3\lambda_4 + 4\lambda_3^2+ 2\lambda_4^2
 + 2\lambda_{2S}^2 + 12\lambda_2Y_t^2 - 12Y_t^4 \,.\notag
\end{align}
On the other hand, any scenario with $\lambda_{1,2,34}\sim 10^{-10}$ would endanger Higgs phenomenology since these small self-couplings could not reproduce a $125$ GeV Higgs. Therefore, we can safely neglect inflationary realizations where only 2HDM-fields are involved. This is the main reason to solely focus on PQI and PQTHI.     

\section{2hdSMASH at the Matching Scale}
\label{app:matching_scale}

In this section we show that the full high energy theory of 2hdSMASH reduces to a softly broken $U(1)$-symmetric $\nu$2HDM low-energy theory where the extra $U(1)_{L}$-symmetry can be associated with lepton number caused by a technically natural limit for $Y_N,Y_{\nu},\lambda_{12S}\to 0$, cf. \cite{Clarke:2015bea}. We consider the limit $\lambda_{1S,2S,12S}, Y_{N,\nu}\to 0$ corresponding to the enhanced Poincaré symmetry $\mathcal{G}_{P}^{\nu}\times\mathcal{G}_{P}^{\text{2HDM}}\times \mathcal{G}_{P}^{S}$ which protects the electroweak scale from large radiative corrections, cf. \cite{Foot:2013hna,Clarke:2015bea}. We will motivate these consideration by matching 2hdSMASH at the matching scale $m_{s}$ to its low-energy theory where we show that tiny portal couplings will prove to protect the electroweak scale. Therefore, we consider the equation of motion of $s$ at zero momentum at $m_{s}$:
\begin{align}
\dfrac{\partial V}{\partial s}=0 ~~\Rightarrow ~~& s^{2}=-\frac{h_{1}^2 \lambda_{1S}}{\lambda_{S}}+\frac{2 h_{1} h_{2} \lambda_{12S} }{\lambda_{S}}-\frac{h_{2}^2 \lambda_{2S}}{\lambda_{S}}-\frac{2 M_{SS}^{2}}{\lambda_{S}}\,.
\label{eq:zero_momentum_rho}
\end{align}
By considering now the scalar potential with $s$ at zero momentum, i.e. integrating $s$ out, we obtain the threshold corrected scalar potential:
\small{
\begin{align}
V&=\frac{h_{1}^2}{2}\left(M_{11}^{2}-\frac{\lambda_{1S} M_{SS}^{2}}{\lambda_{S}}\right)+\frac{h_{2}^2}{2} \left(M_{22}^{2}-\frac{\lambda_{2S} M_{SS}^{2}}{\lambda_{S}}\right)+\frac{\lambda_{12S} M_{SS}^{2}}{2\lambda_{S}}h_{1} h_{2} \left( \frac{h_{1}^2 \lambda_{1S}+h_{2}^2 \lambda_{2S}}{M_{SS}^{2}}+2\right)\label{eq:scalar_pot_with_threshold_corr}\\
&+\frac{h_{1}^4}{8} \left(\lambda_{1}-\frac{\lambda_{1S}^2}{\lambda_{S}}\right)+\frac{h_{2}^4}{8}\left(\lambda_{2}-\frac{\lambda_{2S}^2}{\lambda_{S}}\right)+\frac{h_{1}^2 h_{2}^2}{4}\left(\lambda_{34}-\frac{\lambda_{1S} \lambda_{2S}+2 \lambda_{12S}^2}{\lambda_{S}}\right)-\frac{M_{SS}^{2}}{2 \lambda_{S}}\,.\notag
\end{align}}
\normalsize
The third quadratic term of eq. \eqref{eq:scalar_pot_with_threshold_corr} constitutes a non-linear term given by the coupling $\lambda_{12S}$. By taking into account that $h_{1}^2 \lambda_{1S}+h_{2}^2 \lambda_{2S}\ll M_{SS}^{2}\sim m_{s}^{2}$ we can approximate the terms as follows:
\begin{align}
\frac{\lambda_{12S} M_{SS}^{2}}{2\lambda_{S}}h_{1} h_{2} \left( \frac{h_{1}^2 \lambda_{1S}+h_{2}^2 \lambda_{2S}}{M_{SS}^{2}}+2\right)\approx \frac{\lambda_{12S} M_{SS}^{2}}{\lambda_{S}}h_{1} h_{2}\,.
\end{align}
Therefore, the scalar potential can be approximated to: 
\small{
\begin{align}
V&\simeq\frac{h_{1}^2}{2}\left(M_{11}^{2}-\frac{\lambda_{1S} M_{SS}^{2}}{\lambda_{S}}\right)+\frac{h_{2}^2}{2} \left(M_{22}^{2}-\frac{\lambda_{2S} M_{SS}^{2}}{\lambda_{S}}\right)+\frac{\lambda_{12S} M_{SS}^{2}}{\lambda_{S}}h_{1} h_{2}\\
&+\frac{h_{1}^4}{8} \left(\lambda_{1}-\frac{\lambda_{1S}^2}{\lambda_{S}}\right)+\frac{h_{2}^4}{8}\left(\lambda_{2}-\frac{\lambda_{2S}^2}{\lambda_{S}}\right)+\frac{h_{1}^2 h_{2}^2}{4}\left(\lambda_{34}-\frac{\lambda_{1S} \lambda_{2S}+2 \lambda_{12S} ^2}{\lambda_{S}}\right)-\frac{M_{SS}^{4}}{2 \lambda_{S}}\,.\notag
\end{align}}
\normalsize
The corresponding threshold corrections for the matching of 2hdSMASH to its effective low energy theory can be read off:
\begin{align}
&\bar{M}_{11,22}^{2}=\left(M_{11,22}^{2}-\frac{\lambda_{1S,2S} M_{SS}^{2}}{\lambda_{S}}\right)\,,\\
&\bar{M}_{12}^{2}=2 \lambda_{12S} M_{SS}^{2}\,,\\
&\bar{\lambda}_{1,2}=\left(\lambda_{1,2}-\frac{\lambda_{1S,2S}^2}{\lambda_{S}}\right)\,,\\
&\bar{\lambda}_{34}=\left(\lambda_{34}-\frac{\lambda_{1s} \lambda_{2s}+2 \epsilon ^2}{\lambda_{s}}\right)\,,
\end{align}
which represents a softly broken $U(1)$-symmetric 2HDM. The tadpole equation for $M_{SS}^{2}$ is calculated by:
\begin{align}
\dfrac{\partial V}{\partial s}\overset{!}{=}0 ~~\Rightarrow~~ M_{SS}^{2}=\frac{1}{2} \left(-\lambda_{1S} v_{1}^2+2 v_{1} v_{2} \lambda_{12S} -\lambda_{2S} v_{2}^2-\lambda_{S} v_{S}^2\right)\,.
\end{align} 
The VEVs $v_{1,2,S}$ follow a hierarchy, namely  $v_{S}^2\gg v_{1,2}^{2}$. Hence, the tadpole equation $M_{SS}^{2}$ can be approximated as follows:
\begin{align}
M_{SS}^{2}=\frac{v_{S}^2}{2} \left(-\lambda_{1S} \frac{v_{1}^2}{v_{S}^2}+2 \frac{v_{1} v_{2}}{v_{S}^2} \lambda_{12S} -\lambda_{2S} \frac{v_{2}^2}{v_{S}^2}-\lambda_{S} \right)\overset{v_{S}^2\gg v_{1,2}^{2}}{\approx} -\frac{\lambda_{S} v_{S}^2}{2}\,.
\end{align}
Finally, the threshold corrected scalar potential of the effective low-energy theory is acquired and reads:
\small{
\begin{align}
V&\simeq\frac{h_{1}^2}{2} \left(M_{11}^{2}+\frac{\lambda_{1S} v_{S}^2}{2}\right)+\frac{h_{2}^2}{2}\left(M_{22}^{2}+\frac{\lambda_{2S} v_{S}^2}{2}\right)-\frac{h_{1} h_{2} \lambda_{12S} v_{S}^2}{2}\\
&+\frac{h_{1}^4}{8}\left(\lambda_{1}-\frac{\lambda_{1S}^2}{\lambda_{S}}\right)+\frac{h_{2}^4}{8} \left(\lambda_{2}-\frac{\lambda_{2S}^2}{\lambda_{S}}\right)+\frac{h_{1}^2 h_{2}^2}{4}  \left(\lambda_{34}-\frac{\lambda_{1S} \lambda_{2S}+2 \lambda_{12S}^2}{\lambda_{S}}\right)\,,\notag
\end{align}}
\normalsize
where the effective mass term $m_{s}^{2}v_{S}^2\propto \lambda_{S} v_{S}^4$ was omitted. The following coupling associations for the effective low-energy theory are thus given by:
\begin{align}
&m_{11,22}^{2}=\left(M_{11,22}^{2}+\frac{\lambda_{1S,2S} v_{S}^2}{2}\right),\notag\\
&m_{12}^{2}=\lambda_{12S} v_{S}^2,\notag\\
&\bar{\lambda}_{1,2}=\left(\lambda_{1,2}-\frac{\lambda_{1S,2S}^2}{\lambda_{S}}\right),\label{eq:coupling_association_2hdm_limit}\\
&\bar{\lambda}_{34}=\left(\lambda_{34}-\frac{\lambda_{1S} \lambda_{2S}+2 \lambda_{12S}^2}{\lambda_{S}}\right),\notag\\
&M_{SS}^{2}\simeq -\frac{\lambda_{S} v_{S}^2}{2},\notag
\end{align} 
where $M_{SS}^{2}$ is related to the aforementioned matching-scale via $m_{s}\simeq\sqrt{-2 M_{SS}^{2}}$.  The threshold corrections are negligible, i.e. $\lambda_{iS}^{2}/\lambda_{S}\ll 1$. Therefore, the effective low energy theory of a softly broken $U(1)$-symmetric 2HDM has approximately the following scalar potential $V^{\nu 2\text{HDM}}$:
\begin{align}
V^{\nu 2\text{HDM}}\simeq &m_{11}^{2}|\Phi_{1}|^{2}+m_{22}^{2}|\Phi_{2}|^{2}-m_{12}^{2}\left(\Phi_{1}^{\dagger} \Phi_{2}+\Phi_{2}^{\dagger} \Phi_{1}\right)\\
&+\frac{\lambda_{1}}{2}|\Phi_{1}|^{4}+\frac{\lambda_{2}}{2}|\Phi_{2}|^{4}+\lambda_{3}\left|\Phi_{1}\right|^{2}\left|\Phi_{2}\right|^{2}+\lambda_{4}\left(\Phi^{\dagger}_{1}\Phi_{2}\right)\left(\Phi^{\dagger}_{2}\Phi_{1}\right)\,.\notag
\end{align}

\section{Coleman-Weinberg Potential}
\label{app:coleman_weinberg}
The Coleman-Weinberg potential is an RG-improved potential which is given by, cf. Ref. \cite{Weinberg:1973am},
\begin{align}
V_{\text{CW}}(\phi_i)&=\dfrac{1}{64\pi^{2}}\\
&\times\left(\sum_{b} g_{b}m_{b}^{4}(\phi_i)\left[\log\left(\frac{m_{b}^{2}(\phi_i)}{\Lambda^{2}}\right)-\frac{3}{2}\right]-\sum_{f} g_{f}m_{f}^{4}(\phi_i)\left[\log\left(\frac{m_{f}^{2}(\phi_i)}{\Lambda^{2}}\right)-\frac{5}{2}\right]\right)\notag
\end{align}
where $g_{f/b}$ are the degrees and $m_{f/b}$ are the masses of fermions/bosons. The sum is performed over all scalars $S=\left\lbrace h_{1},h_{2},s,h_{1}^{+},h_{2}^{+},a_{1},a_{2},a_{S}\right\rbrace$, Fermions\footnote{We only consider the top quark and Majorana neutrino contributions since their masses are the largest amongst the fermions. However, we note that one has to include all fermions in order to give a complete description, which is beyond the scope of what we are considering.} $F=\left\lbrace t, N_i \right\rbrace$ and transverse (longitudinal) vectors $V_{T,L}=\left\lbrace Z_{T,L}, W^{\pm}_{T,L}\right\rbrace$ with the following degrees of freedom:
\begin{align}
&g_{Z_{T}}=2 ~~,~~g_{Z_{L}}=1 ~~,~~ g_{W^{\pm}_{T}}=4 ~~,~~g_{W^{\pm}_{L}}=2 ~~,~~ g_{t}=12 ~~,~~ g_{h_{1}}=g_{h_{2}}=g_{s}=1\\
&g_{h_{1}^{+}}=g_{h_{2}^{+}}=2~~,~~g_{a_{1}}=g_{a_{2}}=g_{a_{S}}=1~~,~~g_{N}=6\,.\notag
\end{align}
The masses of $S$, $F$ and $V$ are given by:
\begin{align}
&m_{Z}^{2}=\dfrac{1}{4}\left(h_{1}^{2}+h_{2}^{2}\right)\left(g_{1}^{2}+g_{2}^{2}\right)~~~,~~~ m_{W^{\pm}}^{2}=\dfrac{1}{4}\left(h_{1}^{2}+h_{2}^{2}\right)g_{1}^{2}~~~,~~~m_{t}^{2}=\dfrac{Y_{t}}{2}h_{2}^{2},\\
& m_{h_{1,2}}^{2}=\dfrac{3 h_{1,2}^2 \lambda_{1,2}+h_{2,1}^2 \lambda_{34}+s^2 \lambda_{1S,2S}}{2}~~~,~~~m_{s}^{2}=\dfrac{h_{1}^2 \lambda_{1S}-2 h_{1} h_{2} \lambda_{12S}+h_{2}^2 \lambda_{2S}+3 s^2 \lambda_{S}}{2},\notag\\
&m_{a_{1,2}}^{2}=\frac{1}{2} \left(h_{1,2}^2 \lambda_{1,2}+h_{2,1}^2 (\lambda_{3}+\lambda_{4})+h_{3}^2 \lambda_{1S}\right)~~,~~m_{h_{1,2}^{+}}^{2}=\frac{1}{2} \left(h_{1,2}^2 \lambda_{1,2}+h_{2,1}^2 \lambda_{3}+h_{3}^2 \lambda_{1S}\right)\,,\notag\\
&m_{a_s}^{2}=\frac{1}{2} \left(h_{1}^2 \lambda_{1S}+2 h_{1} h_{2} \lambda_{12S}+h_{2}^2 \lambda_{2S}+s^2 \lambda_{S}\right)\,.\notag
\end{align}


\renewcommand{\bibname}{References}
\bibliographystyle{unsrt}
\bibliography{ref1,ref}
\end{document}